\documentclass[12pt]{article}

\usepackage{epsf}
\usepackage{latexsym,euscript}
\usepackage[dvips]{graphicx}
\usepackage{amsmath}
\usepackage{amsfonts}
\usepackage{amssymb, epsfig}
\usepackage{float}
\usepackage{array}

\usepackage[dvipsnames]{xcolor}

\graphicspath{{figures/}}

\usepackage[utf8]{inputenc}

\numberwithin{equation}{section}

\usepackage{color,soul}

\allowdisplaybreaks

\DeclareMathOperator{\arcsinh}{arcsinh}

\begin{document}

\begin{center}
{\Large \textbf{One-dimensional QCD at finite density \\
		and its 't Hooft-Veneziano limit} }

\vspace*{0.3cm}

\vspace*{0.6cm}
\textbf{O.~Borisenko${}^{\rm a b}$\footnote{email: oleg@bitp.kyiv.ua},
V.~Chelnokov${}^{\rm c}$\footnote{email: chelnokov@itp.uni-frankfurt.de},
S.~Voloshyn${}^{\rm b}$\footnote{email: s.voloshyn@bitp.kyiv.ua},
P.~Yefanov${}^{\rm d}$\footnote{email: pavlo.yefanov@cvut.cz}}

\vspace*{0.5cm}

{\large \textit{${}^{\rm a}$ INFN Gruppo Collegato di Cosenza, Arcavacata di Rende, 87036 Cosenza, Italy}} \\ 
{\large \textit{${}^{\rm b}$ N.N.Bogolyubov Institute for Theoretical
Physics, National Academy of Sciences of Ukraine, 03143 Kyiv, Ukraine}} \\
\vspace{0.3cm}
{\large \textit{${}^{\rm c}$ Institut f\"ur Theoretische Physik, Goethe-Universit\"at
Frankfurt, 60438 Frankfurt am Main, Germany}} \\
\vspace{0.3cm}
{\large \textit{${}^{\rm d}$ Czech Institute of Informatics, Robotics and Cybernetics, Czech Technical University in Prague, 
16000 Prague, Czech Republic}}
\end{center}

\vspace*{0.5cm}

\begin{abstract}
An exact solution of one-dimensional lattice gauge theory at finite temperature and non-zero chemical potential is reviewed for the gauge groups $G=Z(N),U(N),SU(N)$ for all values of $N$ and the number of fermion flavors $N_f$.
Calculated are the partition function, free energy, the Polyakov loop expectation values, baryon density, quark condensate, meson and baryon correlation functions. Detailed analysis of the exact solutions is done for $N=2,3$ with one and two fermion flavors. In the large $N_f$ limit we uncover the Roberge-Weiss phase transition and discuss its remnants at finite $N_f$.  
In the case of $N_f$ degenerate flavors we also calculate 1) the large $N$ limit,
2) the large $N_f$ limit and 3) the 't Hooft-Veneziano limit of all models. The critical behavior of the models in these limits is studied and the phase structure is described in details. A comparison of all limits with $U(3)$ and $SU(3)$ QCD is also performed.
In order to achieve these results we explore several representations of the partition function of one-dimensional QCD obtained and described in the text.
\end{abstract}

\newpage

\tableofcontents

\newpage

\section{Introduction}
\label{sec_intro}

\subsection{Motivation}

Exactly solvable models play an important role in theoretical physics.
Exact solutions of the two-dimensional Ising model and the spherical model provide relevant examples 
 for quantum field theory and condensed matter physics. These solutions were used both in many physical applications and in developing analytical and numerical methods. Such solutions allow the determination of the exact phase diagram, calculation of the critical indices and establishing the universality class of the model. Moreover, they can serve as a basis for studying more complicated theories. {\it E.g.}, the spherical model is used to calculate the large $N$ asymptotic expansion of various lattice and continuum $O(N)$ models. Many other exact solutions of two-dimensional models are described in a famous book by R.~J.~Baxter \cite{baxter_book}.

Non-perturbative regularization of gauge theories by K.~Wilson in the form of a lattice gauge theory (LGT) \cite{wilson_lgt} is widely recognized as the most important tool for obtaining qualitative and/or rigorous analytical as well as quantitative numerical results in many areas, where the gauge field dynamics plays a crucial role. A gauge theory in the form of LGT can be regarded as a certain statistical-mechanical model, and any exact solution of such model would greatly contribute to our understanding of 1) the dynamics of gauge fields and 2) the phase structure of gauge theories as a function of bare coupling constants, masses, temperature and other parameters. So far we know only two LGTs that are solved exactly: two-dimensional pure LGT ({\it i.e.}, without matter fields) and one-dimensional lattice QCD.

Consider the following integral over the $SU(N)$ group
\begin{eqnarray}
 Z =	\int dU  \ e^{\frac{1}{2} h_+{\rm Tr} U + \frac{1}{2} h_- {\rm Tr} U^{\dagger} }
= \sum_{q=-\infty}^{\infty}	\left ( \frac{h_+}{h_-} \right )^{q N}
\underset{1\le i,j\le N}{\det} \ I_{i-j+q}\left ( \sqrt{h_+ h_-} \right ) \ .
\label{2dim_lgt}
\end{eqnarray}
If $h_+=h_-=\frac{2N}{g^2}$, this integral describes two-dimensional pure gauge LGT in the thermodynamic limit. The right-hand side of this expression gives an exact solution convenient to use at small values of $N$. The exact solution in the large $N$ 't Hooft limit was derived in \cite{gross_witten,wadia}. This is the famous Gross-Witten-Wadia (GWW) solution which had a big impact on the development of the theory of matrix integrals and many other applications.
If $h_{\pm}=h e^{\pm \mu}$, where $h$ is a function of quark mass and $\mu$ is the baryon chemical potential, the integrand appears in many Polyakov loop models when the static quark determinant is expanded at large quark masses, {\it i.e.} $\mu$ is fixed and $h\to 0$. The large $N$ limit of the corresponding integral was derived recently in Ref.\cite{un_vs_sun}. One of the most essential conclusions of \cite{un_vs_sun} was that the large $N$ limit turns out to be in general different for the $U(N)$ and $SU(N)$ groups. In particular, they differ in the presence of the chemical potential when the group matrices appear with  different weights in the integrand. Unfortunately, practically all studies of the large $N$ limit of matrix models deal with integrals over the $U(N)$ group and, thus, cannot be directly applied to the physically relevant $SU(N)$ models, 
see for example Ref.\cite{un_group_integral} and references therein. 

It is important to emphasize that the full static determinant in high-dimensional QCD coincides with one-dimensional QCD. The static approximation itself can be useful in several situations like the high density limit and/or limit of large quark masses.
Moreover, the majority of dual formulations obtained so far at finite baryon density and possessing positive Boltzmann weight suitable for numerical simulations have been derived in the static approximation for the full quark determinant \cite{spin_flux1,duals_lgt,dual_abelian}.

Summarizing, our motivation to study one-dimensional QCD is as follows:
\begin{itemize}
\item
One-dimensional QCD is one of very few LGTs that can be solved exactly and as such it
is important to have deep knowledge of its properties.
\item
Static quark determinant appears as a main building block in the finite-temperature QCD
in the strong coupling and high density approximations
\cite{philipsen_14, philipsen_quarkyon_19,hv_lat21,hv_pl21}. Static determinant coincides with one-dimensional QCD.
\item
As an exactly solvable model one-dimensional QCD is widely used in testing many computational methods and verifying some new physical ideas
\cite{sign_langevin,sign_subset_13,sign_num,sign_thimble}.
\item
Some forms of mean-field approximations lead to calculations of certain expectation values like the Polyakov loop over an ensemble defined by the static quark determinant \cite{ilgenfritz_85}.
\item
Group integrals, which appear in one-dimensional QCD, have a deep relation to many important problems in mathematical physics (listed below).
\item
The 't Hooft large $N$ limit \cite{Hooft_74}, the limit of a large number of flavors $N_f$ and the 't Hooft-Veneziano limit \cite{Veneziano_76} exhibit rich and interesting phase structure even in one-dimensional QCD which deserves thorough investigation.
\item
$U(N)$ and $SU(N)$ QCD at finite temperature and non-zero chemical potentials appear
to be different in the large $N$ and the 't Hooft-Veneziano limits even in the approximation of Eq.(\ref{2dim_lgt}) \cite{un_vs_sun,hv_pl21}.
We expect the same holds for the exact static determinant.

\end{itemize}

There are several analytic results in the literature devoted to one-dimensional QCD and relevant for our work. The partition function of $SU(N)$ QCD with one fermion flavor and the quark condensate at finite baryon chemical potential were calculated for the first time in \cite{bilic_88}. Most analytic results obtained so far on one-dimensional QCD can be found in \cite{1d_qcd_07}. In particular, the partition function of $U(N)$ QCD with an arbitrary number of flavors has been expressed in the form of a certain determinant or in the form of a sum over permutations (this formula is listed below). Also, exact expressions for $SU(N)$ QCD with one and two flavors at finite chemical potentials have been computed. These solutions made it possible to evaluate the average sign factor and discuss the severity of the sign problem at non-zero chemical potential. 
Next, we would like to mention Ref.\cite{qcd_cont_sphere}, where 
the 't Hooft-Veneziano limit was studied in the low-temperature and high-density limits of the continuum QCD in a small hyperspherical box. In these limits the partition function of Ref.\cite{qcd_cont_sphere} coincides with the partition function of the reduced model studied in this paper. Finally, a family of $U(N)$ matrix models has been thoroughly investigated in Refs.
\cite{russo_20,russo_tiers_20, santilli_20}. Some of them describe partition functions of one-dimensional $U(N)$ QCD. The 't Hooft-Veneziano limit of the massless \cite{russo_20,russo_tiers_20} and full models \cite{santilli_20} have been calculated. The results presented here are obtained by different methods and fully agree with \cite{santilli_20} for $U(N)$ QCD. 

Despite this progress, a number of interesting problems remain open. In this article we present a comprehensive study of many aspects of one-dimensional QCD with gauge groups $Z(N)$, $U(N)$ and $SU(N)$ at finite chemical potential. In all these cases we calculate the free energy, the quark condensate, the particle density and the expectation value of the Polyakov loop. In some cases we also discuss computation of meson and baryon correlations. The central point of these investigations is the derivation of different limits of one-dimensional QCD: 1) the limit of a large number of colors; 2) the limit of a large number of flavors; 3) the 't Hoot-Veneziano limit. Especially interesting and rich is the 't Hoot-Veneziano limit. Here, we uncover a non-trivial phase structure and describe the critical behavior in detail.

\subsection{Notations and conventions}

Let $ U\in G=Z(N), U(N), SU(N)$ and $N$, $N_f$ be the number of colors and flavors, respectively.
Lattice sites are denoted by $t$, $t\in [1,N_t]$ with $N_t$ - the lattice extension and $a$ is the lattice spacing. The following partition function
describes one-dimensional QCD at non-zero particle density (equivalent to the static QCD at finite temperature in the strong coupling limit)
\begin{eqnarray}
\label{PF_1dqcd_def}
Z_G(m,\mu; N_t,N,N_f)  \equiv  Z \ = \
\int_G \ \prod_{t=1}^{N_t} \ dU(t)  \ \prod_{f=1}^{N_f} \ B_q(m_f,\mu_f) \ .
\end{eqnarray}
The integration is performed over the normalized invariant Haar measure on $G$.
When $G=Z(N)$ the integration is replaced by a summation over configurations of
$Z(N)$ gauge field. To simplify notations we use for link variables $U(t,t+1)\equiv U(t)$ due to a one-to-one correspondence between sites and links pointing in positive direction. The Boltzmann weight $B_q(m_f,\mu_f)$ is a result of integration over fermion fields 
\begin{eqnarray}
B_q(m_f,\mu_f) &=& \int \ \prod_{t=1}^{N_t} \ \prod_{i=1}^{N} \ d\psi^i(t)
d\overline{\psi}^i(t) \ \exp \left [ \sum_{t,t^{\prime}}
\overline{\psi}^i(t) {\cal{M}}_f^{ij}(t,t^{\prime}) \psi^j(t^{\prime}) \right ] \nonumber \\
&\times& \exp \left [ \sum_t \left ( \eta_f^i(t)\overline{\psi}^i(t) +
\bar{\eta}_f^i(t) \psi^i(t)  \right ) \right ]  \ .
\label{Bq_def}
\end{eqnarray}
Periodic (anti-periodic) boundary conditions on the gauge (fermion) fields are assumed.
In order to compute various correlation functions the sources $\bar{\eta},\eta$ have been
added to the action. They should be put to zero after taking the corresponding derivatives.
The fermion matrix $\cal{M}$ reads
\begin{equation}
{\cal{M}}_f^{ij}(t,t^{\prime}) \ = \ \tilde{m}_f \delta_{i,j} \delta_{t,t^{\prime}} +
\frac{1}{2} \left ( e^{\tilde{\mu}_f} U^{ij}(t) \delta_{t,t^{\prime}-1} -
e^{-\tilde{\mu}_f} U^{+,ij}(t^{\prime}) \delta_{t,t^{\prime}+1} \right) \ ,
\label{matrixM_def}
\end{equation}
where $\tilde{m}_f = am_f^{ph}, \tilde{\mu}_f = a\mu_f^{ph}$ are dimensionless mass and
chemical potential of $f$th fermion flavor. Integration gives
\begin{equation}
B_q(m_f,\mu_f) = \mbox{Det} {\cal{M}}_f^{ij}(t,t^{\prime}) \
\exp\left [ \sum_{t,t^{\prime}} \bar{\eta}_f^i(t) {\cal{M}}_f^{-1,ij}(t,t^{\prime})
\eta_f^j(t^{\prime}) \right ] \ .
\label{generating_func}
\end{equation}
In one dimension both the determinant and the inverse matrix can be evaluated exactly (see
Appendix \ref{ferm_matr_app}). The Boltzmann weight takes the following form for vanishing
sources and even $N_t$
\begin{equation}
B_q(m_f,\mu_f) \ = \
A_f \ {\rm det} \left [ 1 + h_+^f U \right ] \ {\rm det}
\left [ 1 + h_-^f U^{\dagger} \right ] \ ,
\label{Zf_stag}
\end{equation}
where the remaining determinants are taken over group indices and
$U$ is the Polyakov loop variable
\begin{equation}
U \ = \ \prod_{t=1}^{N_t} \ U(t) \ .
\label{PL_def}
\end{equation}
The constants are given by
\begin{equation}
A_f \ = \ 2^{-N N_t} \ h_f^{-N} \ , \ h_{\pm}^f \ = \ h_f e^{\pm \mu_f}  \ , \
h_f \ = \ e^{-m_f} \ ,
\label{hpm_stag}
\end{equation}
where $m_f= N_t \arcsinh \tilde{m}_f$ and $\mu_f=N_t\tilde{\mu}_f$.
Near the continuum limit one can write
\begin{equation}
h_{\pm}^f \ = \ e^{- \beta m_f^{ph} \pm \beta \mu_f^{ph}} \ , \
\beta = a N_t \ - \ {\mbox{inverse \ temperature}} \ .
\label{hpm_cont}
\end{equation}
In what follows we use sometimes notation $A=\prod_{f=1}^{N_f} A_f$.
Throughout the paper we calculate the following observables:

\begin{itemize}
\item
the free energy
\begin{equation}
F \ = \ \frac{1}{N N_f} \ \ln Z \ ,
\label{fren_def}
\end{equation}
\item
the Polyakov loop expectation value
\begin{equation}
W(r) = \frac{1}{N} \langle {\mbox {Tr}} U^r \rangle  = \frac{A}{Z} \ \int_G \ \prod_{t=1}^{N_t} \
dU(t) \  \frac{1}{N} {\mbox {Tr}} \left ( \prod_{t=1}^{N_t} \ U(t)  \right )^r \
\prod_{f=1}^{N_f} \ B_q(m_f,\mu_f) \ ,
\label{PLexp_def}
\end{equation}
\item
the (dimensionless) quark condensate
\begin{equation}
\sigma_f \ = \ \frac{1}{N N_t} \ \frac{\partial \ln Z}{\partial \tilde{m}_f} \ = \
\frac{1}{N} \ \frac{1}{\sqrt{1+\tilde{m}_f^2}} \ \frac{\partial \ln Z}{\partial m_f} \ , \
\sigma = \frac{1}{N_f} \ \sum_{f=1}^{N_f} \sigma_f \ ,
\label{quark_cond_def}
\end{equation}
\item
the (dimensionless) particle density of $f$th flavor
\begin{equation}
B_f \ = \ \frac{1}{N N_t} \ \frac{\partial \ln Z}{\partial \tilde{\mu}_f} \ = \
\frac{1}{N} \ \frac{\partial \ln Z}{\partial \mu_f} \ , \
B = \frac{1}{N_f} \ \sum_{f=1}^{N_f} B_f \ .
\label{particle_den_def}
\end{equation}

\end{itemize}
More complicated observables like the meson-meson and the baryon--anti-baryon correlation
functions
\begin{eqnarray}
\label{meson_corr_def}
\Sigma_f(t,t^{\prime}) = \left \langle \ \sigma_f(t) \sigma_f(t^{\prime}) \ \right \rangle \ , \\
\label{baryon_corr_def}
Y_f(t,t^{\prime}) = \left \langle \ B_f(t) \overline{B}_f(t^{\prime}) \ \right \rangle
\end{eqnarray}
can be calculated by differentiating the generating function (\ref{generating_func}) with
respect to sources $\bar{\eta},\eta$. Here, the composite meson and baryon fields are
\begin{eqnarray}
\label{meson_def}
\sigma_f(t) &=& \frac{1}{N} \ \sum_{i=1}^{N} \ \overline{\psi}_f^i(t) \psi_f^i(t) \ ,  \\
\label{baryon_def}
B_f(t) &=& \frac{1}{N!} \ \epsilon^{i_1\ldots i_N} \ \psi_f^{i_1}(t)\ldots \psi_f^{i_N}(t) \ , \\
\label{antibaryon_def}
\overline{B}_f(t) &=& \frac{1}{N!} \ \epsilon^{i_N\ldots i_1} \
\overline{\psi}_f^{i_1}(t)\ldots \overline{\psi}_f^{i_N}(t) \ .
\end{eqnarray}
In general, in this paper we calculate only gauge-invariant quantities. Such quantities do not
depend on the choice of the gauge. The partition function and local observables we compute
without gauge fixing. Non-local observables which require knowledge of the inverse fermion
matrix will be evaluated in the static gauge where the inverse matrix can be easily found.
Finally, it is assumed for simplicity that $N_t$ is even.

\section{Representations of the partition function}
\label{sec_PF_repr}

In this Section we present several different but equivalent representations of the partition function of one-dimensional QCD (\ref{PF_1dqcd_def}). There are many such representations known in the literature. With an obvious change of variables $Z$ is cast into the form of the matrix integral over $G$
\begin{eqnarray}
\label{PF_1dqcd_int}
Z  = A \int_G \ dU  \ \prod_{f=1}^{N_f} \ {\rm det} \left [ 1 + h_+^f U \right ] \
{\rm det} \left [ 1 + h_-^f U^{\dagger} \right ] \ .
\end{eqnarray}
This expression is a starting point for the next derivations. Note, due to the symmetry
\begin{equation}
Z(h_+,h_-) \ = \ Z(h_-,h_+) \ = \ Z(h_-^{-1},h_+^{-1})
\label{PF_1dqcd_sym}
\end{equation}
it is sufficient to study the model in the region $0 \leq h_+,h_- \leq 1$.
Let us emphasize that all results described below are straightforwardly applicable to QCD in higher dimensions in the static approximation for $N_f$ flavors of staggered fermions. With minimal modifications the results are also valid for the Wilson fermions. Namely, {\it e.g.} for $N_f$ degenerate Wilson flavors one should take for constants
\begin{equation}
A = (2\kappa)^{- 2 N N_f N_t} \ , \	
h_{\pm} = \left ( 2\kappa \ e^{\pm \tilde{\mu}} \right )^{N_t} \ , \
\kappa = \frac{1}{2\tilde{m} + 2 d + 2\cosh\tilde{\mu}}
\label{hpm_wilson}
\end{equation}
and replace $N_f\to 2 N_f$ in Eq.(\ref{PF_1dqcd_int}) and following formulas.

\subsection{Finite-temperature model}

{\bf I}. First of all, the matrix integral (\ref{PF_1dqcd_int}) can be rewritten as an integral over the eigenvalues of $U$ as \cite{bilic_88}
\begin{eqnarray}
\label{PF_1dqcd_int2}
Z  = A \int_G \ dU  \ \prod_{f=1}^{N_f} \ \prod_{k=1}^N
\left [ 1 + h_+^f e^{i\omega_k} \right ] \  \left [ 1 + h_-^f e^{-i \omega_k} \right ] \ ,
\end{eqnarray}
where the reduced Haar measure, {\it e.g.} for $G=SU(N)$, is given by
\begin{equation}
\int_{SU(N)} dU \ldots = \sum_{q=-\infty}^{\infty} \frac{1}{N!}
\int_0^{2\pi}\prod_{i=1}^N \frac{d\omega_i}{2\pi} \
\prod_{i<j} 4 \sin^2\left ( \frac{\omega_i - \omega_j}{2} \right ) \
e^{i q \sum_{i=1}^N \omega_i} \ \ldots  \ .
\label{haar_sun}
\end{equation}
If $G=U(N)$ one should take the only term $q=0$ in the last formula.

{\bf II}. Another useful and compact representation for the partition function
has been derived by us in Ref.\cite{duals_lgt}
\begin{equation}
Z = A \sum_{q=-N_f}^{N_f} \ \sum_{\sigma} \  s_{\sigma}(H_+) s_{N^q \sigma}(H_-) \ ,
\label{PF_1dqcd_sumpart}
\end{equation}
where $H_{\pm}=(h_{\pm}^1,\cdots,h_{\pm}^{N_f})$ and $s_{\sigma}(X)$ is the Schur function.
The summation over $\sigma$ runs over all partitions such that $\sigma_1\leq N$ and the length
$l(\sigma)$ of the partition satisfies $l(\sigma)\leq N_f$. For $N_f$ degenerate flavors
(\ref{PF_1dqcd_sumpart}) gives
\begin{equation}
Z = A \sum_{q=-N_f}^{N_f} e^{\mu q N} \sum_{r=0}^{N (N_f-|q|)} h^{2r+N|q|} \
\sum_{\sigma \vdash\ r} \  s_{\sigma}(1^{N_f}) s_{N^{|q|} \sigma}(1^{N_f}) \ .
\label{PF_1dqcd_sumpart_deg}
\end{equation}
The corresponding expression for the Polyakov loop expectation value in the representation $\lambda=(\lambda_1\geq\lambda_2\geq \ldots \geq \lambda_N=0)$ reads
\begin{eqnarray}
W(\lambda)  =  \frac{A}{Z} \ \sum_{q=-N_f}^{N_f}  \sum_{\alpha,\sigma} \
C^{\sigma +q^N}_{\lambda \ \alpha} \
s_{\alpha^{\prime}}(H_+) s_{\sigma^{\prime}}(H_-) \ ,
\label{pl_sun_1}
\end{eqnarray}
where $C^{\nu}_{\lambda \ \alpha}$ are the Littlewood-Richardson coefficients and
$\alpha^{\prime},\sigma^{\prime}$ are representations dual to $\alpha, \sigma$.
The nominator of the last expression can be considered as coefficients $C_{\lambda}(H_+,H_-)$ of the character expansion of the $SU(N)$ partition function
\begin{equation}
Z = \sum_{\lambda} \ C_{\lambda}(H_+,H_-) \ \chi_{\lambda}(U) \ .
\label{Z_char_exp}
\end{equation}

{\bf III}. In the case of $N_f$ degenerate flavors partition function
(\ref{PF_1dqcd_int2}) simplifies to
\begin{eqnarray}
\label{PF_1dqcd_deg}
Z \ = \ A \ \int_G \ dU  \ \prod_{n=1}^{N} \
\left ( 1 + h_+ e^{i\omega_n} \right )^{N_f} \
\left ( 1 + h_- e^{-i\omega_n} \right )^{N_f}  \ .
\end{eqnarray}
This can be equivalently presented for $G=SU(N)$ as
\begin{align}
\label{PF_form_1}
Z &= A \ \left ( \prod_{n=1}^{N}  \ \sum_{k_n,l_n=0}^{N_f} \ h_+^{k_n} \
h_-^{l_n} \ \binom{N_f}{k_n} \ \binom{N_f}{l_n} \right ) \ Q(k_n,l_n) \ ,  \\
\label{group_integral}
Q(k_n,l_n) &= \int_G \ dU  \ \prod_{n=1}^{N} \ e^{i (k_n-l_n) \omega_n}
%= {} \nonumber \\ 
%{} &
= \sum_{q=-\infty}^{\infty} \ \sum_{\sigma \in S_N} \ \frac{1}{N!} \; \underset{1\leq i,j\leq N}{\mbox{det}} \
\delta_{i-j+k_{\sigma_i}-l_{\sigma_i} + q, 0} \ .
\end{align}
Performing summation over $k_n$ produces
\begin{eqnarray}
Z \ = \ A \ \sum_{q=-N_f}^{N_f} \ \left ( \prod_{n=1}^{N}  \
\sum_{l_n=0}^{N_f} \ h_+^{l_n+q} \
h_-^{l_n} \ \binom{N_f}{l_n} \right ) \ \underset{1\le i,j\le N}{\det} \binom{N_f}{l_i-i+j+q}  \ .
\label{PF_form_2}
\end{eqnarray}
Denoting $L_i=l_i-i+q$ the determinant is evaluated as \cite{determinant_calc}
\begin{eqnarray}
\underset{1\le i,j\le N}{\det} \binom{N_f}{L_i+j} = \frac{G(N+N_f+1)}{G(N_f+1)} \
\frac{\prod_{1\le i<j\le N} (L_i-L_j)}{\prod_{i=1}^N (L_i+N)! (N_f-L_i-1)!} \ ,
\label{binom_det}
\end{eqnarray}
where $G(X)$ is the Barnes function. Partition function becomes
\begin{eqnarray}
Z = A \ \frac{G(N+N_f+1)}{G(N_f+1)} \ \sum_{q=-N_f}^{N_f} \
h^{N |q|} \ e^{\mu N q} \ \ Q_{N,N_f}(q) \ ,
\label{PF_form_3}
\end{eqnarray}
\begin{eqnarray}
Q_{N,N_f}(q) = \sum_{l_1,\ldots,l_N=0}^{N_f} \ h^{2l_1+\ldots +2 l_N} \
\frac{\prod_{1\le i<j\le N} (l_i-l_j+j-i) \prod_{i=1}^N \binom{N_f}{l_i}}
{\prod_{i=1}^N (l_i-i+q+N)! (N_f-l_i+i-q-1)!} .
\label{Zq_def_1}
\end{eqnarray}

{\bf IV.} Instead of computing the determinant in Eq.(\ref{PF_form_2}) one can sum up over
$l_n$ variables. This leads to the following representation
\begin{gather}
Z = 2^{-N N_t} (h^{-1}-h)^{N N_f}\sum_{q=-N_f}^{N_f} e^{\mu q N} \
\underset{1\le i,j\le N}{\det} \ \frac{(N_f)! \ P^{i-j+q}_{N_f}(t)}{(N_f+i-j+q)!} \ ,
\label{PF_1dqcd_det_leg}
\end{gather}
where $P_n^y(t)$ is the associated Legendre function and $t=\coth m=\frac{1+h^2}{1-h^2}$.
This is a generalization of the integral in Eq.(\ref{2dim_lgt}) for arbitrary quark masses. Indeed, taking the uniform asymptotics of the associated Legendre function at large mass ($h\to 0$), Eq.(\ref{Legendre_asymp_uniform}), one recovers Eq.(\ref{2dim_lgt}). This form of the partition function turns out to be very useful in deriving the large $N_f$ and the 't Hooft-Veneziano limits.

An equivalent form of the character expansion coefficients $C_{\lambda}(H_+,H_-)$ in (\ref{Z_char_exp}) for $N_f$ degenerate flavors can be derived in the same manner and reads
\begin{gather}
C_{\lambda}(m,\mu) = 2^{-N N_t} (2\sinh m)^{N N_f}\sum_{q=-N_f}^{N_f}
e^{\mu q N - \mu \sum_j\lambda_j} \
\underset{1\le i,j\le N}{\det} \
\frac{(N_f)! \ P^{\lambda_j+i-j+q}_{N_f}(t)}{(N_f+\lambda_j+i-j+q)!} \ .
\label{Z_char_exp_2}
\end{gather}

The expectation value of the Polyakov loop $W(r)$ can be derived from
Eqs.(\ref{PLexp_def}) and (\ref{PF_1dqcd_det_leg})
\begin{gather}
	\label{sun_pl_gen}
	W(r)=\frac{1}{N}\frac{\sum_{q=-N_f+1}^{N_f}\sum_{k=N-r+1}^N\underset{1\le i,j\le N}{\det}\frac{e^{(q-r\delta_{j,k})\mu}P^{i-j+q-r\delta_{j,k}}_{N_f}(\coth m)}{(N_f+i-j+q-r\delta_{j,k})!}}{\sum_{q=-N_f}^{N_f}\underset{1\le i,j\le N}{\det}\frac{e^{q\mu}P^{i-j+q}_{N_f}(\coth m)}{(N_f+i-j+q)!}}  \ .
\end{gather}

{\bf V.} Extension of the representation (\ref{PF_1dqcd_det_leg}) to non-degenerate flavors can be obtained directly from (\ref{PF_1dqcd_int2}) by using the following group integration formula
\begin{eqnarray}
\label{gen_sun_int}
\int_G dU \ \prod_{k=1}^N \ F\left ( e^{i\omega_k} \right ) \ = \
\sum_{q=-\infty}^\infty \ \underset{1\le i,j\le N}{\det} \
\left ( \int_0^{2\pi} \frac{d\omega}{2\pi} \ e^{i (i-j+q) \omega} \
F\left ( e^{i\omega} \right ) \right ) \ .
\end{eqnarray}
This leads to the following representation of the partition function
\begin{eqnarray}
	\label{PF_sun_def}
	Z &=& \sum_{q=-N_f}^{N_f} Z_q \ \ , \ \
	Z_q = 2^{-N N_f N_t} \ \underset{1\le i,j\le N}{\det}T_{i-j+q} \ , \\
	\label{Tk_def}
	T_k &=& 2^{N_f} \ \int_0^{2 \pi} \frac{d\phi}{2\pi} \ e^{i k \phi} \
	\prod_{f=1}^{N_f} \left ( \cosh m_f + \cos(\phi - i \mu_f) \right ) \ .
\end{eqnarray}
For the $U(N)$ model $Z=Z_0$.

{\bf VI.} Partition function for the $U(N)$ model can also be presented
in the form \cite{1d_qcd_07}
\begin{eqnarray}
Z \ = \ \sum_{\sigma\in S_{2N_f}/S_{N_f\times S_{N_f}}} \
\prod_{f=1}^{N_f} \prod_{f^{\prime}=1}^{N_f} \
\frac{e^{N m_{\sigma(+f)}}}{1-\exp[m_{\sigma(-f^{\prime})} - m_{\sigma(+f)}]} \ ,
\label{Z_un_general}
\end{eqnarray}
where $m_{\pm f}=\pm m_f=\pm N_t \arcsinh \tilde{m}_f$ and the sum over $\sigma$ runs over permutations that interchange positive and negative masses.

{\bf VII}. The determinantal representation establishing a connection of $Z$ with the generating function of plane partitions in $N\times (N_f-q)\times (N_f+q)$ box was derived in Ref.\cite{plane_part}.

{\bf VIII}. All previous representations have been derived by first integrating out fermion degrees of freedom. Another route is to perform the first integration over gauge fields. This leads to the following expression for the $U(N)$ partition function \cite{sun_integral}
 \begin{eqnarray}
 	Z = \int \ \prod_{t=1}^{N_t} \ \prod_{i=1}^{N} \prod_{f=1}^{N_f}
 	d\psi_f^i(t) d\overline{\psi}_f^i(t) \ e^{\sum_f \sum_t \tilde{m}_f \sigma_{f f}(t) }
 	\nonumber  \\
 	\prod_t \sum_{s=0}^{N N_f} \  4^{-s} \ \sum_{\lambda\vdash\ s} \
 	\frac{d^2(\lambda)}{(s!)^2} \
 	\frac{s_{\lambda^{\prime}}(\Sigma (t))}{s_{\lambda}(1^N)} \ .
 	\label{meson_repr_un}
 \end{eqnarray}
 Here, $\lambda^{\prime}$ is a representation dual to $\lambda$, $d(\lambda)$ is a dimension of $\lambda$ representation and the composite meson fields are
 \begin{equation}
 	\sigma_{f f^{\prime}}(t) \ = \ \sum_{i=1}^N \
 	\bar{\psi}^{i}_f(t)\psi^i_{f^{\prime}}(t) \ ,
 \end{equation}
 \begin{equation}
 	\Sigma_{f_1f_2}(t) \ = \
 	\sum_{f=1}^{N_f} \sigma_{f_1f}(t)\sigma_{ff_2}(t+1) \ .
 \end{equation}
A similar form for $SU(N)$ theory can be easily obtained using the result for the one-link integral derived in \cite{sun_integral}. In this case an evaluation of the partition function reduces to the problem of counting the number of ways the one-dimensional chain can be covered with multi-component monomer and dimer configurations.

Throughout the paper we use the first five representations listed above. The last three representations are given for completeness and to stress the mathematical richness
of one-dimensional QCD.

\subsection{Thermodynamic and large mass limits}

We define the thermodynamic limit (TL) as
\begin{equation}
F \ = \ \frac{1}{N N_f} \ \lim_{N_t\to\infty} \ \frac{1}{N_t} \ \ln Z  \ ,
\label{thermod_limit_def}
\end{equation}
all other variables being fixed. This is also a zero-temperature limit of the model.
The TL of the interacting model coincides with the TL of the free fermion model
and does not depend on the boundary conditions. Starting from, {\it e.g.} free boundary
conditions one can gauge away all gauge matrices $U(t)$ by a change of variables
\begin{equation}
\begin{cases}
\psi(t) \to \psi^{\prime}(t) = \prod_{\tau=1}^{t-1} \ U^{\dagger}(\tau) \ \psi(t)    \ ,     \\
\overline{\psi}(t) \to \overline{\psi}^{\prime}(t) = \overline{\psi}(t)
\prod_{\tau=t-1}^1 \ U(\tau) \ .
\end{cases}
\label{gauge_fix}
\end{equation}
This leaves a free fermion model which we consider in the next Section.
Hence, one finds in the TL the following simple answer for the free energy
\begin{equation}
F \ = \ \frac{1}{N_f} \  \left ( \sum_{f=1}^{N_f} \ g_f - \ln 2 \right ) \ , \
g_f \ = \
\begin{cases}
\arcsinh \tilde{m}_f \ , \
\arcsinh \tilde{m}_f  \geq | \tilde{\mu}_f | \ , \\
| \tilde{\mu}_f | \ , \
\arcsinh \tilde{m}_f  \leq | \tilde{\mu}_f |   \ .
\end{cases}
\label{thermod_limit_res}
\end{equation}
In the limit of large bare fermion mass $\tilde{m}_f$ ($h_{\pm}^f\to 0$) one can easily
find from (\ref{PF_1dqcd_int2}) and (\ref{hpm_stag})
\begin{equation}
Z \ = \ A \approx \left ( \prod_{f=1}^{N_f} \ \tilde{m}_f \right )^{N N_t} \ .
\label{large_mass_lim}
\end{equation}

\section{Free fermion model}

In this Section we study the free model and compute both local observables and
correlation functions defined in the end of Sec.\ref{sec_intro}.
In addition to being interesting on its own right the results of this Section can be used as an extra check of the calculations in the interacting models: in the low-temperature region both the free energy and all expectation values should converge to the corresponding values of the free fermion model.

\subsection{Partition function and local observables}

As follows from Eq.(\ref{PF_1dqcd_int2}) the partition function of the free fermion
model is given by
\begin{equation}
Z \ = \ 2^{-N N_f N_t} \ \prod_{f=1}^{N_f} 2^N \left [ \cosh m_f  + \cosh \mu_f \right ]^N \ .
\label{PF_free_fermion}
\end{equation}
From here one can derive the free energy
\begin{equation}
F \ = \ -N_t \ln 2 + \frac{1}{N_f} \ \sum_{f=1}^{N_f}
\ln \left [ 2 \cosh m_f  + 2 \cosh \mu_f \right ] \ . 
\label{fren_free_ferm}
\end{equation}
The quark condensate and the particle density acquire simple forms
\begin{equation}
\sigma_f \ = \ \frac{1}{\sqrt{1+\tilde{m}_f^2}} \
\frac{\sinh m_f}{\cosh m_f  + \cosh \mu_f} \ ,
\label{quark_cond_free_ferm}
\end{equation}
\begin{equation}
B_f \ = \ \frac{\sinh \mu_f}{\cosh m_f  + \cosh \mu_f} \ .
\label{particle_den_free_ferm}
\end{equation}

\subsection{Fermion correlators}

It is straightforward to calculate the correlation function of free fermion fields
\begin{equation}
\frac{1}{N} \ \sum_{k=1}^N \left \langle \overline{\psi}^k(t) \psi^k(t^{\prime}) \right \rangle
\ = \ \frac{1}{N} \ \sum_{k=1}^N {\cal{M}}_k^{-1}(t,t^{\prime}) \ .
\label{free_ferm_corr}
\end{equation}
Using Eq.(\ref{M_inverse}) and assuming $\tau = t-t^{\prime}\geq 0$ one obtains
\begin{eqnarray}
\label{free_ferm_corr_res}
&&\frac{1}{N} \ \sum_{k=1}^N \left \langle \overline{\psi}^k(t) \psi^k(t^{\prime}) \right \rangle
= \frac{e^{- \tau \tilde{\mu}}}
{\left [ \cosh N_t \arcsinh \tilde{m}+ \cosh N_t \tilde{\mu}  \right ] \ \sqrt{1+\tilde{m}^2}} \\
&&\times
\begin{cases}
\sinh (N_t-|\tau|) \arcsinh \tilde{m} - e^{N_t \tilde{\mu}}\sinh|\tau| \arcsinh \tilde{m} \ , \
\tau   \ - \mbox{even} \ ,  \\
\cosh (N_t-|\tau|) \arcsinh \tilde{m} + e^{N_t \tilde{\mu}}\cosh|\tau| \arcsinh \tilde{m} \ , \
\tau   \ - \mbox{odd} \ .
\end{cases} \nonumber
\end{eqnarray}
Meson-meson (\ref{meson_corr_def}) and baryon-baryon (\ref{baryon_corr_def}) correlations are of the form
\begin{eqnarray}
\label{meson_corr_free_ferm}
&&\Sigma(t,t^{\prime}) = \left ( \frac{1}{N} \sum_{k=1}^N {\cal{M}}_k^{-1}(0) \right )^2 +
\frac{1}{N^2} \sum_{k=1}^N {\cal{M}}_k^{-1}(t,t^{\prime})
{\cal{M}}_k^{-1}(t^{\prime},t) \ , \\
\label{baryon_corr_free_ferm}
&&Y(t,t^{\prime}) =  \frac{1}{N!} \ \det \mathcal{M}^{-1}(t,t') \ .
\end{eqnarray}
To simplify notations we omit the flavor index and write $m=m^{ph}$, $\mu=\mu^{ph}$ in what follows. In the finite-temperature limit $a\to 0$, $N_t\to\infty$ such that $a N_t=\beta$ one finds
\begin{eqnarray}
\Sigma(\tau) &=& (2\cosh \beta m + 2\cosh\beta\mu)^{-2} 
\bigg [ \sinh^2\beta m + 
\frac{2 \sinh(\beta-|\tau|)m \sinh|\tau|m \cosh\beta\mu}{N}
 \nonumber \\ 
\label{meson_corr_free_res}
&-& \frac{\sinh^2(\beta-|\tau|)m+\sinh^2|\tau|m}{N} \bigg  ] \ , \\
\label{baryon_corr_free_res}
Y(\tau) &=&
\frac{1}{N!}\left(\frac{e^{-\tau\mu}\sinh(\beta-|\tau|)m-e^{(\operatorname{sign}(\tau)\beta-\tau)\mu}\sinh|\tau|m}{2\cosh\beta m+2\cosh\beta\mu}\right)^N .
\end{eqnarray}
In the zero temperature limit $\beta\to\infty$ (corresponding to the TL) it gives for the connected part of the meson correlation if $m>\mu$
\begin{gather}
\label{meson_corr_limit}
	- \ln \mid \Sigma_c(\tau) \mid = 2m|\tau| \ .
\end{gather}
Similar limit for the baryon correlation takes the form
\begin{gather}
\label{baryon_corr_limit}
- \ln \mid Y(\tau>0) \mid = \begin{cases}  N(m+\mu) \tau \ , & m>\mu \ ,  \\
 N(\mu-m) \tau \ , & \mu>m \ .
	\end{cases}
\end{gather}

\section{Z(N) model}
\label{zn_model}

We turn now to the interacting models and start from the simplest case of one-dimensional
$Z(N)$ QCD. The $Z(N)$ model can be obtained from the $SU(N)$ one by replacing
\begin{equation}
U_{ij} \to \delta_{ij} \ \exp\left [  \frac{2\pi i}{N} \ s  \right ] \ , \
s = 0,\ldots,N-1 \ .
\label{zn_variable}
\end{equation}
To the best of our knowledge $Z(N)$ QCD in one dimension has not been studied before.
Taking into account the importance of the center subgroup in QCD we think it is worth to solve exactly also this model.
For simplicity we consider in this Section only the case of $N_f$ degenerate flavors.
From now on we shall omit an unessential constant factor $2^{-N N_f N_t}$ in the definition
of $A_f$ (\ref{hpm_stag}).
The partition function (\ref{PF_1dqcd_deg}) of $Z(N)$ QCD takes the form
\begin{eqnarray}
\label{PF_1dZN}
Z \ = \ \frac{e^{N N_f m}}{N} \ \sum_{s=0}^{N-1} \
\left ( 1 + h_+ e^{\frac{2\pi i}{N} \ s} \right )^{N N_f} \
\left ( 1 + h_- e^{- \frac{2\pi i}{N} \ s} \right )^{N N_f} \ ,
\end{eqnarray}
which is convenient to study the model at small $N$ and arbitrary $N_f$. Summing up over
$s$ with the help of Eq.(\ref{Legendre_sum}) one gets
\begin{gather}
Z = (2\sinh m)^{NN_f} \ \sum_{q=-N_f}^{N_f}\frac{(NN_f)!}{(N(N_f+q))!} \
e^{qN\mu} \ P^{qN}_{NN_f}(\coth m) \ ,
\label{PF_1dZN_res}
\end{gather}
where $P_l^m(x)$ is the associated Legendre polynomial (for its definition and properties,
see Appendix \ref{leg_func_app}). This representation is useful to study the opposite case
of small $N_f$ and arbitrary $N$.
The Polyakov loop, the baryon density and the quark condensate\footnote{Here and below we omit the factor $\sqrt{1+\tilde{m}^2}$ which vanishes in the continuum limit anyway.}
are given by, respectively
\begin{align}
\label{pl_zn}
W(r) &= \frac{(2\sinh m)^{NN_f}}{Z} \ e^{-\mu r} \ \sum_{q=-N_f+1}^{N_f} \frac{(NN_f)! \  e^{qN\mu}}{(N(N_f+q)-r)!} \ P_{NN_f}^{qN-r}(\coth m) \ , \\
%\end{align}
%\begin{gather}
\label{density_zn}	
B &=  \frac{1}{N_f} \ \langle \ q  \ \rangle  \ , \\
%\end{gather}
%\begin{align}
\label{condensate_zn}
\sigma &= \frac{(2\sinh m)^{NN_f}}{Z}\sum_{q=-N_f+1}^{N_f}\frac{(N N_f)! \ e^{qN\mu}}{(N(N_f+q)-1)!}P_{NN_f-1}^{qN}(\coth m) \ .
\end{align}
The expectation value in Eq.(\ref{density_zn}) refers to the partition function (\ref{PF_1dZN_res}). Eqs.(\ref{PF_1dZN})-(\ref{condensate_zn}) allows to study $Z(N)$ QCD if either $N$ or $N_f$ is not too big. {\it E.g.}, one finds for the Polaykov loops for $N=2,3$ and arbitrary $N_f$
\begin{gather}
	W_{N=2} =\frac{1-\left(\frac{\cosh m - \cosh \mu}{\cosh m + \cosh
		\mu}\right)^{2N_f}}{1+\left(\frac{\cosh m - \cosh \mu}{\cosh m + \cosh
		\mu}\right)^{2N_f}} \ , \
	W_{N=3}=\frac{1+2\Re \ e^\frac{i2\pi}{3}\left(\frac{\cosh m +
		\cosh\left(\mu+\frac{i2\pi}{3}\right)}{\cosh m + \cosh
		\mu}\right)^{3N_f}}{1+2\Re \ \left(\frac{\cosh m +
		\cosh\left(\mu+\frac{i2\pi}{3}\right)}{\cosh m + \cosh
		\mu}\right)^{3N_f}}
\label{plzn_n2_n3}
\end{gather}
and similar simple formulas can be easily written for other observables.
Several explicit expressions of the partition function for particular values of $N$ and $N_f$ are given in Appendix \ref{znpf_list}.

\begin{figure}[htb]
	\centerline{
	\resizebox{.95\textwidth}{!}{%
	\epsfysize=7.5cm  \epsfbox{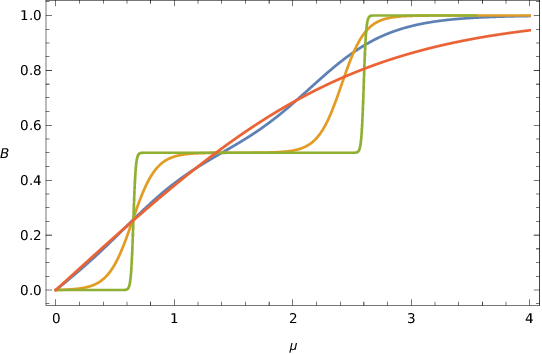} \epsfysize=7.5cm
		\epsfbox{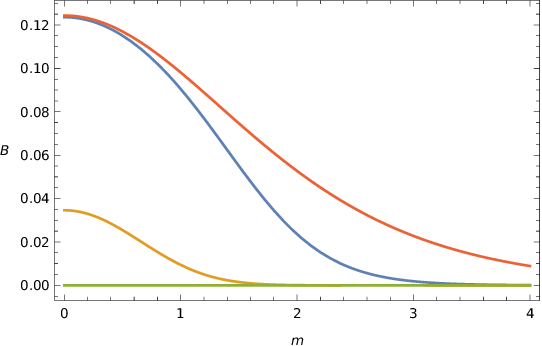} }}
	\caption{Baryon density as a function of $\mu$ for $m=1$ (left) and as a function of $m$ for $\mu=0.25$ (right). Blue line: $N=3,N_f=2$.  Orange line: $N=10,N_f=2$. Green line: $N\to\infty,N_f=2$. Red line: $N_f\to\infty$.}
	\label{Fig:zn_dens1}
\end{figure}

\begin{figure}[htb]
	\centerline{\resizebox{.95\textwidth}{!}{%
	\epsfysize=5cm  \epsfbox{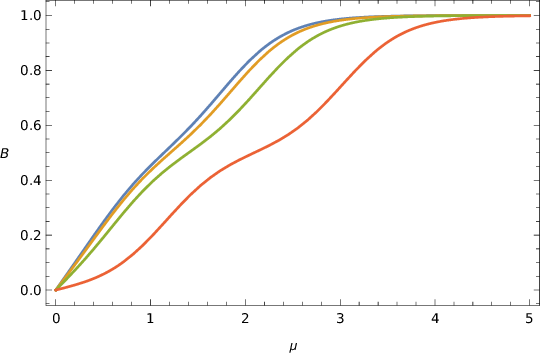} \epsfysize=5cm
		\epsfbox{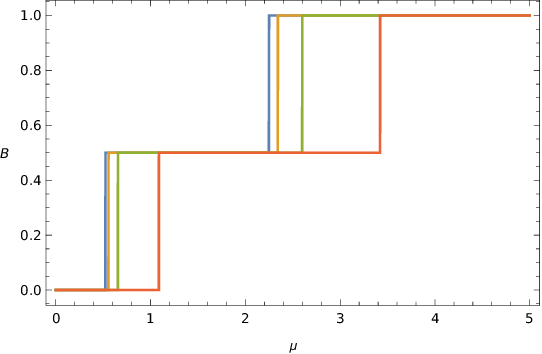} \epsfysize=5cm  \epsfbox{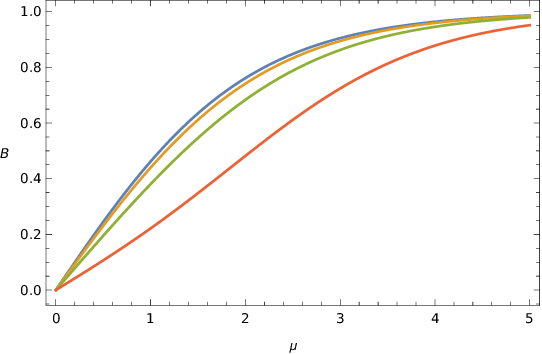}}}
	\caption{Baryon density as a function of $\mu$ for $m=0$ (blue), $m=0.5$ (orange),  $m=1$ (green) and $m=2$ (red). Left panel: $N=3,N_f=2$. Central panel: $N\to\infty,N_f=2$. Right panel: $N_f\to\infty$.}
	\label{Fig:zn_dens2}
\end{figure}

\begin{figure}[htb]
	\centerline{\resizebox{.95\textwidth}{!}{%
	\epsfysize=5cm  \epsfbox{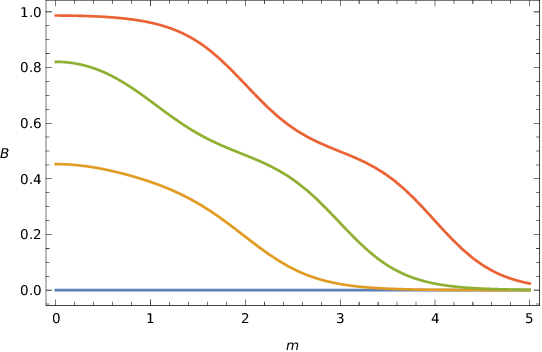} \epsfysize=5cm
		\epsfbox{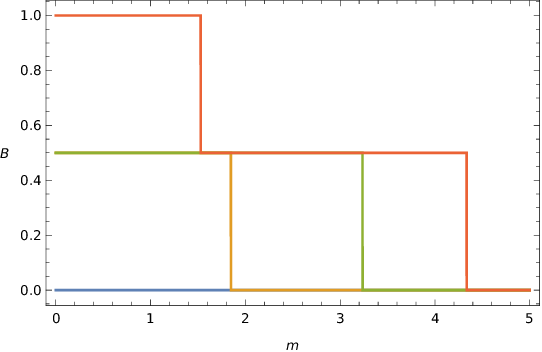} \epsfysize=5cm  \epsfbox{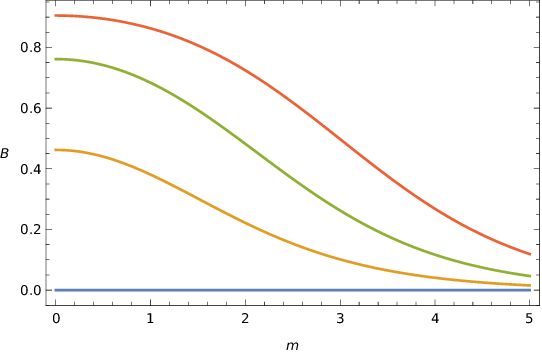}}}
	\caption{Baryon density as a function of $m$ for $\mu=0$ (blue), $\mu=1$ (orange), $\mu=2$ (green) and $\mu=3$ (red). Left panel: $N=3,N_f=2$. Central panel: $N\to\infty,N_f=2$. Right panel: $N_f\to\infty$.}
	\label{Fig:zn_dens3}
\end{figure}

\begin{figure}[htb]
	\centerline{\resizebox{.95\textwidth}{!}{%
	\epsfysize=7.5cm  \epsfbox{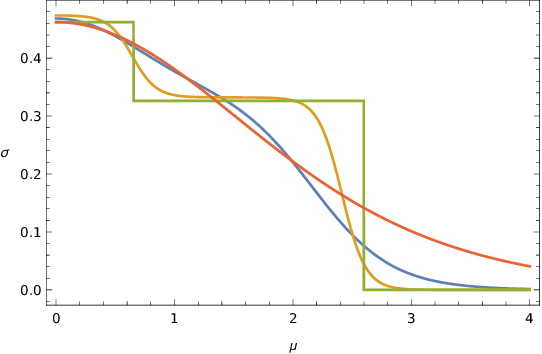} \epsfysize=7.5cm
		\epsfbox{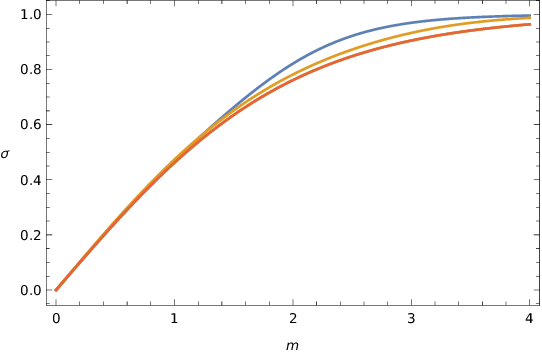} }}
	\caption{Quark condensate as a function of $\mu$ for $m=1$ (left) and as a function of $m$ for $\mu=0$ (right). Blue line: $N=3,N_f=2$.  Orange line: $N=10,N_f=2$. Green line: $N\to\infty,N_f=2$. Red line: $N_f\to\infty$.}
	\label{Fig:zn_cond1}
\end{figure}

\begin{figure}[htb]
	\centerline{\resizebox{.95\textwidth}{!}{%
	\epsfysize=5cm  \epsfbox{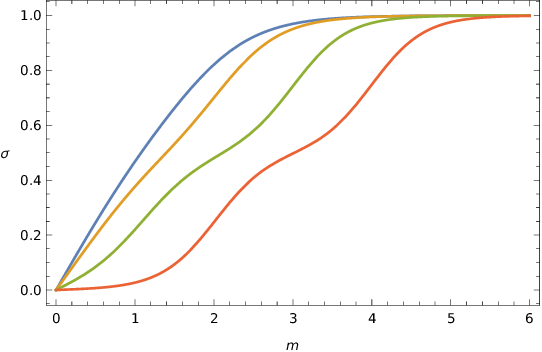} \epsfysize=5cm
		\epsfbox{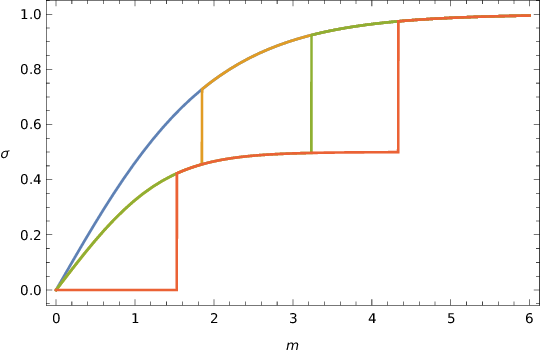} \epsfysize=5cm  \epsfbox{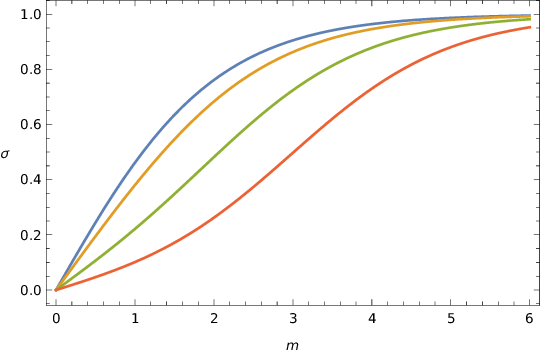}}}
	\caption{Quark condensate as a function of $m$ for $\mu=0$ (blue), $\mu=1$ (orange), $\mu=2$ (green), $\mu=3$ (red). Left panel: $N=3,N_f=2$. Central panel: $N\to\infty,N_f=2$. Right panel: $N_f\to\infty$.}
	\label{Fig:zn_cond2}
\end{figure}

\begin{figure}[htb]
	\centerline{\resizebox{.95\textwidth}{!}{%
	\epsfysize=5cm  \epsfbox{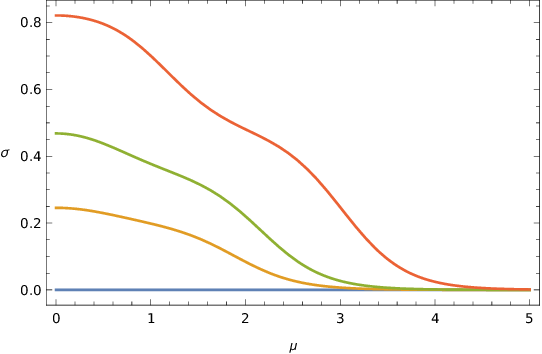} \epsfysize=5cm
		\epsfbox{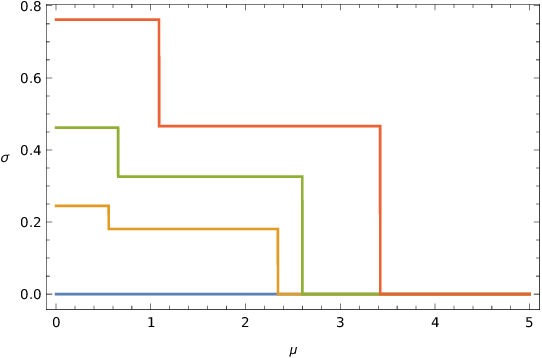} \epsfysize=5cm  \epsfbox{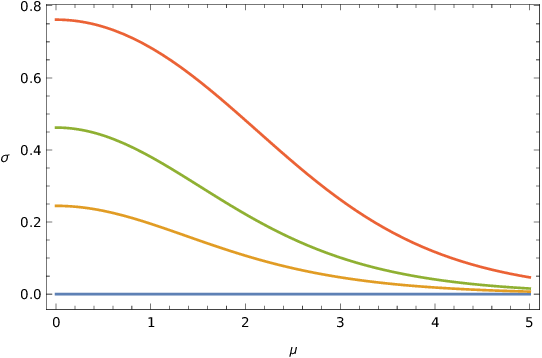}}}
	\caption{Quark condensate as a function of $\mu$ for $m=0$ (blue), $m=0.5$ (orange), $m=1$ (green), $m=2$ (red). Left panel: $N=3,N_f=2$. Central panel: $N\to\infty,N_f=2$. Right panel: $N_f\to\infty$.}
	\label{Fig:zn_cond3}
\end{figure}

The main conclusion one can draw from these explicit expressions is that $Z(N)$ QCD does not exhibit any critical behavior at finite $N$ and/or $N_f$. Let us now inspect the limiting properties of the model. This can be done either by using the uniform asymptotic expansion of the Legendre function (\ref{Legendre_asymp_all_order}) or by using the Poisson resummation formula. In the latter case Eq.(\ref{PF_1dZN}) can be presented as
\begin{eqnarray}
\label{PF_1dZN_poisson}
Z = \sum_{q=-N_f}^{N_f} \int_0^{2\pi} \ \frac{d\phi}{2\pi} \left ( 2\cosh m + 2 \cos(\phi - i \mu) \right )^{N N_f} e^{i N q \phi} \ .
\end{eqnarray}
One can deduce two different limiting behaviors from this representation.
\begin{enumerate}
\item
$N\to\infty$, $N_f$ is fixed and $\mu >0$:
\begin{eqnarray}
\label{PF_1dZN_largeN}
Z &=& \sum_{q=0}^{N_f} \left ( 2 e^{\mu q/N_f} \ Q(q/N_f) \right )^{N N_f} \ ,  \\
Q(x) &=& \left ( \frac{\cosh m+\sqrt{x^2\sinh^2m + 1}}{1-x^2} \right )
\  \left ( \frac{\sqrt{x^2\sinh^2m + 1}-x \cosh m}{1+x} \right )^x \ .  \nonumber
\end{eqnarray}
In this case one can observe a typical threshold behavior similar to that for the quark condensate found in \cite{bilic_88} in the limit $N_t\to\infty$. Moreover, in our case we find exactly $N_f$ jumps in the behavior of the baryon density and the quark condensate with varying the chemical potential or the mass. These jumps correspond to different values of $q$ which maximize the  summand in Eq.(\ref{PF_1dZN_largeN}).

\item
$N_f\to\infty$, $\forall$ $N$:
the model reduces to the free fermion model, Eq.(\ref{PF_free_fermion}). It corresponds to the fact that the leading contribution in this limit comes from the term $s=0$ in (\ref{PF_1dZN}). The quark condensate and the baryon density behave like those of the free model, Eqs.(\ref{quark_cond_free_ferm}) and (\ref{particle_den_free_ferm}).
However, the Polyakov loop becomes $W=1$ for all values of $m$ and $\mu$.
This might look strange at first glance as at least in the limit of large mass one would expect that the Polyakov loop vanishes. This phenomenon is due to the fact that two limits, $m\to\infty$ and $N_f\to\infty$, do not commute. Indeed, this non-commutativity is seen from Eq.(\ref{plzn_n2_n3}): if the limit $m\to\infty$ is taken first the Polyakov loop does vanish.
	
\end{enumerate}
In order to illustrate the behavior found above we plot the particle density and the quark condensate below. 
Figs.\ref{Fig:zn_dens1}, \ref{Fig:zn_dens2}, \ref{Fig:zn_dens3} show the threshold transitions and the convergence to these transitions with growing $N$ for the baryon density. Figs.\ref{Fig:zn_cond1}, \ref{Fig:zn_cond2}, \ref{Fig:zn_cond3} show the same behavior for the quark condensate. In all cases we plot these quantities for two flavors $N_f=2$ and in the limit $N_f\to\infty$. Indeed, one observes two jumps when $N$ gets sufficiently large and $N_f=2$ as explained above. When $N_f\to\infty$ both quantities converge to the corresponding quantities of the free fermion model. Another interesting observation is that when $\mu$ is fixed the baryon density vanishes with the mass increasing (right panel of Fig.\ref{Fig:zn_dens1} and Fig.\ref{Fig:zn_dens3}). When the mass is fixed, the quark condensate vanishes with the chemical potential increasing, thus implying an effective approximate restoration of the chiral symmetry (left panel of Fig.\ref{Fig:zn_cond1} and Fig.\ref{Fig:zn_cond3}).


In Sec.\ref{HV_limit} we study two more limits of the $SU(N)$ model, namely the heavy-dense limit and the massless limit. For completeness and with the goal of comparison
we would like to present here the corresponding limits for the $Z(N)$ model.
In the heavy-dense limit $h\to 0$, $\mu\to\infty$ such that $h_+={\rm const}$,  $h_-=0$ the partition function reads
\begin{equation}
Z = e^{N N_f m} \ \sum_{q=0}^{N_f} \ \binom{N N_f}{q N} \ e^{N(-m+\mu)q} \ .
\label{ZNpf_reduced}
\end{equation}
In the large $N_f$ limit the free energy equals
\begin{equation}
F = \ln \left [ e^m  +  e^{\mu}  \right ] \ .
\label{ZNfren_reduced}
\end{equation}
In the massless limit $h=1$ and the partition function reads
\begin{equation}
Z = 1 + 2 \sum_{q=1}^{N_f} \ \binom{2 N N_f}{N(N_f+q)} \ \cosh\mu N q  \ .
\label{ZNpf_massless}
\end{equation}
In the large $N_f$ limit one gets the free energy of the massless free fermion model, Eq.(\ref{fren_free_ferm}).

\clearpage

\section{U(N) and SU(N) models}
\label{un_sun_model}

In this Section we consider the $U(N)$ and $SU(N)$ models at finite number of colors or  flavors. In the $SU(N)$ case we also discuss the large $N_f$ limit. 
To derive results for models with $N_f$ degenerate flavors we use the representations (\ref{PF_1dqcd_det_leg}), (\ref{sun_pl_gen}) and sometimes the representation (\ref{PF_1dqcd_sumpart_deg}) to check results.
For non-degenerate flavors the representation (\ref{PF_sun_def})-(\ref{Tk_def}) will be used. As before, we omit the constant $2^{-N N_f N_t}$.

\begin{figure}[htb]
	\begin{center}
	\resizebox{.95\textwidth}{!}{%
	\includegraphics[height=0.32\textwidth]{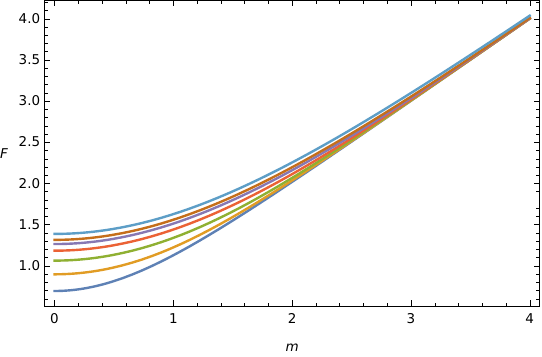}
	\includegraphics[height=0.32\textwidth]{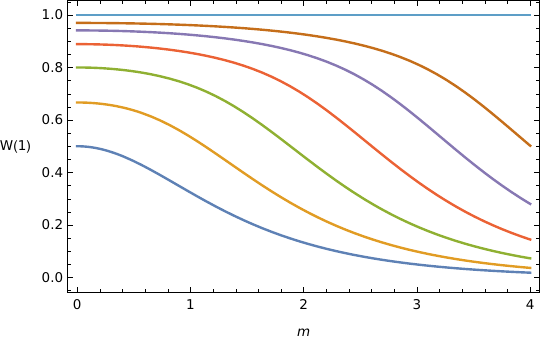}
	\includegraphics[height=0.32\textwidth]{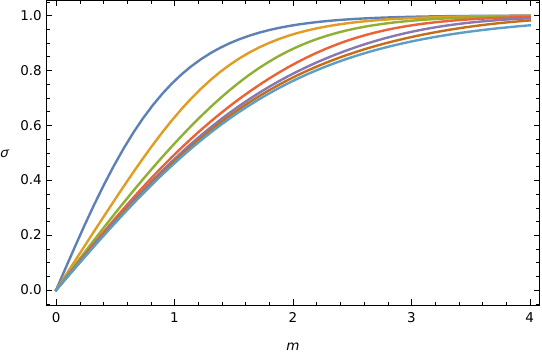}}
	\end{center}
	\caption{Free energy (left), Polyakov loop (middle) and quark condensate (right) as functions of mass for $U(1)$ at various $N_f$: 1 (dark blue), 2 (yellow), 4 (green), 8 (red), 16 (violet), 32 (brown), $\infty$ (light blue).}
	\label{u1_fig1}
\end{figure}

Let us start from the Abelian $U(1)$ model with $N_f$ degenerate flavors. Clearly, the partition function and invariant observables do not depend on the chemical potential.  Eq.(\ref{PF_1dqcd_det_leg}) gives for the partition function and for the Polyakov loop in the representation $r$
\begin{eqnarray}
\label{u1_pf}
Z &=& (2\sinh m)^{N_f} \ P_{N_f}\left(\coth m\right) \ , \\
\label{u1_pl}
W(r) &=& e^{-\mu r} \ \frac{(N_f)!}{(N_f+r)!} \  \frac{P^{r}_{N_f}\left(\coth m\right)}{P_{N_f}\left(\coth m\right)} \ .
\end{eqnarray}
The particle density vanishes while the quark condensate reads
\begin{equation}
\sigma = \frac{P_{N_f-1}\left(\coth m\right)}{P_{N_f}\left(\coth m\right)}  \ .
\label{u1_ferm_cond}
\end{equation}
Large $N_f$ asymptotic expansion at fixed mass follows from Eq.(\ref{Legendre_asymp})
\begin{equation}
Z = \frac{(2\cosh \frac{m}{2})^{2N_f+1}}{2\sqrt{\pi N_f}} \ , \
W(r) = e^{-\mu r} \ , \ \sigma = \tanh\frac{m}{2} \ .
\label{u1_asymp}
\end{equation}
The uniform expansion valid at large masses can be easily obtained from
Eq.(\ref{Legendre_asymp_uniform}).
To illustrate the smooth behavior of the $U(1)$ model we show in Fig.\ref{u1_fig1} the free energy, the Polyakov loop and the quark condensate as functions of mass for various  $N_f$.
\begin{figure}[htb]
	\begin{center}
	\resizebox{.95\textwidth}{!}{%
	\includegraphics[height=0.32\textwidth]{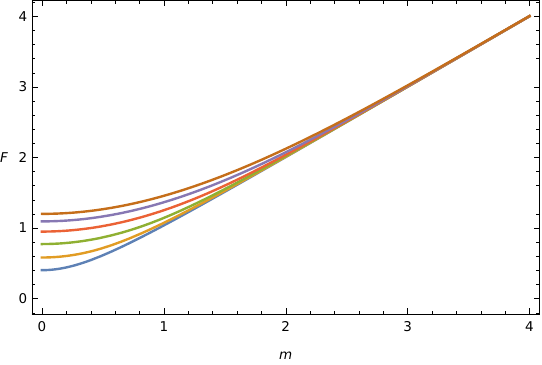}
	\includegraphics[height=0.32\textwidth]{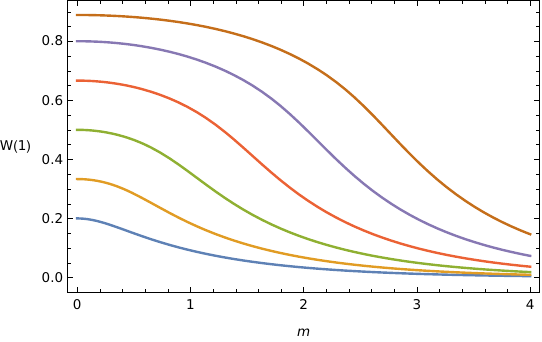}
	\includegraphics[height=0.32\textwidth]{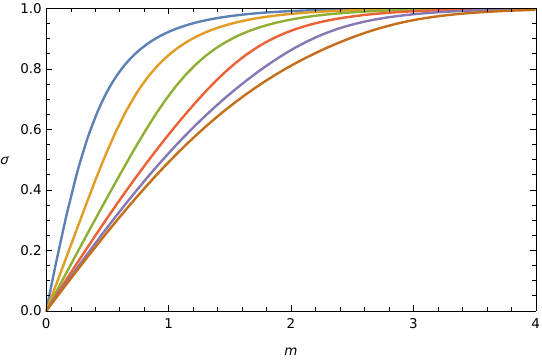}}
	\end{center}
	\caption{Free energy (left), Polyakov loop (middle) and quark condensate (right) as functions of mass for $U(4)$ at various $N_f$: 1 (dark blue), 2 (yellow), 4 (green), 8 (red), 16 (violet), 32 (brown).}
	\label{u4_fig1}
\end{figure}
Similar expressions for the $U(2)$ model have the form
\begin{gather}
	Z=(2\sinh m)^{2N_f}\left[(P_{N_f}(\coth m))^2-(N_f+1)^{-2}(P_{N_f}^1(\coth m))^2\right] \ , 
\end{gather}
\begin{gather}
	W(1) = \frac{P_{N_f}^1(\coth m)}{2(N_f+1)}\frac{P_{N_f}(\coth m)-(N_f+1)^{-1}(N_f+1)^{-2}P_{N_f}^2(\coth m)}{(P_{N_f}(\coth m))^2-(N_f+1)^{-2}(P_{N_f}^1(\coth m))^2} \ , 
\end{gather}
\begin{gather}
	\sigma=2\frac{N_f P_{N_f-1}(\coth m)P_{N_f}(\coth m)-(N_f+1)^{-1}P_{N_f-1}^1(\coth m)P_{N_f}^1(\coth m)}{(P_{N_f}(\coth m))^2-(N_f+1)^{-2}(P_{N_f}^1(\coth m))^2} .
\end{gather}
More explicit expressions of the partition functions of non-Abelian $U(N\geq 2)$ models are collected in Appendix \ref{unpf_list}. Though, in general, these expressions are more complicated, the qualitative behavior of non-Abelian models is very similar to the $U(1)$ model. To illustrate this we show in Fig.\ref{u4_fig1} the plots of the free energy, the Polyakov loop and the quark condensate for the $U(4)$ model.

\begin{figure}[htb]
	\begin{center}
	\resizebox{.95\textwidth}{!}{%
	\includegraphics[height=0.32\textwidth]{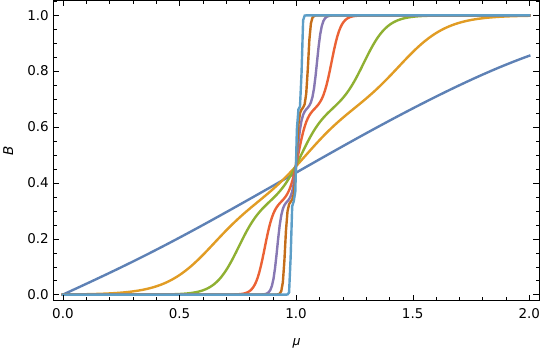} 
	\includegraphics[height=0.32\textwidth]{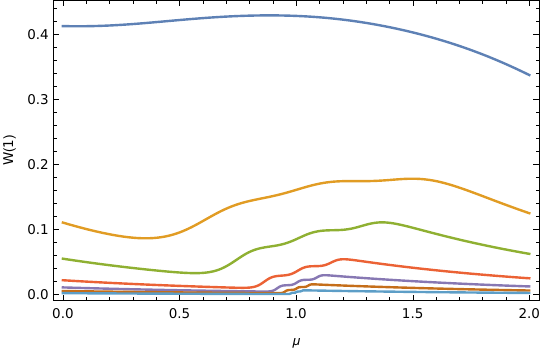} 
	\includegraphics[height=0.32\textwidth]{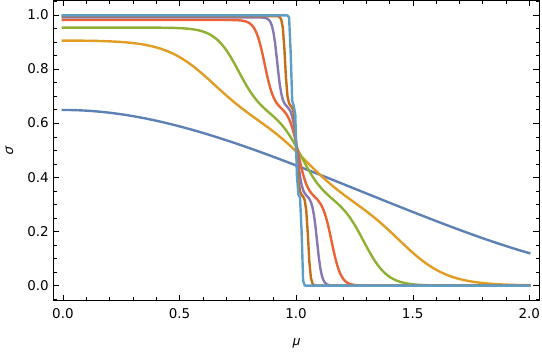} }
	\end{center}
	\caption{Baryon density (left), Polyakov loop (middle) and quark condensate (right) at $m=1$ as functions of chemical potential for $SU(N)$ at various $N$:
	3 (dark blue), 10 (yellow), 20 (green), 50 (red), 100 (violet), 200 (brown), 500 (light blue), $N_f=3$.}
	\label{sun_nf3_mu}
\end{figure}

\begin{figure}[htb]
 \begin{center}
	\includegraphics[width=0.32\textwidth]{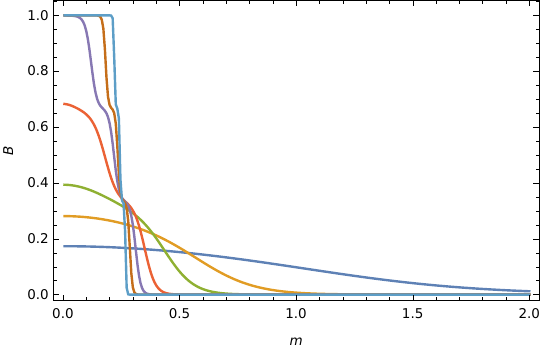} 
	\includegraphics[width=0.32\textwidth]{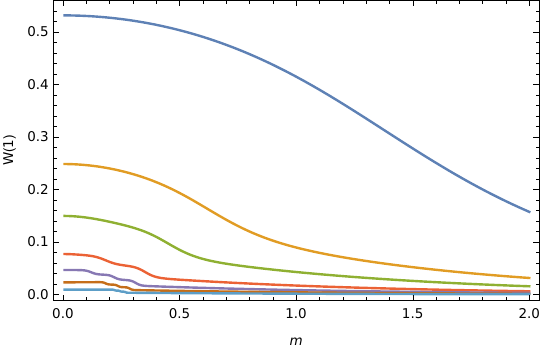} 
	\includegraphics[width=0.32\textwidth]{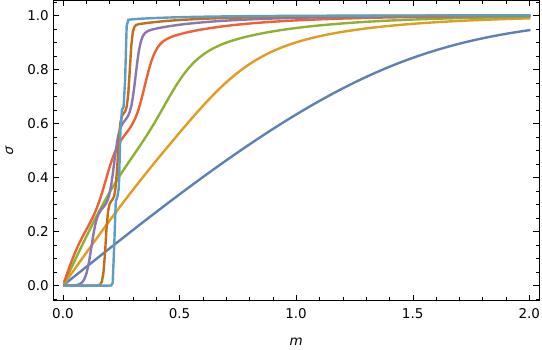} 
	\end{center}
	\caption{Baryon density (left), Polyakov loop (middle) and quark condensate (right) at $\mu = 0.25$ as functions of mass for $SU(N)$ at various $N$:
	3 (dark blue), 10 (yellow), 20 (green), 50 (red), 100 (violet), 200 (brown), 500 (light blue), $N_f=3$.}
	\label{sun_nf3_m}
\end{figure}

\begin{figure}[htb]
	\begin{center}
	\includegraphics[width=0.32\textwidth]{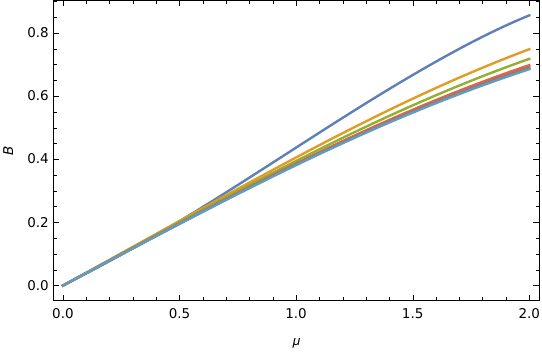} 
	\includegraphics[width=0.32\textwidth]{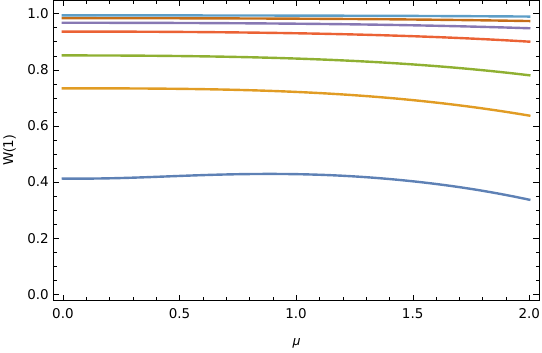} 
	\includegraphics[width=0.32\textwidth]{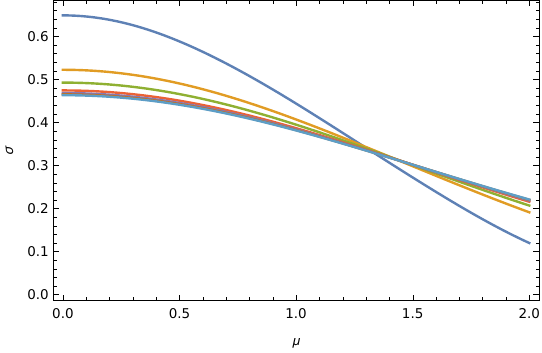} 
	\end{center}
	\caption{Baryon density (left), Polyakov loop (middle) and quark condensate (right) at $m=1$ as  functions of chemical potential for $SU(3)$ at various $N_f$:
	3 (dark blue), 10 (yellow), 20 (green), 50 (red), 100 (violet), 200 (brown), 500 (light blue).}
	\label{su3_mu}
\end{figure}

\begin{figure}[htb]
 \begin{center}
	\includegraphics[width=0.32\textwidth]{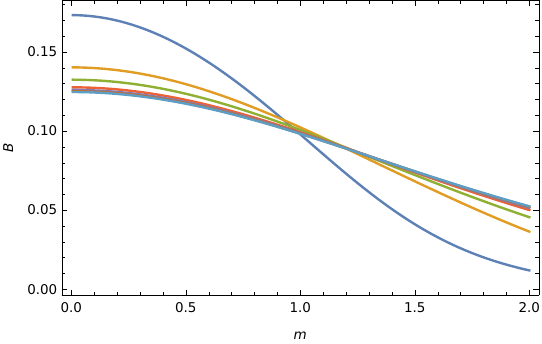} 
	\includegraphics[width=0.32\textwidth]{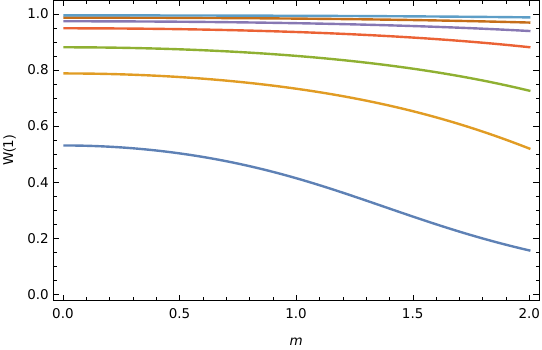} 
	\includegraphics[width=0.32\textwidth]{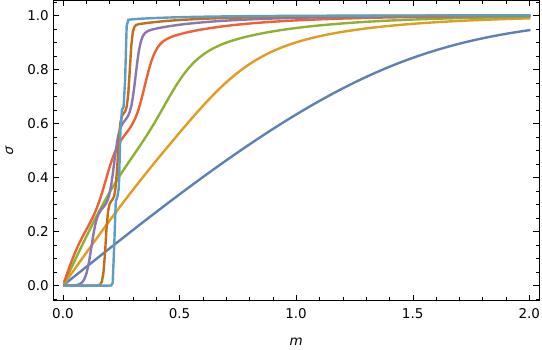} 
	\end{center}
	\caption{Baryon density (left), Polyakov loop (middle) and quark condensate (right) at $\mu = 0.25$ as functions of mass for $SU(3)$ at various $N_f$:  3 (dark blue), 10 (yellow), 20 (green), 50 (red), 100 (violet), 200 (brown), 500 (light blue).}
	\label{su3_m}
\end{figure}

A convenient form of the $SU(N)$ partition function valid for all $N$ and  $N_f$ reads
\begin{equation}
\label{sun_pf_genform}
Z(N,N_f) = Z_0(N,N_f) + 2 \sum_{q=1}^{N_f-1} \ Z_q(N,N_f)
\cosh q N \mu  + 2 \cosh N N_f \mu \ , 
\end{equation}
where $Z_q(N,N_f)$ is defined in Eq.(\ref{PF_sun_def}). 
Various explicit expressions of the $SU(N)$ partition functions are 
presented in Appendix \ref{sunpf_list}. They can be used to calculate 
the free energy, the baryon density and the quark condensate. In order to compute the Polaykov loop expectation value it is more convenient to use the  Eq.(\ref{sun_pl_gen}). To derive the small $h$-expansion, which we will need later, it is even better to start from the representation (\ref{PF_form_3}) which gives directly the small $h$-expansion.  We find 
\begin{eqnarray}
\label{Wr_sun_res}
&&W(r) =  e^{-\mu r} \ h^{r} \ \frac{G(N+N_f+1)}{N G(N_f+1)} \\
&&\times \left(  \sum_{k=1}^{N} 
\frac{\prod_{1\le i<j\le N} (r (\delta_{i,k}-\delta_{j,k})+j-i) }
{\prod_{i=1}^N (r \delta_{i,k}-i+N)! (N_f- r\delta_{i,k}+i-1)!} + {\cal{O}}(h^{2}) \right) \ .  \nonumber 
\end{eqnarray}
Let us now look at the behavior of different observables in the $SU(N)$ model. 
At fixed $N_f$ and large $N$ one can see $N_f$ ``smooth jumps'' in the baryon density and the quark condensate corresponding to the change of $q$ variable that gives the dominant contribution in Eq.~(\ref{sun_pf_genform}). The Polyakov loop dependence on $m$ and $\mu$ shows irregularities at the positions of the ``jumps''. When $N$ grows the ``jumps'' become more pronounced and in the limit $N \to \infty$ the positions of all ``jumps'' go to the point $m = \mu$ (see Figs.~\ref{sun_nf3_mu},\ref{sun_nf3_m}).
At fixed $N$ and large $N_f$ the change between the values of $q$ is smooth, resulting in a more regular behavior of the observables
(see Figs.~\ref{su3_mu},\ref{su3_m}).

\begin{figure}[htb]
	\begin{center}
	   \includegraphics[width=0.5\textwidth]{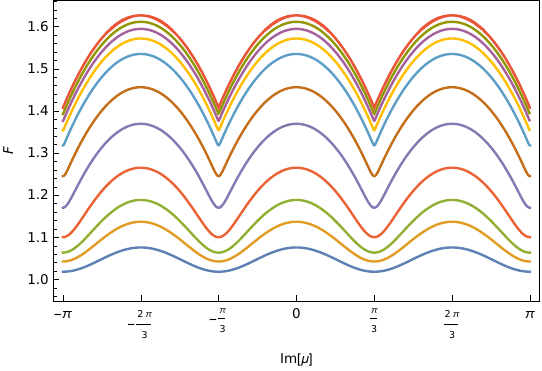} 
	\end{center}
	\caption{Free energy as a function of the imaginary $\mu$ for $m=1$, $N=3$, $N_f = 1,2,3,5,10,20,50,100,200,500,\infty$ (from bottom to top).}
	\label{SU3_Roberge_Weiss}
\end{figure}

In the case of a degenerate quark mass $m_1 = m_2$, it is useful to define the quark and the isospin chemical potentials as 
\begin{equation}
    \mu_1 = \mu_q + \mu_I \ , \quad \mu_2 = \mu_q - \mu_I \ .
\end{equation}
Note that in this notation $\mu_q = \frac{\mu_B}{3}$, where $\mu_B$ is the baryon chemical potential. Since in this case $\mu_1 \neq \mu_2$ we use (\ref{PF_sun_def}) to calculate the partition function and the observables.
For finite $N$ and $N_f$ the partition function is analytic in $m$ and $\mu$, so the model does not exhibit any phase transition. This changes drastically in the large $N_f$ limit. In order to see this, consider arbitrary complex values of $\mu_q$, $\mu_I$. Let us remind that the $4d$ QCD for $\mu_I = 0$ undergoes a first order Roberge-Weiss phase transition at $\mu_q = \frac{2 (k + 1) i \pi}{N}$ when the temperature is large enough \cite{roberge_weiss}. 
This transition is restored in the large $N_f$ limit of one-dimensional theory. Indeed, the saddle point of the integrand in Eq.~(\ref{PF_1dqcd_int2}) appears at $\omega_k = \frac{2 \pi n}{N}$, thus the free energy in the large $N_f$ limit for purely imaginary $\mu_q$ becomes 
\begin{equation}
F = \max_{n = 0..N-1}  \ln \left(1 + h^2 + 2 h \cos\left(\frac{2 \pi n}{N} - i \mu_q\right)\right) \ .
\end{equation}
This results in a first order transition at $\mu_q = \frac{\pi i (2 n + 1)}{N}$. The corresponding free energies are shown in Fig.~\ref{SU3_Roberge_Weiss} for $N=3$ and various values of $N_f$. One sees how the non-analytic behavior restores with increasing $N_f$.  

\begin{figure}[htb]
	\begin{center}
	\includegraphics[height=0.32\textwidth]{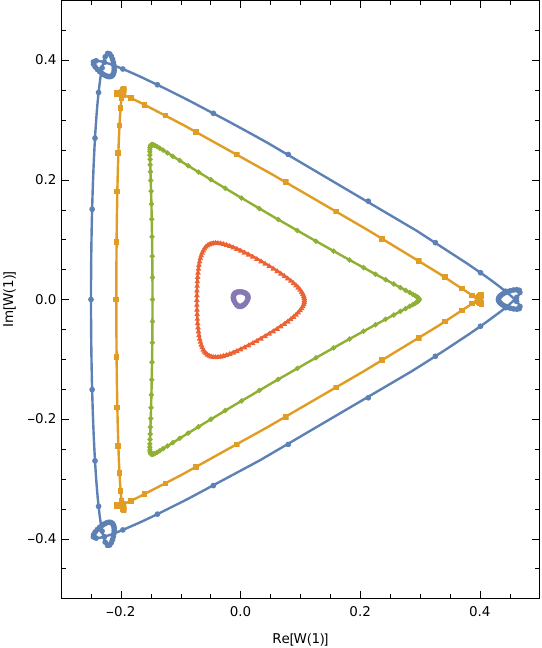} 
	\includegraphics[height=0.32\textwidth]{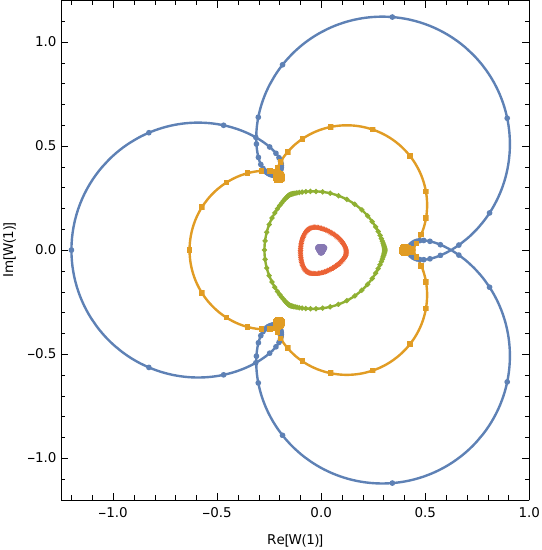} 
	\includegraphics[height=0.32\textwidth]{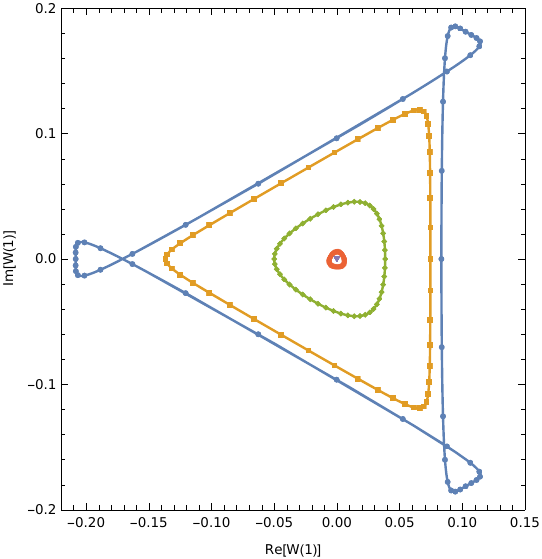} 
	\end{center}
	\caption{Parametric plots of the Polyakov loop in the $SU(3)$ model for $N_f=2$ at complex chemical potential. 
		The parameter $\mu$ in each plot changes from $-\pi$ to $\pi$, dots are set at equal distances in $\mu$ to show the speed of the Polyakov loop change with $\mu$ (100 dots on each line).
		Left panel: $\mu_q = i \mu,\ \mu_I = 0$. Central panel: $\mu_q = 0.2 + i \mu,\ \mu_I = 0$. Right panel: $\mu_q = i \mu,\ \mu_I = i \frac{\pi}{2}$. $m = 0$ (blue), $m = 0.5$ (yellow), $m=1$ (green), $m = 2$ (red), $m = 4$ (violet). }
	\label{SU3_imaginary_potential_parametric_polyakov}
\end{figure}

\begin{figure}[htb]
	\begin{center}
		\includegraphics[width=0.32\textwidth]{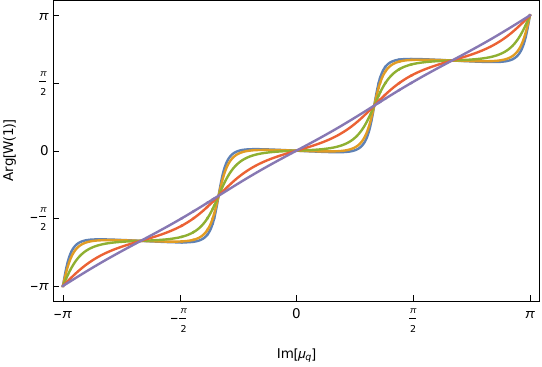} 
		\includegraphics[width=0.32\textwidth]{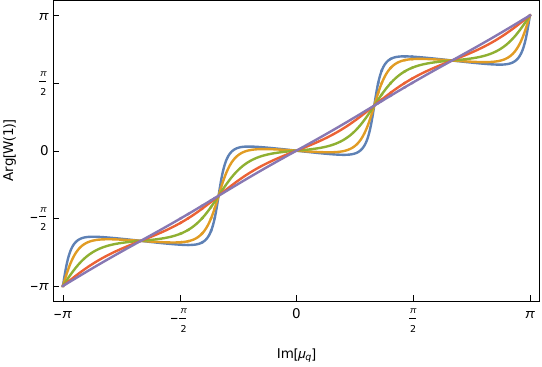} 
		\includegraphics[width=0.32\textwidth]{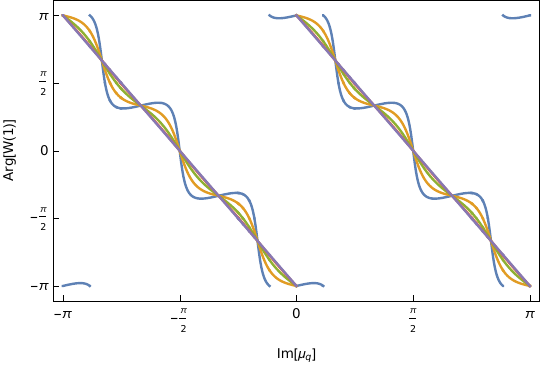} 
	\end{center}
	\caption{Argument of the Polyakov loop for the same parameters as in Figure~\ref{SU3_imaginary_potential_parametric_polyakov}.}
	\label{SU3_imaginary_potential_argument_polyakov}
\end{figure}

While the free energy is analytic at finite $N_f$, the behavior of the Polyakov loop shows the remnants of the Roberge-Weiss transition. This is shown 
on the left plots in Figs.~\ref{SU3_imaginary_potential_parametric_polyakov} and \ref{SU3_imaginary_potential_argument_polyakov}: at small $m$ the Polyakov loop moves around three $Z(3)$ symmetric points with the transition region between points becoming smaller as $m$ decreases. Hence, instead of the first order transition in $4d$ QCD we see a smooth change between Roberge-Weiss phases
that gets sharper at smaller $m$. 

A similar picture can be seen when the quark chemical potential has nonzero real part (middle plots in Figs.~\ref{SU3_imaginary_potential_parametric_polyakov} and \ref{SU3_imaginary_potential_argument_polyakov}). The points corresponding to the Roberge-Weiss phases and the argument of the Polyakov loop do not show significant difference, but the trajectory of the parametric plot connecting the Roberge-Weiss shapes moves significantly outwards, resulting in the absolute values of the Polyakov loop that is larger than 1 for some parameters. 

A generalization of the Roberge-Weiss phase diagram to nonzero $\mu_I$ can be found in \cite{imaginary_isospin_lattice,imaginary_isospin}.
In this case the phase plane (the phase torus, to be precise) is divided into 6 regions corresponding to 3 Roberge-Weiss phases. 
In one-dimensional model adding $\mu_I = i \pi$ is equivalent to a shift in $\pi$ of $\mu_q$ and taking $\mu_I = i \frac{\pi}{2}$. 
This change in $\mu_q$ allows us to cross all 6 regions in the $4d$ QCD phase diagram as it results in 6 "plateau" regions on the 
$\mathrm{arg}\ W(1)$ plot and the Polyakov loop becomes $i \pi$-periodic in $\mu_q$. This is demonstrated in the right plots in Figs.~\ref{SU3_imaginary_potential_parametric_polyakov} and \ref{SU3_imaginary_potential_argument_polyakov}.

Overall we see that the Roberge-Weiss phase transition exists in one-dimensional QCD in the large $N_f$ limit. At finite $N_f$ there is no phase transition but one can still see the remnants of the large $N_f$ phase structure.

\begin{figure}[htb]
	\begin{center}
		\includegraphics[width=0.45\textwidth]{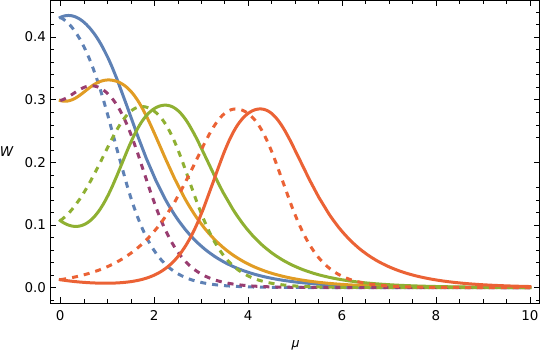} 
		\includegraphics[width=0.45\textwidth]{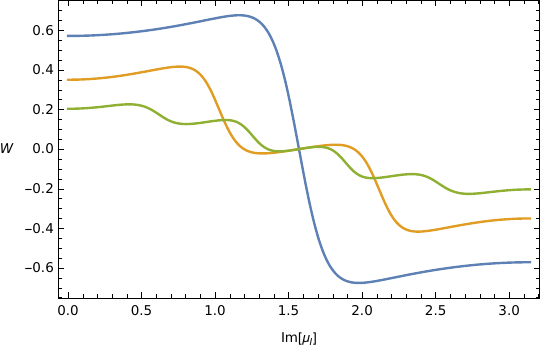} 
	\end{center}
	\caption{Left panel: Polyakov loop and its conjugate (dashed line) as a function of the baryon chemical potential for $N=3$, $N_f=2$ and various $m$: $m=0$ (blue), $m = 1$ (yellow), $m = 2$ (green), $m = 4$ (red). Right panel: Polyakov loop as a function of imaginary isospin potential and vanishing baryon potential for $N_f=2$, $m=0$ and various $N$: $N=2$ (blue), $N = 4$ (yellow), $N=8$ (green).}
	\label{un_fig3_PL}
\end{figure}

Finally, let us comment briefly on the dependence of the Polyakov loop on the chemical potentials.  
At nonzero baryon chemical potential the model loses the invariance with respect to the complex conjugation of the gauge field, so 
in general $W(r) \neq W(-r)$. Both the Polyakov loop and its conjugate remain real but not necessarily equal. The dependence of the Polyakov loop and its conjugate on the chemical potential $\mu$ is shown in the left panel of Fig.~\ref{un_fig3_PL} for $r=1$. For nonzero isospin chemical potential (and zero baryon chemical potential), the symmetry between $W(r)$ and $W(-r)$ is restored. The dependence of the Polyakov loop on the isospin chemical potential is shown in the right panel of Fig.~\ref{un_fig3_PL}.

\newpage

\section{The 't Hooft-Veneziano limit}
\label{HV_limit}

This Section is devoted to the derivation of the 't Hooft-Veneziano limit of one-dimensional QCD. The limit is defined as
\begin{equation}
\label{HV_def}
N \to \infty \ , \ N_f \to \infty  \ \ \mbox{such \ that} \ \
\frac{N_f}{N} = \kappa \ \ \mbox{is \ kept \ fixed} \ .
\end{equation}
This limit turns out to be the most interesting region. Here, the model exhibits a non-trivial phase structure which will be described in detail both for $U(N)$ and $SU(N)$ QCD with degenerate flavors. We shall also calculate the limits $N\to\infty$, $N_f$ is fixed and $N_f\to\infty$, $N$ is fixed as limits $\kappa\to 0$ and $\kappa\to\infty$, correspondingly.

\subsection{U(N) model}
\label{un_model}

To derive the t' Hooft-Veneziano limit for the $U(N)$ model we use the orthogonal polynomial method \cite{orthog_polynom} in conjunction with Eqs.(\ref{PF_1dqcd_det_leg}) and (\ref{sun_pl_gen}). Details of the derivation are presented in Appendix \ref{orthog_pol_app}. For the free energy we find the following answer
\begin{equation}
	F =
	\begin{cases}
		- \ln h - \kappa \ \ln (1-h^2)  \ , \  h < \frac{1}{2 \kappa +1} \ ,  \\
		- k \ln 4 k -\frac{(k+1)^2}{\kappa} \ln (k+1) + \frac{(2 k+1)^2}{2 \kappa}
		\ln (2 k+1)   \\
		+ \frac{2 k+1}{2 \kappa}  \ln \frac{1}{4}(1+h)(1+ h^{-1})\ , \
		\frac{1}{2 \kappa +1} <  h < 2 \kappa +1  \ , \\
		\ln h - \kappa \ \ln (1-h^{-2}) \ , \ h > 2 \kappa +1  \ .
	\end{cases}
\label{UNfren_HVlimit}
\end{equation}
The expectation values of the Polyakov loop are found to be
\begin{gather}
	W(1,\mu) = W(-1,-\mu) =
	\begin{cases}
			\kappa h \ e^{-\mu} \ , \
			& h < \frac{1}{2 \kappa +1}  \ , \\
		e^{-\mu}\left(1-\frac{2+ h + h^{-1}}{4(1+\kappa )}\right) \ , \
		& \frac{1}{2 \kappa +1} <  h < 2 \kappa +1     \ , \\
		\kappa h^{-1}  \ e^{-\mu} \ , \
		& h > 2 \kappa +1 \ .
	\end{cases}
\label{UNPL_HVlimit}
\end{gather}
We have also used the symmetries of the partition function, Eq.(\ref{PF_1dqcd_sym}), to extend results to the region $h>1$ which formally corresponds to a non-physical region of negative masses. These results were checked for small $h$ (large fermion mass) and for $h=1$ (massless fermions) by making use of the representation (\ref{PF_1dqcd_sumpart_deg}).

\begin{figure}[H]
	\centerline{
	\resizebox{.95\textwidth}{!}{%
	{\epsfysize=6.6cm \epsfbox{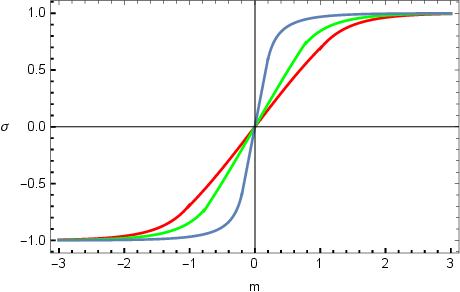}}
	{\epsfysize=6.6cm \epsfbox{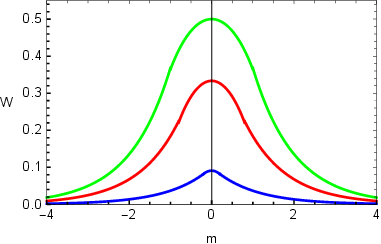}}}}
	\caption{Plots of the quark condensate (left) and the Polyakov loop (right) as functions of mass for $\kappa = 0.1$ (blue), $\kappa = 0.5$ (red),
	$\kappa = 1$ (green). }
	\label{hv_un_conden_pl}
\end{figure}

For all values of $\kappa \ne 0$ there is a third order phase transition at the critical points
\begin{equation}
	h= 1/(2 \kappa +1) \ \ , \ \  h = 2 \kappa+1 \ .
	\label{hv_un_critpoint}
\end{equation}
The third derivative of the free energy exhibits a finite jump
\begin{equation}
\Delta F^{\prime\prime\prime} = \frac{(1 + 2 \kappa)^5}{4 \kappa^2 (1 + \kappa)} \ .
\label{hv_un_jump}
\end{equation}
As expected, the free energy does not depend on the baryon chemical potential, therefore the baryon density vanishes identically. The result for the quark condensate we present as
\begin{gather}
\sigma =
\begin{cases}
	\kappa (1-\coth m) +1 \ , \
	& m > \ln(2 \kappa +1)  \ , \\
	\frac{2 \kappa +1 }{2 \kappa }\tanh \frac{m}{2} \ , \
	& - \ln(2 \kappa +1) <  m < \ln(2 \kappa +1) \ , \\
	-\kappa (1+\coth m) -1 \ , \
	& m < - \ln(2 \kappa +1) \ .
\end{cases}
\label{hv_un_sigma}
\end{gather}
In Fig.\ref{hv_un_conden_pl} we show the behavior of the quark condensate and the Polyakov loop as a function of mass for various values of $\kappa$ and $\mu=0$. When $\kappa \to 0$ the quark condensate approaches a threshold transition at $m=0$.

\begin{figure}[H]
	\begin{center}
	\resizebox{.95\textwidth}{!}{%
		\includegraphics[height=0.48\textwidth]{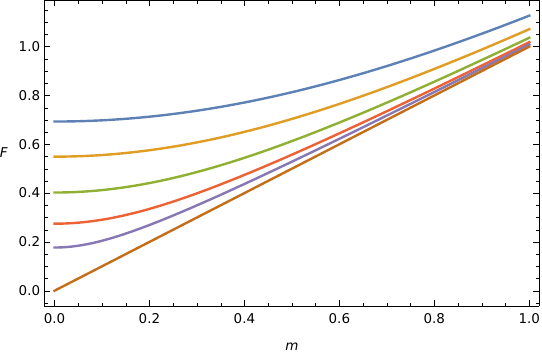} 
		\includegraphics[height=0.48\textwidth]{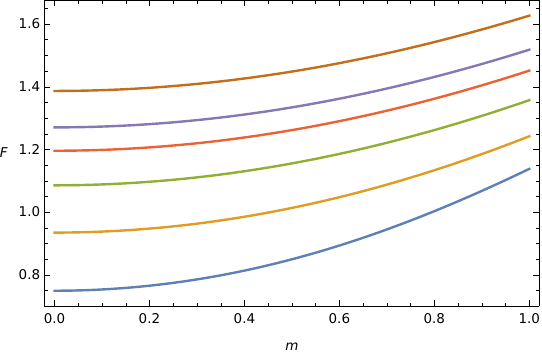}}
	\end{center}
	\caption{Left panel: Convergence of $U(N)$ free energy to the large $N$ limit (lower brown line) at fixed $N_f=1$ for $N=1$ (blue), $N=2$ (yellow), $N=4$ (green), $N=8$ (red), $N = 16$ (violet). Right panel: Convergence of $U(N)$ free energy to the large $N_f$ limit (upper brown line) at fixed  $N=2$ for $N_f=2$ (blue), $N_f = 4$ (yellow), $N_f = 8$ (green), $N_f = 16$ (red), $N_f = 32$ (violet).}
	\label{un_fren_largeN_Mf}
\end{figure}

Eq.(\ref{UNfren_HVlimit}) allows us to calculate two different limits as follows.
\begin{enumerate}
	\item
$N_f$ is fixed, $N\to\infty$. This corresponds to $\kappa\to 0$ limit. One finds
	\begin{equation}
		F =
		\begin{cases}
			- \ln h   \ , \ h < 1 \ ,  \\
			\ln h  \ , \ h > 1 \ .
		\end{cases}
		\label{hv_un_largeN}
	\end{equation}
It follows the quark condensate exhibits a threshold transition at $m=0$. The left panel of Fig.\ref{un_fren_largeN_Mf} shows the convergence of the free energy to this limiting behavior.
	
	\item
$N$ is fixed, $N_f\to\infty$. This corresponds to $\kappa\to \infty$ limit. One finds
	\begin{equation}
		F = - \ln h + 2 \ln (1+h)  \ .
		\label{hv_un_largeNf}
	\end{equation}
This limit can be obtained directly from the asymptotic expansion of the Legendre function  in Eq.(\ref{Legendre_asymp_uniform}). No transition occurs in this limit. The right panel of Fig.\ref{un_fren_largeN_Mf} shows the convergence of the free energy to this limiting behavior.
	
\end{enumerate}

\subsection{SU(N) model}
\label{sun_model}

The orthogonal polynomial method is not very efficient for the $SU(N)$ model except for the case of one fermion flavor. In particular, it seems to be a very non-trivial problem to construct a set of orthogonal polynomials for arbitrary values of $q$ as $N_f\to\infty$. Therefore, to study $SU(N)$ QCD in the 't Hooft-Veneziano limit we used the representation (\ref{PF_1dqcd_sumpart_deg}) and checked results with the help of Eqs.(\ref{PF_form_3}) and (\ref{Zq_def_1}). Even in this case we could not construct the exact solution for the full model. Exact solutions we have found in two cases: 1) for the reduced model corresponding to vanishing value of $h_+$ or $h_-$; 2) for the model with  massless fermions, $h=1$. For the full model we employed the strategy used by us in \cite{un_vs_sun}, namely starting from Eq.(\ref{PF_1dqcd_sumpart_deg}) we obtain an effective action as a power series in $h$ to the very high order and then calculate the final sum over $q$ by a saddle-point method in the 't Hooft-Veneziano limit. While this approach does not give an exact solution, it turns out to be sufficient to reveal the phase diagram of the model.

\subsubsection{Reduced and massless models}
\label{reduced_sun}

\begin{figure}[htb]
	\centerline{
	\resizebox{.95\textwidth}{!}{%
	\epsfysize=7cm \epsfbox{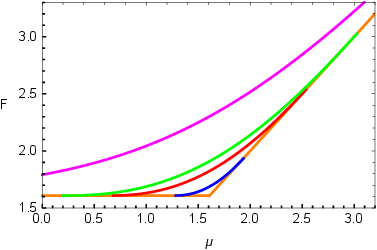}
	\epsfysize=7cm  \epsfbox{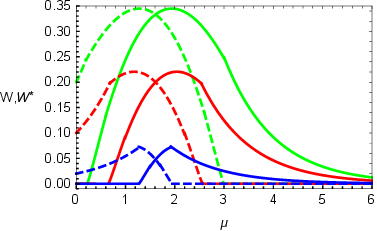}}}
	\caption{\label{Fig:sun_red_fren} Left panel: the free energy as a function of $\mu$ at $h=0.2$ and $\kappa=0$ (orange), $\kappa=0.1$ (blue), $\kappa=0.5$ (red), $\kappa=1$ (green), $\kappa=\infty$ (magenta). Right panel: Polyakov loops $W(1), W^{*}(1)$ (dashed lines) as functions of $\mu$ for $\kappa = 0.1,0.5, 1$.}
\end{figure}

\begin{figure}[htb]
	\centerline{
	\resizebox{.95\textwidth}{!}{%
	\epsfysize=5.0cm  \epsfbox{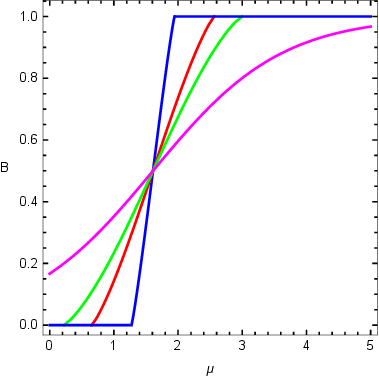} 
	\epsfysize=5.0cm  \epsfbox{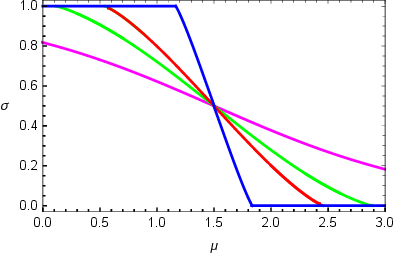} }}
	\caption{\label{Fig:sun_red_bar_ch} The baryon density (left) and the quark condensate (right) as functions of $\mu$ at $h=0.22$ and $\kappa=0.1, 0.5, 1, \infty$. The color legend is as in Fig.\ref{Fig:sun_red_fren}.}
\end{figure}

The reduced model corresponds to the heavy-dense limit of the quark determinant, therefore even if it is an approximation it describes an important physical limit. Let us consider, for definiteness, the limit $h_-=0$
\begin{eqnarray}
\label{PF_1dqcd_reduced}
Z  = A \int_{G} \ dU  \ \prod_{k=1}^N
\left [ 1 + h_+ e^{i\omega_k} \right ]^{N_f} \ , \ A = h^{-N N_f} = e^{N N_f m} \ .
\end{eqnarray}
If $G=U(N)$, the integration is trivial and $Z=A$.
If $G=SU(N)$, Eq.(\ref{PF_1dqcd_sumpart}) takes a particularly simple form
\begin{eqnarray}
\label{PF_sun_red1}
Z  =  A \sum_{q=0}^{N_f}  h_+^{N q} \ s_{0}(1^{N_f}) \ s_{N^q}(1^{N_f})
=  A \sum_{q=0}^{N_f} \ h_+^{N q} \ C_{N,N_f}(q)  \  , \\
\label{CNNf_def}
	C_{N,N_f}(q) = \frac{G(N+1) G(N+N_f+1) G(q+1) G(N_f+1-q)}{G(N_f+1) G(N+q+1) G(N+N_f+1-q)} \ ,
\end{eqnarray}
where $G(X)$ is the Barnes function. To calculate the t' Hooft-Veneziano limit we take the asymptotics of the Barnes function, introduce a new variable $u = q/N $ and replace summation over $q$ by the integration over $u$ in the large $N$ limit. This results in
the following representation
\begin{eqnarray}
\label{PF_sun_red2}
Z  &=&  N h^{- N N_f} \int_{0}^{\kappa}  d u \
e^{ N N_f \left ( \frac{u}{\kappa} \ln h_+ + f(u, \kappa) \right ) }  \  ,  \\
\label{f_hk_def}
f(u, \kappa) &=& \frac{1}{2 \kappa} \bigl[ u^2 \ln u - (1 + u)^2 \ln (1 + u) + (\kappa - u)^2 \ln (\kappa - u)   \\
&+& (1 + \kappa)^2 \ln (1 + \kappa) - \kappa^2 \ln \kappa - (1 + \kappa - u)^2
\ln (1 + \kappa - u) \bigr] \ . \nonumber
\end{eqnarray}
The last integral is calculated by the saddle-point approximation.
The saddle-point equation
\begin{eqnarray}
\label{sun_saddle_red}
h_+ = \frac{(\kappa - u)^{\kappa - u}(1 + u)^{1 + u}}{(1 + \kappa - u)^{1 + \kappa - u} u^{u} }
\end{eqnarray}
has no solution if $h_+$ is small. The maximum of the integrand in (\ref{PF_sun_red2}) is achieved for $u=0$ and the partition function equals $U(N)$ partition function. There is no $\mu$-dependence in this regime, so the baryon density vanishes.
When $h_+$ ({\it i.e.}, $\mu$) grows non-trivial solutions appear. They can be found around $u\sim 0$ and $u\sim \kappa$
\begin{eqnarray}
\label{sun_red_u_solution}
u_s = \frac{z_1}{W_{-1} \left(\frac{\kappa \, z_1}{e (\kappa+1)}\right)} +
{\cal{O}}\left ( z_1^2 \right ) \ , \
u_s =  \kappa -\frac{z_2}{W_{-1}\left(\frac{\kappa \, z_2}{e (\kappa+1)}\right)} +
{\cal{O}}\left ( z_2^2 \right ) \ ,
\end{eqnarray}
where $W_{-1}(x)$ is a lower branch of the Lambert function and
\begin{eqnarray}
\label{sun_red_z1}
z_1 &=& \kappa \ln \kappa - (1 + \kappa) \ln (1 + \kappa) - \ln h_+  \ ,  \\
\label{sun_red_z2}
z_2 &=& \kappa \ln \kappa - (1 + \kappa) \ln (1 + \kappa) + \ln h_+   \ .
\end{eqnarray}
Critical lines of the reduced model are found to be
\begin{equation}
\label{sun_red_critline}
z_1 \ = \ 0 \ \ , \ \ z_2 \ = \ 0 \ .
\end{equation}
There are two lines of phase  transitions given by
\begin{equation}
\label{sun_red_full_crit}
\kappa \ln \kappa - (1 + \kappa) \ln (1 + \kappa) \pm  \ln h  e^{\mu} = 0 \ .
\end{equation}
The free energy satisfies the equality
$F[h, u, \kappa] = F[1/h, \kappa - u, \kappa]+ \ln h$ and can be written in three phases as
\begin{equation}
F_1(u=0) = m \ , \ F_2(u_s) \ , \  F_3(u=\kappa) =  \mu \ .
\label{fren_sun_red}
\end{equation}
$F_2(u_s)$ can be expressed in terms of $z_1$ or $z_2$ using solutions  (\ref{sun_red_u_solution}). The simpler way is to express the free energy in terms
of the $u$ variable. Its physical meaning follows from (\ref{PF_sun_red2}): the expectation value of $u$ gives the baryon density. Then, using Eq.(\ref{sun_saddle_red}) one finds
\begin{eqnarray}
F_2(u) &=& m + 
\frac{1}{2 \kappa} \biggl [ \left(\kappa^2-u^2\right) \ln (\kappa-u)- \kappa^2 \ln \kappa -\left((\kappa+1)^2- u^2\right) \ln (\kappa-u+1) \nonumber  \\
\label{fren_sun_F2}
&+& (\kappa+1)^2 \ln (\kappa+1)- u^2 \ln u -\left(1-u^2\right) \ln (u+1) \biggr ] 
\ .
\end{eqnarray}
Using the same strategy one finds for the expectation value of the Polyakov loop
\begin{eqnarray}
\label{sun_red_pl1}
W(1) &=&  \frac{h^{- N N_f}}{Z} \ \sum_{q=1}^{N_f} h_+^{N q-1} \ C_{N , N_f}(q) \ \frac{q}{N_f + N - q} \ , \\
\label{sun_red_pl1_con}
W^{*}(1) &=&  \frac{h^{- N N_f}}{Z}  \ \sum_{q=0}^{N_f-1} h_+^{N q+1} \ 
C_{N , N_f}(q) \ \frac{(N_f- q)}{N+ q} \ .
\end{eqnarray}
In the 't Hooft-Veneziano limit this results in
\begin{eqnarray}
\label{sun_red_pl1_res}
W(1) = h_+^{-1} \ \left \langle  \frac{u}{1 + \kappa - u} \right \rangle \ =  \begin{cases}
	0 \ , \ u=0 \ , \\
	h_+^{-1} \ \frac{u_s}{1 + \kappa - u_s} \ , \  0<u<\kappa \ , \\
	\kappa \ h_+^{-1} \ , \ u=\kappa \ .
\end{cases} \\
\label{sun_red_pl1_con_res}
W^{*}(1) =  h_+ \ \left \langle  \frac{\kappa - u}{1 + u} \right \rangle =
\begin{cases}
\kappa \ h_+ \ , \ u=0 \ , \\
h_+ \ \frac{\kappa - u_s}{1 + u_s} \ , \  0<u<\kappa \ , \\
0 \ , \ u=\kappa \ .
\end{cases}
\end{eqnarray}
Here, the expectation values refer to the partition function (\ref{PF_sun_red2}) and $u_s$ is given by Eq.(\ref{sun_red_u_solution}). Phase transition along critical lines (\ref{sun_red_critline}) is of a third  order.
In the large $N$ limit ($\kappa  \to 0$) the middle phase shrinks to zero and the 3rd order phase transition turns into a threshold transition.
The large $N_f$ limit ($\kappa\to\infty$) can be calculated from Eq.(\ref{PF_sun_red2}).
There is one saddle-point solution in this limit $u_s=\frac{h_+}{1+h_+}$ and the free energy becomes
\begin{equation}
\label{sun_red_latgeNf}
F = - \ln h + \ln (1+h_+) \ .
\end{equation}
Remembering that $h_+=h e^{\mu}$ we can plot the free energy and all observables as functions of the chemical potential. 
Plots of the free energy and the Polyakov loop are shown in Fig.\ref{Fig:sun_red_fren}. 
Plots of the baryon density and the quark condensate are shown in Fig.\ref{Fig:sun_red_bar_ch}.
Both the  baryon density and the quark condensate exhibit the approach to the threshold transition around $\mu\approx -\ln h$ with $\kappa$ decreasing.

We turn now to the massless model $h=1$. Starting from Eq.(\ref{PF_form_2}) and  using
\begin{eqnarray}
	\label{binom_sum}
	\sum_{l=0}^{N_f} \ \binom{N_f}{l} \binom{N_f}{l-i +j +q} = \binom{2 N_f}{N_f-i+j +q}
\end{eqnarray}
we obtain
\begin{eqnarray}
\label{sun_h1_pf1}
Z = \sum_{q=-N_f}^{N_f} e^{\mu q N} \
\underset{1\le i,j \le N}{\det} \ \binom{2 N_f}{N_f-i+j + q} =
\sum_{q=-N_f}^{N_f} e^{\mu q N} \ K_{N,N_f}(|q|) \ ,  \\
\label{sun_h1_Cnnf}
K_{N,N_f}(q) = \frac{ G(N+ 2 N_f+1) G(N+1) G(N_f-q+1) G(N_f+q+1) }
{ G(2 N_f+1) G(N+ N_f- q+1) G(N+ N_f+ q+1) } \ .
\end{eqnarray}
Proceeding as in the case of the reduced model we end up with the following partition function in the t' Hooft-Veneziano limit
\begin{eqnarray}
	\label{sun_h1_pf2}
&&Z  = N \int_{- \kappa}^{\kappa}  d u \
	e^{ N N_f \left ( \frac{u}{\kappa} \mu + g(u, \kappa) \right ) }  \  ,  \\
&&g(u, \kappa) = \frac{1}{2 \kappa } \bigl [ (1 + 2 \kappa)^2 \ln (1 + 2 \kappa)+(\kappa-u)^2 \ln (\kappa-u)  + (\kappa+u)^2 \ln (\kappa+u) \nonumber  \\
\label{g_h1_def}
&& - 4 \kappa^2 \ln 2 \kappa - (\kappa-u+1)^2 \ln (\kappa-u+1) - (\kappa+u+1)^2
	\ln (\kappa+u+1) \bigr ]  \ .
\end{eqnarray}
There are three types of solutions of the saddle-point equation. At $u \simeq 0$ we have
\begin{equation}
\label{sun_h1_us1}
u_s^{(1)} = \frac{\mu}{2 (\ln (\kappa+1) -\ln \kappa)} +
{\cal{O}}\left ( \mu^2 \right ) \ .
\end{equation}
Near $u \simeq \pm \kappa $ we have
\begin{equation}
\label{sun_h1_us2}
u_s^{(2)} = \pm \kappa \mp \frac{z \pm \mu }{W_{-1}\left[\frac{2 \kappa (z \pm \mu) }{e (2 \kappa +1)}\right] } + {\cal{O}}\left ( (z \pm \mu)^2 \right ) \ ,
\end{equation}
where $z=  2 \kappa \ln 2 \kappa  - (2 \kappa +1) \ln (2 \kappa +1) $.
One finds a third order phase transition along the critical line
\begin{equation}
	z \pm \mu =0 \ .
	\label{sun_h1_critline}
\end{equation}
Above the critical line one has $u_s^{(3)} = \pm \kappa$ and the free energy
equals $F=  |\mu|$.

\begin{figure}[H]
	\centerline{\epsfxsize=7.0cm \epsfbox{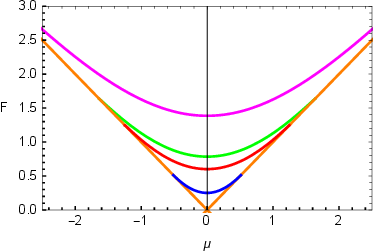} \epsfxsize=7.0cm \epsfbox{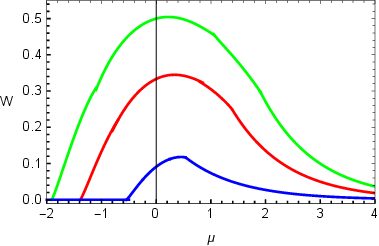} }
	\caption{\label{Fig:sun_h1_fren} Left panel: the free energy of the massless model at fixed $\kappa=0,0.1, 0.5, 1,\infty$ as a function of $\mu$. Right panel: the Polyakov loop $W(1)$  at fixed $\kappa=0.1,0.5,1$ as a function of $\mu$. Color legend: $\kappa=0$ (orange), $\kappa=0.1$ (blue), $\kappa=0.5$ (red), $\kappa=1$ (green), $\kappa=\infty$ (magenta).}
\end{figure}

For the expectation value of the Polyakov loop we find
\begin{eqnarray}
\begin{cases}
W (1) \\ W^{*}(1)
\end{cases}
= \frac{e^{ \mp \mu}}{Z} \ \int_{- \kappa}^{\kappa} du \
e^{ N N_f \left ( \frac{u}{\kappa} \mu + g(u, \kappa) \right ) } \
\frac{\kappa \pm  u }{1+\kappa  \mp u } =
e^{ \mp \mu} \ \frac{\kappa \pm  u_s^{(i)} }{1+\kappa  \mp u_s^{(i)} } \ .
\label{sun_h1_pl}
\end{eqnarray}
Plots of the free energy and the Polyakov loop are displayed in Fig.\ref{Fig:sun_h1_fren}.

\begin{figure}[H]
	\centerline{ \epsfxsize=7.0cm \epsfbox{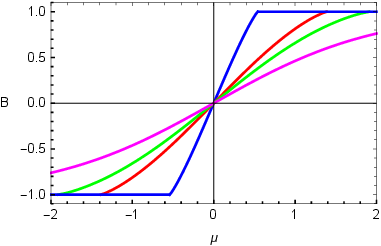} }
	\caption{\label{Fig:sun_h1_bar_dens} The baryon density of the massless model  at fixed $\kappa=0.1, 0.5, 1, \infty$ as a function of $\mu$. The color legend is as in Fig.\ref{Fig:sun_h1_fren}.}
\end{figure}

The baryon density in three phases equals $B=\frac{u_s^{(i)}}{\kappa}$. The chiral condensate of the massless model is zero due to the symmetry $m \to -m$
of the free energy (see equations for the full model in \ref{sunpf_list}).
Plot of the baryon density is displayed in Fig.\ref{Fig:sun_h1_bar_dens}. 
In the large $N$ limit ($\kappa =0$) the maximum of the integrand in (\ref{sun_h1_pf2}) is reached for $u=\kappa$ and the free energy equals $F= |\mu|$. In the large $N_f$ limit ($\kappa \to \infty$) the solution is simple
$u=\kappa \tanh \frac{\mu}{2}$ which leads to the following free energy
\begin{equation}
F = 2 \ln \left [ 2 \cosh \frac{\mu }{2} \right ] \ .
\label{sun_h1_largeNf}
\end{equation}

\subsubsection{Phase diagram of the full model}
\label{sun_full_hv}

In order to reveal the phase structure of the full $SU(N)$ model we have, in addition to the heavy-dense and massless limits, to study the region $h<1$ for arbitrary chemical potential $\mu$. This can be done as follows.
First, Eq.(\ref{PF_1dqcd_sumpart_deg}) is re-written as
\begin{eqnarray}
\label{sun_full_pf1}
Z = A \sum_{q=-N_f}^{N_f} e^{\mu q N} \sum_{r=0}^{N N_f} h^{2r+N|q|} \
E_{N,N_f}(r,|q|) \ , \\
\label{EN_Nf_coeff_def}
E_{N,N_f}(r,q) =
\sum_{\sigma \vdash\ r} \  s_{\sigma}(1^{N_f}) s_{N^{|q|} \sigma}(1^{N_f}) \ .
\end{eqnarray}
Second, we use a convenient representation
\begin{eqnarray}
\label{EN_Nf_coeff}
E_{N,N_f}(r,q) =  C_{N,N_f}(q) \ B_{N,Nf} (r, q) \ ,
\end{eqnarray}
where the coefficients $C_{N,N_f}(q)$ are given by Eq.(\ref{CNNf_def}) and we compute the coefficients $B_{N,N_f}(r,q)$ by a "brute force" method from (\ref{EN_Nf_coeff_def}) calculating the sum over all partitions $\sigma$ of $r$. Finally, the sum over $r$ in (\ref{sun_full_pf1}) is presented in 
the 't Hooft-Veneziano limit as
\begin{eqnarray}
\label{sun_sum_r}
\sum_{r=0}^{N N_f} h^{2r} \ B_{N,Nf} (r, q) =
\exp\left [ N N_f \sum_{k=1} h^{2k} \ C_k(u,\kappa) \right ] \ , \ u=\frac{q}{N} \ .
\end{eqnarray}
We have calculated the first nine $B_{N,Nf} (r, q)$ and $C_k(u,\kappa)$ coefficients.
Since these coefficients become very awkward with increasing $r$ we give explicit expressions only for the first few coefficients in Appendix \ref{coeff_ck_app}.

Collecting all formulas together we find in the 't Hooft-Veneziano limit the following  expression for the partition function
\begin{eqnarray}
\label{sun_pf_HVlimit}
Z &=&  N \  \ \int_0^{\kappa} \ e^{N N_f S_{eff}(h,\mu,\kappa;u)} \ du \ ,  \\
\label{sun_Seff_def}
S_{eff}(h,\mu,\kappa;u) &=& - \ln h + \frac{u}{\kappa} \left ( \mu  + \ln h \right )
+ f(u,\kappa) + P(h,u,\kappa) \ ,  \\
\label{Phuk_sun}
P(h,u,\kappa) &=& \sum_{k=1} h^{2k} \ C_k(u,\kappa) \ .
\end{eqnarray}
We have assumed that $\mu \geq 0$. Then, in the large $N$
limit the dominant contribution to the integral over $u$ comes from the region $u\geq0$.
The function $f(u,\kappa)$ is defined in Eq.(\ref{f_hk_def}). The free energy in
the 't Hooft-Veneziano limit follows from the last expression and equals
\begin{equation}
\label{sun_full_fren}
F = - \ln h + \frac{u_s}{\kappa} \left ( \mu  + \ln h \right )
+ f(u_s,\kappa) + P(h,u_s,\kappa) \ ,
\end{equation}
where $u_s$ is a solution of the saddle-point equation
\begin{equation}
\label{sun_saddle_eq}
\mu  + \ln h + \kappa \
\frac{\partial f(u,\kappa) + \partial P(h,u,\kappa)}{\partial u} = 0 \ .
\end{equation}
If $\mu$ is sufficiently small, there is no solution to this equation, the maximum of
the effective action $S_{eff}$ is reached at $u=0$. The free energy does not depend on
the chemical potential and reduces to the $U(N)$ free energy.  Next, we proceed as in the case of the reduced model and find solutions around $u=0$ and $u=\kappa$.  Using explicit expressions for coefficients $C_k(u,\kappa)$ we obtain the following small $u$ expansion for the function $P(h,u,\kappa )$
\begin{eqnarray}
P(h,u,\kappa ) &=& - \kappa \ln (1- h^2)- \frac{u}{\kappa} \ f_1(\kappa, h) +
  \frac{u^2}{2\kappa} \ \ln z +
 {\cal{O}} \left ( u^3 \right) \ \label{Phuk_small_u}.
\end{eqnarray}
We have introduced here notation
\begin{eqnarray}
	\label{f_1func}
	f_1(\kappa, h) = \ln \frac{z+1}{\sqrt{z^2 \xi^2 -1 }}- \xi\ln \sqrt{\frac{\xi z+1}{\xi z-1}} - \kappa  \ln \kappa  + (1 + \kappa) \ln (1+ \kappa)  \ , 
\end{eqnarray}
where $z=\sqrt{\frac{1-h^2}{1-h^2\xi^2}}$ and $\xi= 2 \kappa+1$. 
At small $u$, Eq.(\ref{sun_saddle_eq}) takes the form 
\begin{equation}
\label{sun_saddle_eq_small_u}
\mu +\ln h - f_1(\kappa, h)-\kappa  \ln (\kappa )+ (\kappa +1)\ln (\kappa +1) 
+ u \left[ \ln  \frac{ z u \kappa  }{\kappa +1} -1\right] = 0  
\end{equation}
and the solution is given by
\begin{eqnarray}
\label{sun_us_small_u}
&& u_s^{(1)}  = -\frac{ \mu-\mu_{cr}(\kappa ,h)}{W_{-1}\left[-\frac{ z \kappa  }{e ( \kappa +1) } ( \mu -\mu_{cr}( \kappa ,h)) \right]} +
{\cal{O}} \left ( (\mu-\mu_{cr})^2\right) \ ,  \\
\label{sun_mucrit_small_u}
&&\mu_{cr}(\kappa ,h) =  \ln \frac{z+1}{\sqrt{z^2 \xi^2 -1 }}- \xi\ln \sqrt{\frac{\xi z+1}{\xi z-1}} - \ln h  \ .  
%= \ln \sqrt{\frac{z+1}{z -1 }}- \xi\ln \sqrt{\frac{\xi z+1}{\xi z-1}} \ .
\end{eqnarray} 

Analyzing other regions of parameters, $\mu<0$ and $h>1$ reveals the existence of four critical lines determined by the equation
\begin{eqnarray}
	\pm \mu = \mu_{cr}(\kappa, h_{\pm}) \ .
	\label{sun_critline_small_u}
\end{eqnarray}
In all cases the third order phase transition occurs along the critical lines.

In a similar way one constructs the solution around $u= \kappa$ which turns out to be
\begin{eqnarray}
\label{sun_us_u_kappa}
&&u_s^{(2)} = \kappa - \frac{\mu - \overline{\mu_{cr}}(\kappa,h ) }{W_{-1} \left[ \frac{(1+ s)\varepsilon }{e(1+ \varepsilon)} (\mu - \overline{\mu_{cr}}(\kappa,h)) \right]}
+ {\cal{O}} \left ( (\mu-\overline{\mu_{cr}})^2\right) \ ,  \\ 
&& \overline{\mu_{cr}}(\kappa ,h)= \kappa \ln \frac{\sqrt{(1- \varepsilon^2)(s^2- \varepsilon^2)}}{2 (s+1) \varepsilon^2} -(1+ \kappa)\ln \sqrt{\frac{(1+ \varepsilon)(s+ \varepsilon)}{(1- \varepsilon)( s- \varepsilon)}}   \ , 
\label{sun_mucrit_u_kappa}
\end{eqnarray}
where $s=\sqrt{\frac{1+ \sinh^2 m}{1 + \frac{1}{\varepsilon^2}\sinh ^2m}}$, $\varepsilon = \frac{\kappa}{\kappa+1}$. 
The corresponding critical lines are determined by the equation
\begin{equation}
	\pm \mu =\overline{\mu_{cr}}(\kappa, h_{\pm} ) \ .
	\label{sun_critline_u_kappa}
\end{equation}
Needless to say, one finds a third order phase transition along the critical lines.

\begin{figure}[htb]
%\begin{figure}[th]
\centerline{ 
\epsfxsize=7cm \epsfbox{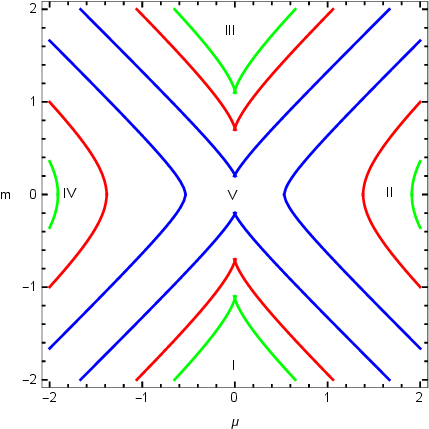} 
\epsfxsize=7cm \epsfbox{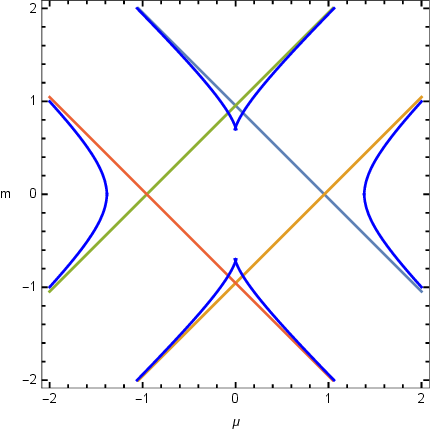} }
\caption{\label{Fig:sun_full} 
Left panel: Phases of the $SU(N)$ model in the 't Hooft-Veneziano limit in the $(\mu-m)$-plane. Blue, red and green lines correspond to $\kappa =0.1,0.5,1$.
Right panel: Phase diagram of the $SU(N)$ model for $\kappa=0.5$ in the $(\mu-m)$-plane and  phase lines of the reduced model as asymptotes. 
See the text for an explanation of different phases.
}
\end{figure}

Combining the results of this study with the heavy-dense and massless limits we can describe the full phase diagram of the $SU(N)$ model in the 't Hooft-Veneziano limit.
This diagram is shown in the left panel of Fig.\ref{Fig:sun_full}.
The free energies in different phases are as follows. $F_I=-F_{III}=-m$, $F_{II}=-F_{IV}=\mu$. $F_V$ depends on $m,\mu$ in a non-trivial way via Eq.(\ref{sun_full_fren}) with approximate solutions for $u_s$ given by Eqs.(\ref{sun_us_small_u}) and (\ref{sun_us_u_kappa}). 
In the 't Hooft limit, $\kappa=0$, the region V collapses to one point. In this limit one can always find a threshold transition. A comparison of the phase diagram of the full model with the ones of the two reduced models, 
neglecting either $h_{+}$ or $h_{-}$, is shown in the right panel of Fig.\ref{Fig:sun_full}. 
Finally, when $\kappa$ increases the region V extends and covers the whole 
$(\mu-m)$-plane in the limit $\kappa\to\infty$, where the free energy reduces to that of the free fermion model. The large $\kappa$ expansion of the free energy 
is given in (\ref{SUN_HV_fren_4}).

\section{Meson and baryon correlations}
\label{meson_corr}

In order to get more insight into the behaviour of the system 
we evaluate the meson-meson and the baryon-anti-baryon correlation functions defined in Eqs.(\ref{meson_corr_def}) and (\ref{baryon_corr_def}), correspondingly. Taking derivatives with respect to the sources $\eta$ and $\bar{\eta}$ in Eq.(\ref{generating_func}) one finds
\begin{gather}
	\label{meson_corr_res}
	\Sigma_f(t,t^{\prime}) =  (ZN^2)^{-1}  \int_G  dU  \prod_{f'=1}^{N_f}
	\det {\cal{M}}_{f'}\times\\
	\times\bigg[(\operatorname{Tr}\mathcal{M}_f^{-1}(t,t))(\operatorname{Tr}\mathcal{M}_f^{-1}(t',t')) - \operatorname{Tr}(\mathcal{M}_f^{-1}(t,t')\mathcal{M}_f^{-1}(t',t))\bigg] \ ,  \nonumber  \\
	\label{baryon_corr_res}
Y_f(t,t^{\prime}) =  (ZN!)^{-1} \int_G dU \  \prod_{f'=1}^{N_f} \rm{Det} {\cal{M}}_{f'} \ \prod_{k=1}^N \mathcal{M}^{-1}_{k,f}(t,t') \ , 
\end{gather}
where an inverse of the matrix $\cal{M}$ is given in Appendix B. In what follows we suppose that $\tau=t-t^{\prime}$ is even. 

Consider first the meson-meson correlation function. In the finite-temperature limit the factors $M_0$ and $M_1$ in (\ref{M_inverse}) become
\begin{equation}
M_0 = 2\sinh(\beta-|\tau|)m, \quad M_1 = -2\sinh |\tau|m  
\label{M0_M1_fin_T}
\end{equation}
and the connected part of $\Sigma^c_f(t,t^{\prime})\equiv\Sigma^c_f(\tau)$ can be written down as 
\begin{align}
\label{meson_cor_general}
&\Sigma^c_f(\tau) = \frac{4\sinh^2\beta m}{N^2}  \ \sum_{k=1}^N 
\Bigg\langle  \frac{b_+(\tau)e^{i\phi_k}+b_-(\tau)e^{-i\phi_k}-b_0(\tau)}{(\mbox{det}\mathcal{M}_k(\tau))^2} \Bigg\rangle \ , \\
\label{b0_bpm}
&b_0(\tau)=\frac{\sinh^2(\beta-|\tau|)m + \sinh^2|\tau|m }{\sinh^2\beta m} \ , \ 
b_{\pm}(\tau)= \frac{\sinh(\beta-|\tau|)m\sinh|\tau|m}{\sinh^2\beta m} \ 
e^{\pm \beta \mu} \ .
\end{align}
Expanding $(\mbox{det}\mathcal{M}_k(\tau))^2$ into the Fourier series one obtains 
\begin{align}
\label{meson_cor_PLexp}
&\Sigma^c_f(\tau) = \frac{1}{N} \ \sum_{n=-N_f}^{N_f} (-1)^n 
(|n|+\coth\beta m) e^{-\beta(m |n|-\mu n)} \\
&\times \Bigg (   b_+(\tau)W(n+1) + b_-(\tau) W(n-1) -
b_0(\tau) W(n) \Bigg )  \ .  \nonumber 
\end{align}
Here we derive the leading term of the meson correlation at large mass.  
The leading behavior of the Polyakov loop average can be calculated from  (\ref{Wr_sun_res}). One gets 
\begin{equation}
W(r) = (-1)^{r+1} \frac{N_f}{N} h^{|r|} e^{- \mu r}+O(h^{|r|+2}) \ , 
1 \leq |r| \leq N \ . 
\label{small-h-trace}
\end{equation}
If $N_f\leq N$ we arrive at the following result for the meson correlation at large mass
\begin{align}
&\Sigma^c_f(\tau)  = \frac{2 e^{\beta  m}}{N \left(e^{2\beta  m}-1\right)^2}   \bigg[ \left(2 \frac{N_f}{N}  -e^{2 \beta  m}+1\right) \cosh (m (2 \tau -\beta )) \nonumber  \\
&+ \frac{N_f}{N}  \left(e^{\beta  m}- e^{- 2 N_f  \beta  m} (3 \cosh (\beta  m)+(2 N_f+1) \sinh (\beta  m))\right) \bigg] \ . 
\end{align}
In the limit $\beta\to\infty$ one obtains the meson mass as
\begin{equation}
\label{meson_mass}
- \frac{1}{|\tau|} \ \ln\mid \Sigma^c_f \mid \ \propto \  2 m \ . 
\end{equation}

Consider now the baryon correlations. One finds from (\ref{baryon_corr_res}) 
for $\tau\geq 0$
\begin{align}
Y_f(\tau) &=\frac{Z_{N_f-1}}{N!Z_{N_f}} \ e^{-N \tau \mu} \ 
\bigg \langle \prod_{k=1}^N \left ( \sinh(\beta-\tau)m - 
e^{\beta\mu}\sinh\tau m e^{i\phi_k} \right ) \bigg \rangle_{N_f-1} \nonumber \\
\label{bar_corr_1}
&= \frac{1}{N!} \  e^{-N \tau \mu} \ \frac{Z_{N_f-1,1}}{Z_{N_f}} \ . 
\end{align}
The expectation value in (\ref{bar_corr_1}) refers to the partition function 
with $N_f-1$ degenerate fermions. The partition function $Z_{N_f-1,1}$ is given by up to an irrelevant constant
\begin{align}
\label{Z_Nf_1}
Z_{N_f-1,1} = 2^{N N_f} \sum_{q=-N_f}^{N_f} \ e^{N q \mu} \
\underset{1\le i,j\le N}{\det}S_{i-j+q} \ ,  \\
\label{Sk_def}
S_k = \sinh(\beta-\tau)m \ T_k - \sinh\tau m \ T_{k+1} \ , 
\end{align}
where $T_k$ is given in Eq.(\ref{Tk_def}) for $\mu=0$ and $N_f-1$ degenerate flavors. In order to derive the small $h$ expansion it is convenient to use 
the representation (\ref{PF_form_3})-(\ref{Zq_def_1})  which takes the form for the partition function $Z_{N_f-1,1}$
\begin{align} 
\label{PF_Nf_1_3}
&Z = A \ \left ( \sinh(\beta - \tau)m \right )^N  \ 
\frac{G(N+N_f)}{G(N_f)} \ \sum_{q=-N_f+1}^{N_f-1} \
h^{N |q|} \ e^{-\beta\mu N q} \ \ Q_{N,N_f-1,1}(q) \ , \\ 
&Q_{N,N_f-1,1}(q) = \sum_{l_1,\ldots,l_N=0}^{N_f-1} \ 
\sum_{s_1,\ldots,s_N=0,1} \ h^{2l_1+s_1+\ldots +2 l_N+s_N} \ 
\left ( - \frac{\sinh\tau m}{\sinh(\beta - \tau)m} \right )^{s_1+\ldots +s_N} 
\nonumber  \\
&\times \frac{\prod_{1\le i<j\le N} (l_i+s_i-l_j-s_j+j-i) 
\prod_{i=1}^N \binom{N_f-1}{l_i}}
{\prod_{i=1}^N (l_i+s_i-i+q+N)! (N_f-l_i-s_i+i-q-2)!}   \ .
\label{Zq_def_Nf_1}
\end{align}
To the leading order in $h$ it gives 
\begin{align}
\label{Q_Nf_1_small_h}
Q_{N,N_f-1,1}(q) &= \frac{G(N+1) G(q+1) G(N_f-1)}{G(N+q+1) G(N+N_f-1)} \\
&\quad \times
\left ( 1- \frac{h N \sinh\tau m}{\sinh(\beta - \tau)m} 
\frac{N_f-2-q}{N+q}  \right ) . \nonumber
\end{align} 
The phase diagram which follows from this representation coincides with the one 
described in Sec.\ref{reduced_sun}. 

Another easily solved limit is the case of one-flavor model, $N_f=1$. One finds 
\begin{eqnarray}
Z_{0,1} = \left ( \sinh(\beta-\tau)m \right )^N + 
(-1)^N \left ( \sinh\tau m \right )^N \ e^{N\beta\mu} \ . 
\label{bar_corr_one_flavor}
\end{eqnarray}
Combining this result with Eq.(\ref{sunN1}) we find baryon masses in the limit of large $\beta$ 
\begin{align}
&- \frac{1}{|\tau|} \ \ln \mid Y_f(\tau>0) \mid \ \propto \ 
 N  (m+\mu) \ , \ m > \mu \ ,  \nonumber \\ 
&- \frac{1}{|\tau|} \ \ln \mid Y_f(\tau>0) \mid \ \propto \ 
 N (\mu-m) \ , \ \mu > m \ . 
\label{bar_corr_large_beta}
\end{align}
In this limit both meson and baryon masses coincide with the corresponding masses of the free fermion model, Eqs.(\ref{meson_corr_limit}) and  (\ref{baryon_corr_limit}). 

\section{Summary and Perspectives}
\label{summary}

One-dimensional QCD in the lattice regularization is one of the few lattice gauge models that can be solved exactly. In the 't Hooft and in the 't Hooft-Veneziano limits the model exhibits a non-trivial phase structure. 
In this paper we have presented a detailed study of one-dimensional QCD both at finite $N, N_f$ and in the large $N$ and/or $N_f$ limits. As a gauge group we have considered $Z(N)$, $U(N)$ and $SU(N)$ groups. All fermion fields were taken in the fundamental representation. Let us briefly summarize our main results: 
\begin{itemize}
\item 
We have collected and described various different but equivalent representations for the partition function of one-dimensional QCD and for the Polyakov loop expectation value in Sec.\ref{sec_PF_repr}. Some of these representations, {\it e.g.} (\ref{PF_1dqcd_sumpart})-(\ref{Z_char_exp}), (\ref{PF_form_3}), (\ref{PF_1dqcd_det_leg})-(\ref{sun_pl_gen}) and (\ref{meson_repr_un}), 
we believe are new. The number of so different representations demonstrates the rich mathematical structure of one-dimensional QCD as one can relate, for example, the unitary integrals to the sum over partitions, to the combinatorics  of the monomer-dimer model, etc.  

\item 
For the first time we have studied one-dimensional $Z(N)$ QCD in  Sec.\ref{zn_model}. The model does not exhibit any critical behavior. Moreover, in the limit of large $N_f$ it reduces to the free fermion model. However, when $N_f$ is fixed and $N\to\infty$ one finds the threshold transitions with $N_f$ jumps in the behavior of the quark condensate and the particle density. 

\item 
One-dimensional $U(N)$ QCD has been considered in Sec.\ref{un_sun_model} at finite $N$. The 't Hooft-Veneziano limit was calculated in Sec.\ref{un_model}. Using the method of the orthogonal polynomials we were able to calculate exact expressions for the free energy and the Polyakov loop. The model exhibits a 3rd order phase transition as the fermion mass varies.   

\item 
In Sec.\ref{un_sun_model} we also studied one-dimensional $SU(N)$ QCD 
both at finite $N$ and in the 't Hooft limit. The model is not critical when $N$ is fixed. When $N$ is sufficiently large and $N_f$ is fixed one observes $N_f$ "smooth jumps" in the behavior of the baryon density, the Polyakov loop and the quark condensate. In the limit $N\to\infty$ these jumps move to the point $m=\mu$, where one finds the threshold transition. In the opposite limit, $N$ is fixed and $N_f\to\infty$, one discovers the Roberge-Weiss transition in the presence of the isospin chemical potential.

\item 
The 't Hooft-Veneziano limit of the $SU(N)$ model was thoroughly investigated in Sec.\ref{sun_model}. Here we derived the exact solution of the model in two limits: 1) the heavy-dense limit and 2) the massless limit. The phase structure of the full model was obtained approximately by using large and small mass expansions up to high orders. The model exhibits a rich phase structure which is described in detail in Sec.\ref{sun_full_hv}. In all cases we find 3rd order phase transitions. When the baryon chemical potential $\mu$ is sufficiently small, the $SU(N)$ free energy in the 't Hooft-Veneziano limit does not depend on $\mu$ and coincides with the $U(N)$ free energy. When $\mu$ grows the line of 3rd order phase transition appears. Above this line the free energy depends on the chemical potential.  $SU(N)$ QCD significantly differs from $U(N)$ QCD in this region. 

\item 
One important mathematical consequence of the present study is that integrals over unitary groups $U(N)$ and $SU(N)$ are different even in the large $N$ limit. This supports our earlier claim on such difference made in Ref.\cite{un_vs_sun}. 

\item 
We have also evaluated some exact and approximate expressions for the meson and baryon correlation functions. The corresponding masses have been calculated in the zero temperature (thermodynamic) limit.

\end{itemize}

Let us address now the following question: which value of $\kappa=N_f/N$ provides the best approximation to $U(3)$ and $SU(3)$ QCD in 
the 't Hooft-Veneziano limit? 

\begin{figure}[H]
	\centerline{\epsfxsize=7.5cm \epsfbox{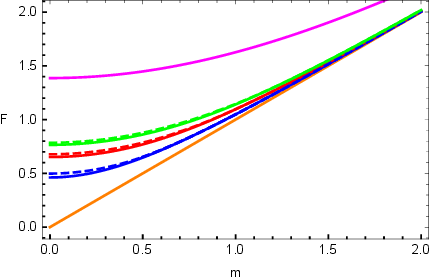}}
	\caption{\label{Fig:un_comparison} Comparison of the $U(3)$ free energy as a function of mass with the 't Hooft-Veneziano limit for $N_f=1$ (blue line), $N_f=2$ (red line) and $N_f=3$ (green line). Dashed lines of the same color correspond to the 't Hooft-Veneziano limit for $\kappa=1/3,2/3,1$, respectively. The lower orange line describes the large $N$ limit. The upper magenta line describes the large $N_f$ limit which coincides with the free fermions.}
\end{figure}

\begin{figure}[H]
	\centerline{\epsfxsize=7.5cm \epsfbox{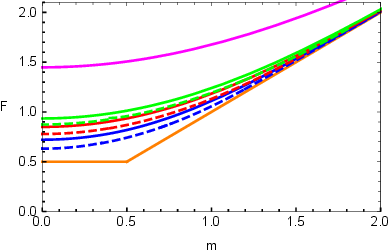}
	 \epsfxsize=7.5cm \epsfbox{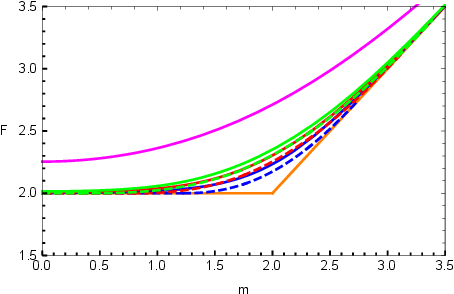} }
	\caption{\label{Fig:sun_comparison} Comparison of the $SU(3)$ free energy as a function of mass with the 't Hooft-Veneziano limit for $N_f=1$ (blue line), $N_f=2$ (red line) and $N_f=3$ (green line). Dashed lines of the same color correspond to the 't Hooft-Veneziano limit for $\kappa=1/3,2/3,1$, respectively. The lower orange line describes the large $N$ limit. The upper magenta line describes the large $N_f$ limit which coincides with the free fermions. Left panel: $\mu=0.5$. Right panel: $\mu=2$. }
\end{figure}

\begin{figure}[H]
	\centerline{\epsfxsize=7.5cm \epsfbox{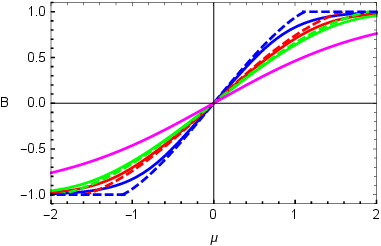}
		\epsfxsize=7.5cm \epsfbox{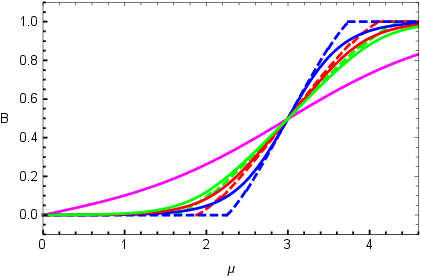} }
	\caption{\label{Fig:sun_density_comparison} Comparison of the $SU(3)$ baryon density with the 't Hooft-Veneziano limit for $N_f=1$ (blue line), $N_f=2$ (red line) and $N_f=3$ (green line). Dashed lines of the same color correspond to the 't Hooft-Veneziano limit for $\kappa=1/3,2/3,1$, respectively. The magenta line corresponds to the large $N_f$ limit which coincides with the free fermions. Left panel: $m=0$. Right panel: $m=3$.}
\end{figure}
Fig.\ref{Fig:un_comparison} shows the $U(3)$ free energy as a function of the mass for $N_f=1,2,3$ degenerate flavors and for the corresponding values of $\kappa$. The same for the $SU(3)$ free energy and the baryon density is shown in Figs.\ref{Fig:sun_comparison}, \ref{Fig:sun_density_comparison}, correspondingly. Inspecting these (and others not shown here) plots one can make a few conclusions: 1) The 't Hooft-Veneziano limit gives much better approximation than the 't Hooft limit alone; 2) The approximation is rather reasonable even for small values of $N_f$; 3) The approximation gets better with $N_f$ and $\kappa$ increasing.

Finally, let us outline some perspectives for future work. 
\begin{itemize} 
\item 
Constructing an exact solution of full $SU(N)$ QCD in the 't Hooft-Veneziano limit remains a challenge. One possible route to achieve this which was not explored here is to use an analog of the monomer-dimer representation (\ref{meson_repr_un}) for the partition function, see Ref.\cite{sun_integral}. 

\item 
An interesting extension of the present work is to study the popular adjoint and scalar QCD in one dimension.  {\it E.g.}, in the case of the adjoint QCD the partition function takes the form of Eq.(\ref{PF_1dqcd_int}), where $U$ belongs to the adjoint representation of $U(N)$ or $SU(N)$ group. In this case the original action is invariant under the global center transformations so that one can address the question if the symmetry can be spontaneously broken. 

\item 
Another natural extension is to study the high dimensional QCD in the strong coupling region with static quarks. Consider the following Polyakov loop model 
\begin{eqnarray}
&&Z = \int \ \prod_x \ dU(x)
\prod_{x,n}\left(1+ \lambda \ {\rm Re}W(x) W^{\dagger}(x+e_n) \right) \nonumber \\  
&& \times \ \prod_x \prod_{f=1}^{N_f} \det  \left[1+ h_+^f U(x) \right]  
\det \left[1 +  h_-^f U^{\dagger}(x)  \right]  \ .
\label{3d_PF_Interact}
\end{eqnarray}
%where $\lambda$ is an effective gauge coupling constant.  
This model cannot be solved exactly at finite $N$ or $N_f$. We can explore the factorization property to construct an exact solution in the 
't Hooft-Veneziano limit. Due to the factorization the mean-field approximation becomes eligible and calculations reduce to the evaluation of the powers of the Polyakov loop traces over the static quark determinant which is one-dimensional QCD. 

\end{itemize}

\section*{Acknowledgements}
O. B. acknowledges support from INFN/NPQCD project.
V. C. acknowledges support by the Deutsche
Forschungsgemeinschaft (DFG, German Research Foundation) through 
the CRC-TR 211 ’Strong-interaction matter under extreme conditions’ – project number 315477589 – TRR 211.
S. V. acknowledges support from the Simons Foundation (Grant No. 1290596).

\newpage

\appendix

\section{List of partition functions}
\label{list_pf_app}

Here we list 1) partition functions $Z(x,y)\equiv Z(N,N_f)$ for various values of $N$ and $N_f$ and 2) the free energies (\ref{fren_def}) in the large $N$ and/or $N_f$ limits. $Z_0(N,N_f)$ in \ref{sunpf_list} denotes the $U(N)$ partition function, $F(N_f)$ is the free energy in the large $N$ limit, $F(N)$ - the free energy in the large $N_f$ limit and $F$ - the free energy in the 't Hooft-Veneziano limit. $P_n^m(x)$ is the associated Legendre polynomial, $U_n(x)$ is the Chebyshev polynomial of the 2nd kind, $G(n)$ is the Barnes function,
${}_pF_q(\ldots)$ is the hypergeometric function and
$W_{-1}(x)$ is a lower branch of the Lambert function.
We shall also use the short-hand notations $t=\coth m$ and
\begin{equation}
\label{tk_sk_rk__def}
m_k = \cosh k m \ , \ s_k = \sinh k m \ ,  \ \mu_k = \cosh k \mu \ , \
U_n(m_1) = \frac{\sinh(n+1)m}{\sinh m} \  .
\end{equation}
%Results for $U(N)$ and $SU(N)$ models have been obtained from %Eqs.(\ref{PF_1dqcd_sumpart_deg}) and (\ref{PF_1dqcd_det_leg}).

\subsection{Z(N) model}
\label{znpf_list}

\noindent
{\bf General expressions:}
\begin{align}
\label{znN1}
&Z(N,1) = (2 s_1)^N P_N(t)+2\mu_N \ .  \\
\label{zN2}
&Z(N,2) = (2s_1)^{2N} \left ( P_{2N}(t) + 2\mu_N \ \frac{(2N)!}{(3N)!} \
P_{2N}^N(t) \right ) + 2\mu_{2N} \ .  \\
\label{znN3}
&Z(N,3) = (2s_1)^{3N} \left ( P_{3N}(t) + 2\mu_N \ \frac{(3N)!}{(4N)!} \ P_{3N}^N(t)
 + 2\mu_{2N} \ \frac{(3N)!}{(5N)!} \ P_{3N}^{2N}(t) \right ) + 2\mu_{3N} \ . \\
\label{zn2Nf}
&Z(2,N_f) = \frac{4^{N_f}}{2} \
\left ( (m_1 + \mu_1)^{2N_f}+(m_1 - \mu_1)^{2N_f} \right ) \ .  \\
\label{zn3Nf}
&Z(3,N_f) = \frac{1}{2} \
\left ( \left ( 2m_1 + 2\mu_1 \right )^{3N_f} +
\Re \ \left ( 2m_1 - \mu_1 +i \sqrt{3(\mu_1^2 - 1)} \right )^{3N_f}
\right ) \ .
\end{align}
\noindent
{\bf Particular values:}
\begin{align}
\label{zn21}
&Z(2,1) = 4 \cosh^2m + 4 \cosh^2\mu  \ .   \\
\label{zn22}
&Z(2,2) = 8 \left ( ( \cosh m - \cosh\mu )^4 +
( \cosh m + \cosh\mu )^4  \right )   \  .  \\
\label{zn23}
&Z(2,3) = 32 \left ( ( \cosh m - \cosh\mu )^6 +
( \cosh m + \cosh\mu )^6  \right ) \  .  \\
 \label{zn31}
&Z(3,1) = 18 m_1 + 2 m_3 + 2\mu_3  \ . \\
\label{zn32}
&Z(3,2) = 400 + 450 m_2 + 72 m_4 + 2 m_6
 + 40 \left ( 9m_1 + 2m_3 \right ) \mu_3 + 2 \mu_6  \ .  \\
\label{zn33}
&Z(3,3) = 4 m_1 \left ( 5018 + 5840 m_2 + 1216 m_4 + 80 m_6 + m_8 \right ) \\
&+ 168 \left ( 84 + 108 m_2 + 27 m_4 + 2 m_6 \right ) \mu_3
+ 96 m_1 \left ( 10 + 7 m_2 \right ) \mu_6 + 2 \mu_9 \ .  \nonumber
\end{align}

\noindent
{\bf Reduced and massless models:}
\begin{align}
\label{zn_pf_reduced}
&Z(N,N_f;h_-=0) = e^{N N_f m} \ \sum_{q=0}^{N_f} \ \binom{N N_f}{q N} \ e^{N(-m+\mu)q} \ .  \\
\label{zn_pf_massless}
&Z(N,N_f;m=0) = 1 + 2 \sum_{q=1}^{N_f} \ \binom{2 N N_f}{N(N_f+q)} \
\cosh\mu N q  \ .
\end{align}
{\bf Limiting behavior:}
\begin{eqnarray}
	\label{Fren_1dZN_largeN}
	&F(N_f) = \ln \left [ 2 e^{\mu q_m/N_f} \ Q(q_m/N_f) \right ] \ .  \\
	\label{Fren_1dZN_largeNf}
	&F(N) = F =  \ln \left [ 2 \cosh m  + 2 \cosh \mu \right ] \ .
\end{eqnarray}
$q_m$ in (\ref{Fren_1dZN_largeN}) maximizes the summand in Eq.(\ref{PF_1dZN_largeN}).

\noindent
{\bf The 't Hooft-Veneziano limit} coincides with the large $N_f$ limit
(\ref{Fren_1dZN_largeNf}).

\subsection{U(N) model}
\label{unpf_list}

{\bf General expressions:}
\begin{align}
	\label{unN1}
	&Z(N,1) = U_N(m_1) \ .  \\
	\label{uN2}
	&Z(N,2) = \frac{1}{(2\sinh m)^2} \ \left [ U_{N+1}^2(m_1) - (N+2)^2  \right ] \ .  \\
	\label{unN3}
	&Z(N,3) =  \frac{1}{(2\sinh m)^6} \ \bigl [ U_{N+2}^3(m_1) -2 (N+3)^3 \cosh m  \ U_{N+1}(m_1) \nonumber  \\
	&+ (N+3)^2 \ (3+2N-(N+2)^2 \sinh^2 m) \ U_{N+2}(m_1)  \bigr ] \ . \\
	\label{un1Nf}
	&Z(1,N_f) = (2\sinh m)^{N_f} \ P_{N_f}\left ( t \right ) \ .  \\
	\label{un2Nf}
	&Z(2,N_f) = (2\sinh m)^{2N_f} \left [ (P_{N_f}(t))^2-(N_f+1)^{-2}
	(P_{N_f}^1(t))^2 \right ]  \ . \\
	\label{un3Nf}
	&Z(3,N_f) = \frac{(2\sinh m)^{3N_f}}{(N_f+1)^3 (N_f+2)^2} \bigl [ (N_f+2) (N_f+1) P_{N_f}(t) - P_{N_f}^2(t) \bigr ]  \\
	&\times \bigl [ (N_f+1) P_{N_f}(t) \left ( (N_f+2) (N_f+1) P_{N_f}(t) +  P_{N_f}^2(t) \right ) - 2(N_f+2)  (P_{N_f}^1(t))^2  \bigr ]
	\ .  \nonumber 
\end{align}
\noindent
{\bf Particular values:}
\begin{align}
	\label{un11}
	&Z(1,1) = 2 \cosh m \ .   \\
	\label{un12}
	&Z(1,2) = 2 \left ( 2 + \cosh 2m \right ) \ . \\
	\label{un13}
	&Z(1,3) = 4 \cosh m \left ( 4 + \cosh 2m \right ) \ .  \\
	\label{un21}
	&Z(2,1) = 1 + 2 \cosh 2m  \  .   \\
	\label{un22}
	&Z(2,2) = 10 + 8 \cosh 2m + 2 \cosh 4m \ .  \\
	\label{un23}
	&Z(2,3) = 65 + 90 \cosh 2m + 18 \cosh 4m + 2 \cosh 6m \ .   \\
	\label{un31}
	&Z(3,1) = 4 \cosh m  \cosh 2m \ .  \\
	\label{un32}
	&Z(3,2) = 20 + 20 \cosh 2m + 8 \cosh 4m + 2 \cosh 6m \ .  \\
	\label{un33}
	&Z(3,3) = 4 \cosh m \left ( 71 + 128 \cosh 2m + 37 \cosh 4m +
	8 \cosh 6m + \cosh 8m \right ) \ .
\end{align}
\noindent
{\bf Reduced and massless models:}
\begin{align}
	\label{un_pf_reduced}
	&Z(N,N_f;h_-=0) = e^{N N_f m} \ .  \\
	\label{un_pf_massless}
	&Z(N,N_f;m=0) = \frac{ G(N+ 2 N_f+1) G(N+1) }{ G(2 N_f+1) } \
	\left ( \frac{G(N_f+1)}{G(N+ N_f+1)} \right )^2 \ .
\end{align}
\noindent
{\bf Limiting behavior:}
\begin{align}
\label{Fren_1dun_largeN}
&F(N_f) =  m \ . \\
\label{Fren_1dun_largeNf}
&F(N) = m + 2 \ln \left ( 1 + e^{-m} \right ) \ .
\end{align}
\noindent
{\bf The 't Hooft-Veneziano limit:}  ($\xi=2\kappa+1$)
\begin{gather}
\label{Fren_1dun_HV}
F = \begin{cases}
\frac{\xi}{\kappa}\ln\cosh\frac{m}{2}+\frac{\xi^2}{2\kappa}
\ln\xi-\frac{(1+\kappa)^2}{\kappa}\ln(1+\kappa)-\kappa\ln4\kappa \ , &m<\ln\xi \ , \\
m - \kappa \ln(1-e^{-2m}) \ ,  &m>\ln\xi \ .
\end{cases}
\end{gather}

\subsection{SU(N) model}
\label{sunpf_list}

Listing $SU(N)$ partition functions we use Eq.(\ref{PF_1dqcd_det_leg}).
For the free energy in the 't Hooft-Veneziano limit we give 4 expansions valid near the critical lines at small and large values of the mass and the chemical potential.

\noindent
{\bf General expressions:}
\begin{align}
\label{sunN1}
&Z(N,1) = Z_0(N,1) + 2 \mu_N  \ .  \\
\label{sunN2}
&Z(N,2) = Z_0(N,2) + 2 Z_1(N,2) \mu_N  + 2 \mu_{2N}  \ . \\
\label{sunN2q1}
&Z_1(N,2) = \frac{1}{(2\sinh m)^2} \ \left [ (N+1) U_{N+2}(m_1) -
(N+3) U_N(m_1)  \right ] \ .   \\
\label{sunN3}
&Z(N,3) = Z_0(N,3) + 2 Z_1(N,3) \mu_N + 2 Z_2(N,3) \mu_{2N} +
2 \mu_{3N}  \ . \\
\label{sunN3q1}
&Z_1(N,3) = \frac{1}{(2\sinh m)^6} \biggl [ (N^2+6N+11) U_{N+1}^2(m_1) \\
&+ (N+2) ((N+1)\cosh 2m -3) U_{N+2}^2(m_1)
+(N+2)(N+4) ((N+3)^2 \sinh^2 m  \nonumber  \\
&+2 - 2(N+1)(N+2)\cosh m \ U_{N+1}(m_1) U_{N+2}(m_1) \nonumber \biggr ]
\ .  \\
\label{sunN3q2}
&Z_2(N,3) = \frac{1}{(2\sinh m)^4} \ \bigl [ 3(N+3) U_{N}(m_1)  \\
&+ (N+1) \left ( (N+2)\cosh 2m -N -5 \right ) U_{N+2}(m_1)  \bigr ] \ .
\nonumber  \\
\label{sun2Nf}
&Z(2,N_f) = \frac{4^{N_f}}{\sqrt{\pi}} \sum_{k=0}^{[N_f/2]} \
( 2\cosh m \cosh \mu )^{2k}
\frac{\Gamma(N_f-k+1/2)}{\Gamma(N_f-k+1)} \
\binom{N_f}{2k} \\
&\times {}_2F_1 \left ( k-N_f-1,2k-N_f; k-N_f-1/2;
- (\cosh 2m + \cosh 2\mu )/2 \right) \ .
\nonumber \\
\label{sun3Nf}
&Z(3,N_f) = Z_0(3,N_f) + 2 \sum_{q=1}^{N_f-1} \ Z_q(3,N_f)
\mu_{3q}  + 2 \mu_{3N_f} \ . \\
\label{sun3NfZq}
&Z_q(3,N_f) =  \left ( \frac{(2\sinh m)^{N_f}N_f!}{(N_f+q)!}  \right )^3 \  \biggl [ (P_{N_f}^q(t))^3 + \frac{(N_f+q)(N_f+q-1)}{(N_f+q+1)^2}
\\
&\times P_{N_f}^{q-2}(t) (P_{N_f}^{q+1}(t))^2 + \frac{(N_f+q)}{(N_f+q+1)(N_f+q+2)} \bigl ( (N_f+q) P_{N_f}^{q+2}(t) (P_{N_f}^{q-1}(t))^2 \nonumber  \\
&-P_{N_f}^q(t) \bigl ( 2(N_f+q+2) P_{N_f}^{q-1}(t) P_{N_f}^{q+1}(t) + (N_f+q-1) P_{N_f}^{q-2}(t) P_{N_f}^{q+2}(t)  \bigr )
\bigr ) \biggr ] \ .   \nonumber
\end{align}

\noindent
{\bf Particular values:}
\begin{align}
\label{sun21}
&Z(2,1) =  1 + 2 \cosh 2m + 2 \mu_2 \ .  \\
\label{sun22}
&Z(2,2) = 10 + 8 \cosh 2m + 2 \cosh 4m +
4( 2 +  3 \cosh 2m ) \mu_2 + 2 \mu_4  \ .  \\
\label{sun23}
&Z(2,3) = 65 + 90 \cosh 2m + 18 \cosh 4m + 2 \cosh 6m \\
&+ 6(15 + 16 \cosh 2m + 4 \cosh 4m) \mu_2 + 6(3 + 4 \cosh 2m) \mu_4
+ 2 \mu_6 \ .  \nonumber  \\
\label{sun31}
&Z(3,1) = 2 \cosh m + 2 \cosh 3m + 2 \mu_3  \ . \\
\label{sun32}
&Z(3,2) = 20 + 20 \cosh 2m + 8 \cosh 4m + 2 \cosh 6m   \\
&+ 8 \cosh m (1 + 4 \cosh 2m) mu_3 + 2 \mu_6 \ .  \nonumber  \\
&Z(3,3) = 540 \cosh m + 330 \cosh 3m + 90 \cosh 5m + 18 \cosh 7m
+ 2 \cosh 9m  \nonumber   \\
\label{sun33}
&+ 4 (82 + 108 \cosh 2m + 45 \cosh 4m + 10 \cosh 6m) \mu_3  \\
&+ 16 \cosh m (2 + 5 \cosh 2m) \mu_6 + 2\mu_9 \ .  \nonumber
\end{align}

\noindent
{\bf Reduced and massless models:}
\begin{align}
&Z(N,N_f;h_-=0) = e^{N N_f m} \
\biggl [ 1 + \frac{G(N+1) G(N+N_f+1)}{G(N_f+1)} \nonumber  \\
&\times \sum_{q=1}^{N_f} \ h_+^{N q} \frac{G(q+1) G(N_f+1-q)}{G(N+q+1) G(N+N_f+1-q)} \biggr ] \ .
\label{sun_pf_reduced}
\end{align}
\begin{align}
&Z(N,N_f;m=0) = Z_0(N,N_f) + 2 \frac{ G(N+ 2 N_f+1) G(N+1) }{ G(2 N_f+1) } 
\nonumber \\
\label{sun_pf_massless}
&\times \sum_{q=1}^{N_f-1} \frac{ G(N_f-q+1) G(N_f+q+1) }{ G(N+ N_f- q+1) G(N+ N_f+ q+1) } \ \cosh N q\mu  + 2 \cosh N N_f \mu \  .
\end{align}

\noindent
{\bf Small mass expansion:}
\begin{align}
\label{sun_pf_small_mass}
&Z(N,N_f;m\approx 0)= \sum_{q=-N_f}^{N_f} \ Z_q(N,N_f;m=0) \
\biggl (  1 + \frac{s \ t}{\left(4N_f^2-1\right)}
\sum_{k=1} C_k \frac{m^{2k}}{(2k)!} \biggr ) \ ,  \\
\label{sun_pf_small_mass_coeff1}
&s = N (N + 2N_f ) \ , \ t= N_f^2- q^2 \ , \ C_1 = 1 \ , \
C_2 = \frac{3 (s+2) (t-2)}{4 N_f^2-9}-2 \ , \nonumber \\
&C_3 = \frac{20 (s+2)  (t-2) (s+6)(t-6)}{\left(4 N_f^2-25\right) \left(4N_f^2-9\right)}+ \frac{10 s t (s+2) (t-2)}{\left(4 N_f^2-9\right)
\left(4 N_f^2-1\right)}  -    \\
& -\frac{15 s t (s+6)  (t-6)}{\left(4 N_f^2-25\right) \left(4 N_f^2-1\right)} - \frac{5 (s+6) (t-6)}{4 N_f^2-25} -\frac{45 (s+2) (t-2)}{4 N_f^2-9} +
\frac{20 s t}{4 N_f^2-1}+16  \ . \nonumber
\end{align}

\noindent
{\bf Limiting behavior:}
\begin{align}
	\label{Fren_1dsun_largeN}
	&F(N_f) =
	\begin{cases}
		m \ , m > | \mu | \ , \\
		| \mu | \ , m < | \mu | \ .
	\end{cases} \\
	\label{Fren_1dsun_largeNf}
	&F(N) = \ln \left [ 2 \cosh m  + 2 \cosh \mu \right ] \ . 
\end{align}

\newpage 

\noindent
{\bf The 't Hooft-Veneziano limit:}

\noindent
{\bf Region I.}  $ h < \frac{1}{2 \kappa +1}, u\approx 0$:
\begin{eqnarray}
F &=&-\ln h - \kappa \ln \left(1-h^2\right)-\frac{(\mu- \mu_{cr}(\kappa,h))^2 }{2 \kappa
W_{-1}\left[-  e^{s (\kappa,h ) } (\mu- \mu_{cr}(\kappa,h))\right]} \nonumber \\
\label{SUN_HV_fren_1}
&\times&\left(1 + \frac{1}{2 W_{-1}\left[- e^{s (\kappa,h ) } (\mu- \mu_{cr}(\kappa,h))\right]}\right) \\
&+& \frac{   \xi^2 z^4+3}{24 \kappa^2 (\kappa +1) z} \ \frac{(\mu- \mu_{cr}(\kappa,h))^3}{ ( W_{-1}\left[-  e^{s (\kappa,h ) } (\mu- \mu_{cr}(\kappa,h))\right])^3} + 
{\cal{O}}\left ( (\mu- \mu_{cr}(\kappa,h))^4 \right ) \ , \nonumber 
\end{eqnarray}
where $s (\kappa,h ) = \ln \frac{ \kappa  z }{( \kappa +1)}-1$, $z=\sqrt{\frac{1-h^2}{1-h^2\xi^2}}$, $\xi= 2 \kappa+1$ and 
\begin{equation}
\mu_{cr}(\kappa ,h)=  \ln \sqrt{\frac{z+1}{z -1 }}- \xi\ln \sqrt{\frac{\xi z+1}{\xi z-1}} \ . 
\label{SUN_HV_mucr_1}
\end{equation}

\noindent
{\bf Region II.} $\frac{1}{2 \kappa +1} <  h < 2 \kappa +1, u\approx 0$:
\begin{align}
&F = \frac{1}{2\kappa}\left ( (1+2\kappa)^2\ln(1+2\kappa) - 2 (1+\kappa)^2\ln(1+\kappa) \right ) - \kappa \ln 4\kappa 
+\frac{(2 k+1)}{2 k}\ln \cosh^2\frac{m}{2} \nonumber \\
&+  \frac{ \mu^2 }{2  \kappa x}
+ \frac{\mu^4}{x^4  }\frac{8 \kappa (2 \kappa +1)  \cosh ^4 \frac{m}{2} (-2 \kappa^2 +(2 \kappa  (\kappa +3)+3) (\cosh m -1))}{3 (\kappa +1)^2 (2 \kappa^2 -(2 \kappa  +1) (\cosh m  -1))^3}+ {\cal{O}}\left (\mu^6 \right ) \ , 
\label{SUN_HV_fren_2}
\end{align} 
where $x=  \ln\frac{2 k^2 -(2 k+1) (\cosh m -1)}{2(\kappa+1)^2}$ 
and $m= - \ln h$. 

\noindent
{\bf Region III.} $\mu >  2 \kappa \ln 2 \kappa  - (2\kappa +1) \ln(2\kappa + 1) , \ u\approx \kappa$:
\begin{align} 
\label{SUN_HV_fren_3}
&F=\mu -\frac{(\mu- \overline{\mu_{cr}}(\kappa,h))^2 }{2 \kappa
W_{-1}\left[e^{ \overline{s}(\kappa ,h)} (\mu - \overline{\mu_{cr}}(\kappa,h))\right]} \left(1 + \frac{1}{2  W_{-1}\left[e^{  \overline{s}(\kappa ,h)} (\mu - \overline{\mu_{cr}}(\kappa,h)) \right]}\right) \\
&+ \frac{2 s^3-3 s^2+2+ 3 \varepsilon^2}{12 \kappa^2 (1 -\varepsilon^2 )} \frac{(\mu- \overline{\mu_{cr}}(\kappa,h))^3}{ ( W_{-1}\left[e^{  \overline{s}(\kappa ,h)} (\mu - \overline{\mu_{cr}}(\kappa,h)) \right])^3}+ {\cal{O}} \left ( (\mu- \overline{\mu_{cr}}(\kappa,h))^4 \right ) \ , \nonumber 
\end{align}
where $\overline{s}(\kappa ,h) =  	\ln \frac{\varepsilon(1+ s) }{1+ \varepsilon} -1$, $s=\sqrt{\frac{1+ \sinh^2 m}{1 + \frac{1}{\varepsilon^2}\sinh ^2m}}$, $\varepsilon = \frac{\kappa}{\kappa+1}$ and 
\begin{equation}
\label{SUN_HV_mucr_3}
\overline{\mu_{cr}}(\kappa ,h)= \kappa \ln \frac{\sqrt{(1- \varepsilon^2)(s^2- \varepsilon^2)}}{2 (s+1) \varepsilon^2} - (1+ \kappa)\ln \sqrt{\frac{(1+ \varepsilon)(s+ \varepsilon)}{(1- \varepsilon)( s- \varepsilon)}}  \ . 
\end{equation}

\noindent
{\bf Region IV.} $\mu_{cr} < \mu < \overline{\mu_{cr}}$, large $\kappa$ expansion: 
\begin{eqnarray}
&&F = \ln [2 \cosh \mu +2 \cosh m ]- \frac{1}{\kappa}\left( \frac{3}{4} +\frac{1}{2} \ln \left[ \kappa  \frac{1+ \cosh m \cosh \mu}{(\cosh m +\cosh \mu )^2} \right] \right)  \nonumber \\ 
&&-\frac{1}{4 \kappa^2}-\frac{\sinh \mu  }{192 \kappa^2 (\cosh \mu  \cosh m+1)^3} \big[ \sinh \mu (13 \cosh 2m +\cosh 4m-4) \nonumber \\
&&+2 \cosh m (4 \sinh 2\mu \cosh 2m +\sinh 3 \mu \cosh m) \big]  + 
{\cal{O}}\left ( \kappa^{-3} \right ) \ . 
\label{SUN_HV_fren_4}
\end{eqnarray}

\section{Fermion matrix}
\label{ferm_matr_app}

The determinant of the fermion matrix (\ref{matrixM_def}) is the gauge invariant object.
Hence, it can depend only on the gauge-invariant combinations of individual link variables $U(t)$.
In one dimension there is only one such combination, namely the Polyakov loop (\ref{PL_def}).
By the change of variables (\ref{gauge_fix}) one can remove gauge fields $U(t)$ from all links
but the last one connecting lattice sites $t=0$ and $t=N_t-1$ where they are grouped into
the Polyakov loop $U$. The group matrix $U$ becomes a new global variable. In the parametrization
$U=V E V^{\dagger}$, where $E$ is a diagonal matrix, the hermitian matrix $V$ can be removed from
the action by a global shift of the fermion fields $\overline{\psi}\to\overline{\psi}V^{\dagger}$,
$\psi\to V\psi$. The diagonal matrix $E$ can be distributed uniformly over all links, so
the fermion matrix takes the following form
\begin{eqnarray}
{\cal{M}}_k =
\begin{pmatrix}
\tilde{m} & \frac{1}{2} e^{\phi_k} & 0 & 0 & . & \frac{1}{2} e^{-\phi_k} \\
-\frac{1}{2} e^{-\phi_k} & \tilde{m} & \frac{1}{2} e^{\phi_k} &0 &.& 0\\
0  &  -\frac{1}{2} e^{-\phi_k}  &   \tilde{m} & \frac{1}{2} e^{\phi_k} & . & 0\\
. & . & . & . & . & . \\
. & . & . & -\frac{1}{2} e^{-\phi_k} & \tilde{m} & \frac{1}{2} e^{\phi_k} \\
-\frac{1}{2} e^{\phi_k} & 0 & 0 & . & -\frac{1}{2} e^{-\phi_k} & \tilde{m}
\end{pmatrix}
\label{static_gauge}
\end{eqnarray}
Here, $\phi_k=\tilde{\mu} + i\omega_k/N_t$.
This is equivalent to a famous static diagonal gauge.

Now, the determinant and the inverse of the fermion matrix can be easily calculated
by using, {\it e.g.} the Fourier transform. One finds
\begin{eqnarray}
\mbox{Det} {\cal{M}} = \prod_{k=1}^N \
\underset{0\leq t,t^{\prime}\leq N_t-1}{\mbox{det}} {\cal{M}}_k =
\prod_{k=1}^N \ \prod_{p=0}^{N_t-1} \left [ \tilde{m} +
i \sin\left ( \frac{2\pi}{N_t}\ p + \frac{\omega_k+\pi}{N_t} -
i \tilde{\mu} \right ) \right ] \ ,
\label{ferm_matr_det}
\end{eqnarray}
\begin{eqnarray}
{\cal{M}}_k^{-1}(t,t^{\prime}) = \frac{1}{N_t} \ \sum_{p=0}^{N_t-1} \
\frac{e^{\frac{2\pi i}{N_t}\tau (p+1/2)}}
{\tilde{m} + i \sin\left ( \frac{2\pi}{N_t}\ p + \frac{\omega_k+\pi}{N_t} -
i \tilde{\mu} \right )} \ , \ \tau = t - t^{\prime} \ .
\label{ferm_matr_inverse}
\end{eqnarray}
Calculating the product and the sum in the last expressions one obtains
\begin{eqnarray}
\mbox{Det} {\cal{M}} = 2^{-N N_t} e^{N m} \ \prod_{k=1}^N
\left [ 1 + h_+^f e^{i\omega_k} \right ] \  \left [ 1 + h_-^f e^{-i \omega_k} \right ] \ ,
\label{ferm_matr_det_res}
\end{eqnarray}
\begin{eqnarray}
\frac{{\cal{M}}_k^{-1}(t,t^{\prime})}{2^{-N N_t+N-1}} &=&
\frac{\exp\left [ -i \frac{\tau}{N_t}\omega_k - \tau \tilde{\mu} \right ]}
{{\mbox{det}} {\cal{M}}_k \ \sqrt{1+\tilde{m}^2}} \
\begin{cases}
M_0 +  e^{N_t \tilde{\mu} + i \omega_k} \ M_1  \ , \ \tau \geq 0 \ , \\
(-1)^{\tau} \ \left [ M_0 + e^{ - N_t \tilde{\mu} - i \omega_k} \ M_1  \right ] \ , \tau < 0 \ ,
\end{cases}  \ ,  \nonumber \\
\label{M_inverse}
M_0 &=& (\sqrt{1+\tilde{m}^2}+\tilde{m})^{N_t-|\tau|} - (-1)^{\tau} (\sqrt{1+\tilde{m}^2}
+\tilde{m})^{-N_t+|\tau|}  \ ,  \\
M_1 &=& (\sqrt{1+\tilde{m}^2}+\tilde{m})^{-|\tau|} -
(-1)^{\tau} (\sqrt{1+\tilde{m}^2}+\tilde{m})^{|\tau|} \ . \nonumber
\end{eqnarray}

\section{Associated Legendre function}
\label{leg_func_app}

In this paper we use the following definition of the associated Legendre function
\footnote{This definition corresponds to the function $LegendreP[l,m,3,x]$ in Mathematica.}
\begin{gather}
P_l^m(x) = \frac{1}{2^l l!}(x^2-1)^{\frac{m}{2}} \
\left(\frac{d}{dx}\right)^{l+m}(x^2-1)^l \ , \ \quad m\ge0 \ , \\
 P_l^{-m}(x) = \frac{(l-m)!}{(l+m)!} \ P_l^m(x) \ .
\label{Legendre_def}
\end{gather}
Representation in terms of a finite sum
\begin{gather}
\label{Legendre_sum}
\sum_{k=0}^{l-m}\binom{l}{k}\binom{l}{k+m} \ x^k =
\frac{l!}{(l+m)!} \ x^{-\frac{m}{2}} \ (1-x)^l  \ P^m_l\left(\frac{1+x}{1-x}\right) \ .
\end{gather}
Integral representation reads
\begin{equation}
P_l^m(x) = \frac{(l+m)!}{l!} \ \int_{0}^{\pi} \frac{d\phi}{\pi} \
\left ( x + \sqrt{x^2-1} \cos\phi \right )^l \ \cos m\phi \ .
\label{Legendre_integral}
\end{equation}
Derivatives are given by
\begin{gather}
(x^2-1)\frac{d}{dx} \ P^m_l(x)=l x \ P^m_l(x)-(l+m) \ P^m_{l-1}(x) \ ,  \\
\frac{d}{dx}\big[(2\sinh x)^l \ P^m_l(\coth x)\big] = (l+m) \ P^m_{l-1}(\coth x) \ .
\label{Legendre_deriv}
\end{gather}
Asymptotics of the associated Legendre polynomial for $l\to\infty$ and $m$ fixed
\begin{gather}
P_l^m(\coth x) = l^m \ \frac{\cosh \frac{x}{2}}{\sqrt{\pi l}} \
\left(\coth\frac{x}{2}\right)^l \ \left (1+{\cal{O}}(l^{-1}) \right ) \ .
\label{Legendre_asymp}
\end{gather}
If $0<m<(1-\delta)l$, $\delta>0$ and $\alpha = \frac{m}{l}$ one can use instead the  expansion
\begin{gather}
		\label{Legendre_asymp_all_order}
	P^m_l(x)=\exp\Bigg(l\bigg(\alpha(\ln l-1)+(1+\alpha)\ln\left(1+\alpha\right)-\\
	-\ln\left(x-\sqrt{x^2-1+\alpha^2}\right)-\alpha\ln\left(\frac{\alpha x+\sqrt{x^2-1+\alpha^2}}{(1-\alpha)\sqrt{x^2-1}}\right)
	+ {\cal{O}}(l^{-1}) \bigg)\Bigg) \ .  \nonumber
\end{gather}
Finally, the following asymptotic expansion
\begin{gather}
P^{m}_l(\cosh \xi) = l^{m} \ \sqrt{\frac{\xi}{\sinh\xi}} \
I_m\left(\left(l+\frac{1}{2}\right)\xi\right) \ \left(1+{\cal{O}}(l^{-1})\right)
\label{Legendre_asymp_uniform}
\end{gather}
holds uniformly in $\xi\in (0,\infty)$. Here, $I_m(x)$ is the modified Bessel function.

\section{Orthogonal polynomial method for U(N)}
\label{orthog_pol_app}

Here we outline the derivation of the 't Hooft-Veneziano limit for the $U(N)$ model. We use the orthogonal polynomial method \cite{orthog_polynom}.
Starting from the representations (\ref{PF_1dqcd_det_leg}) and (\ref{sun_pl_gen})
let us introduce the following set of variables ($n=N$):
\begin{gather}
	\label{new_var_def}
	c_n = Z(n,N_f),\quad h_n = \frac{c_{n+1}}{c_n},\quad f_n=\frac{h_n}{h_{n-1}} \ , \\
	a_{k} = (2\sinh m)^{N_f} \ \frac{N_f!}{(N_f+k)!} \ P^{k}_{N_f}(\coth m) \ , \\
	P_n(z) = \frac{1}{c_n}\det\left|\begin{array}{c}
		\big\{a_{i-j}\big\}_{i,j\in\overline{0,n},\,i\neq n}  \\
		\big\{z^j\big\}_{j\in\overline{0,n}}
	\end{array}\right|  \ , \\
	w(\theta) = (2\cosh m+2\cos\theta)^{N_f}, \quad
	\int_{-\pi}^\pi \frac{d\theta}{2\pi} \ w(\theta) \  P_{n}(e^{i\theta})P_{n'}(e^{-i\theta})=h_n\delta_{n,n'} \ .
\end{gather}
The free energy can be expressed through coefficients $f_k$ as
\begin{gather}
	\label{fren_un_fk}
F = \frac{1}{N_f}\ln h_0 + \frac{1}{N_f}\sum_{k=1}^N\left(1-\frac{k}{N}\right)\ln f_k \ .
\end{gather}
In the 't Hooft-Veneziano limit this gives
\begin{gather}
\label{fren_un_hv_def}
F = 2\ln\left(2\cosh\frac{m}{2}\right) + \int_0^{1/\kappa} (1-\kappa x)\ln f(x)dx
\end{gather}
and the problem is reduced to the determination of the finite function $f(x)=f_{xN_f}$.
One can use an orthogonality of the defined polynomials to retrieve recursion relation for $f_k$ coefficients. To this end it is convenient to consider the following set of integrals
\begin{gather}
	\int_{-\pi}^\pi \frac{d\theta}{2\pi} \frac{dw}{d\theta}(2\cosh m+2\cos\theta)P_{n-1}(e^{i\theta})P_n(e^{-i\theta}) \ , \\
	\int_{-\pi}^\pi \frac{d\theta}{2\pi} e^{i\theta}\frac{dw}{d\theta}(2\cosh m+2\cos\theta)P_{n}(e^{i\theta})P_n(e^{-i\theta}) \ , \\
	\int_{-\pi}^\pi \frac{d\theta}{2\pi} e^{i\theta}\frac{dw}{d\theta}(2\cosh m+2\cos\theta)P_{n-1}(e^{i\theta})P_n(e^{-i\theta}) \ , \\
	\int_{-\pi}^\pi \frac{d\theta}{2\pi} e^{i\theta}\frac{dw}{d\theta}(2\cosh m+2\cos\theta)P_{n-1}(e^{i\theta})P_{n+1}(e^{-i\theta})
\end{gather}
and calculate them both directly and by parts using expansion coefficients of $zP_n$
\begin{gather}
	zP_n = P_{n+1} + R_nP_n + S_nP_{n-1} + \dots \ \ .
\end{gather}
After long manipulations following \cite{orthog_polynom} one gets a system of recursion relations
\begin{align}
\label{rec1_hv}
&(N_f+n)f_n - (N_f+n+1)S_n - n - \sum_{k=0}^{n-1} \left (  2 S_k
+ R_k^2 + 2\cosh m  \ R_k \right )  = 0  \ , \\
%\end{align}
%\begin{eqnarray}
\label{rec2_hv}
&(N_f+n+2)(S_n+S_{n+1}+R_n^2)+
\sum_{k=0}^{n-1} \left ( R_k(R_k+R_n+2\cosh m)+2S_k \right )  \\
&-N_f	+ 2\cosh m (n+1)R_n + k_{n+1}l_n = 0 \ , \nonumber  \\
%\end{eqnarray}
%\begin{eqnarray}
\label{rec3_hv}
&	(N_f+n+1)(R_{n-1}+R_n)f_n - (1-f_n) ( 2n\cosh m + \sum_{k=0}^{n-2}R_k  )
 - R_{n-1} +k_nl_n  = 0 \ , \\
%\nonumber  \\
%\end{eqnarray}
%\begin{align}
\label{rec4_hv}
&	(N_f+n+1)f_nf_{n+1} - \sum_{k=0}^{n} \left ( R_k(R_k-R_{n} + 2\cosh m)+2S_k \right ) \\
&+ (n+1)(2 R_n\cosh m - 1) + k_nl_{n+1} = 0 \ , \ \ \ \
\mbox{where} \nonumber  \\
&l_n = \frac{(-1)^{n-1}}{c_n}\det\left|a_{i-j}\right|_{i,j\in\overline{0,n},\,i\neq n,\,j \neq 2}\propto N_f \ , \
k_n = \frac{(-1)^{n}}{c_n}\det\left|a_{i-j}\right|_{i,j\in\overline{0,n},\,i\neq n,\,j \neq 1}\propto 1 \ .  \nonumber
\end{align}
Up to terms vanishing in the large $N_f$ limit
\begin{equation}
k_{xN_f}l_{xN_f+i}=N_fT(x)(-1)^i  \ , \
R_{xN_f}=R(x) \ , \ S_{xN_f}=S(x)
\end{equation}
and recursion relations can be replaced by a system of integral equations
\begin{align}
\label{rec5_hv}
&(1+x)f(x) - (1+x)S(x) - x
=\int_0^x \left [ 2S(y)+R^2(y)+2\cosh m R(y) \right ] dy \ , \\
\label{rec6_hv}
&(1+x)(2S(x)+R^2(x))-1+2\cosh m x R(x) - T(x)  \\
&=-\int_0^x\left [ R(y)(R(y)+R(x))+2S(y)+2\cosh m R(y) \right ] dy \ ,
\nonumber \\
\label{rec7_hv}
&7(1+x)2R(x)f(x) - x(1-f(x))2\cosh m + T(x)
= (1-f(x))\int_0^x R(y)dy \ , \\
\label{rec8_hv}
&(1+x)f^2(x) + 2x \cosh m R(x) - T(x) - x  \\
&=\int_0^x dy
\left [ R(y)(R(y)-R(x))+2S(y) + 2\cosh m R(y) \right ] \ . \nonumber
\end{align}

\noindent
The function $T(x)$ can be easily excluded and one finds
\begin{eqnarray}
&& R(x) = 1 - f(x) \ , \ S(x) = f(x) (f(x)-1) \ ,  \\
&& f(x) = 1 \lor f(x) = \cosh^2\frac{m}{2} \ \left(1-\frac{C(m)}{(1+x)^2} \right) \ .
\end{eqnarray}
Taking into account boundary conditions $f(0)=0 \land f(\infty)=1$ implies that $f(x)$
has the following form
\begin{gather}
	f(x) = \begin{cases}
		\cosh^2\frac{m}{2}\left(1-\frac{1}{(1+x)^2}\right), & x<\frac{2}{e^m-1} \ , \\
		1, & x>\frac{2}{e^m-1} \ .
	\end{cases}
\end{gather}
Substituting this result into Eq.(\ref{fren_un_hv_def}) we obtain for the free energy
\begin{gather}
	F = 2\ln\left(2\cosh\frac{m}{2}\right) +\!\!\!\!\!\!\!\!\!\! \int\limits_0^{\min\left\{\frac{1}{\kappa},\frac{2}{e^m-1}\right\}}\!\!\!\!\!\!\!\!\!\!(1-\kappa x)\ln\left(\cosh^2\frac{m}{2}\left(1-\frac{1}{(1+x)^2}\right)\right)dx\\
	=\begin{cases}
		\frac{\xi}{\kappa}\ln\cosh\frac{m}{2}+\frac{(\xi)^2}{2\kappa}
		\ln\xi-\frac{(1+\kappa)^2}{\kappa}\ln(1+\kappa)-\kappa\ln4\kappa \ , 
		& m < \ln\xi \\
		m - \kappa \ln(1-e^{-2m}) \ , & m > \ln\xi
	\end{cases}
\end{gather}
The expectation value of the Polyakov loop can be calculated in a similar manner.
With notations
\begin{gather}
	W(1, N=n) = \frac{e^{-\mu}}{n} \ W_n = \frac{e^{-\mu}}{n} \ \frac{\underset{1\le i,j\le n}{\det}a_{i-j+\delta_{j,n}}}{\underset{1\le i,j\le n}{\det}a_{i-j}} \ , \\
	P_n(z) = z^n - z^{n-1}W_n + \dots \ , \\
	W_{n+1} - W_n = R_n \ , \ W_n = \sum_{k=0}^{n-1}R_k
\end{gather}
and repeating the steps described above one finds in the 't Hooft-Veneziano limit
\begin{gather}
	W(1) = \frac{e^{-\mu}}{N} \ \sum_{k=0}^{N-1}R_k \ \Rightarrow \
	\kappa e^{-\mu}\int_0^{1/\kappa}R(x)dx \ .
 \end{gather}
Using $R(x)=1-f(x)$ and calculating the last integral we get
\begin{gather}
\label{pl_un_hv_fin}
W(1)	=\begin{cases}
	e^{-\mu}\left(1-\frac{\cosh^2\frac{m}{2}}{1+\kappa}\right) \ , & \frac{1}{\kappa}<\frac{2}{e^m-1} \ , \\
	\kappa e^{-\mu-m} \ , & \frac{1}{\kappa}>\frac{2}{e^m-1} \ .
\end{cases}
\end{gather}

\section{Coefficients $B_{N,Nf} (r, q)$ and  $C_k(u,\kappa)$}
\label{coeff_ck_app}

In this Appendix we list exact expressions for the first coefficients $B_{N,Nf} (r, q)$
appearing in the representation (\ref{sun_full_pf1})-(\ref{EN_Nf_coeff})  for the $SU(N)$ partition function. These expressions are valid for all values of $N$ and $N_f$.
\begin{equation*}
	B_{N,N_f}(0,q) \ = \  1  \ ,
%\label{QSUN_r0}
\end{equation*}
\begin{equation*}
	B_{N,N_f}(1,q) = N N_f \ \frac{N_f - q}{N + q} \ ,
%\label{C1u}
\end{equation*}
\begin{eqnarray*}
B_{N,N_f}(2,q) =  \frac{N N_f (N_f-q) \left[\left(N^2-1\right)
\left(N_f^2+1\right)-q  (N_f-N+q) (N N_f-1) \right]}{2 (N+q-1) (N+q) (N+q+1)} \ ,
%\label{QSUN_r2}
\end{eqnarray*}
\begin{eqnarray*}
&&B_{N,N_f}(3,q) \ = \ \frac{ N N_f (N_f-q) }{6 y(N^2 y^2-1)(N^2 y^2-4)}
[( N^2-1)( N^2 -4)(1 + N_f^2)(2 + N_f^2)  \\
&&+  q[4 N_f (3 + N_f^2)- N^4 N_f (3 + 2 N_f^2) + N^2 N_f (3 + 4 N_f^2) + 2 N^3 (2 + 3 N_f^2 + N_f^4)  \\
&&- N (8 + 19 N_f^2 + 3 N_f^4)] + q^2 ( N N_f-2) [5 + N^3 N_f + 2 N_f^2
+ N N_f ( N_f^2-1) \\
&&- 4 N^2 ( N_f^2+1)]
+ q^3(2 N - 2 N_f + q) ( N N_f -2) ( N N_f-1) ]  \ , \ y = 1+q/N \ .
%\label{QSUN_r3}
\end{eqnarray*}
Finally, we list below first coefficients $C_k(u,\kappa)$ appearing in the representation of the $SU(N)$ partition function (\ref{sun_pf_HVlimit}) in the 't Hooft-Veneziano limit.
\begin{equation}
	C_1(u,\kappa) =  \frac{\kappa - u}{1 + u} \ ,
	\label{C1_sun}
\end{equation}
\begin{eqnarray}
	C_2(u,\kappa) =  - \frac{  (\kappa - u) }{2 (1 + u)^4}[\kappa^2 u + \kappa u (2 + u)- (1 + u) (1 + u + u^2) ] \ ,
\label{C2_sun}
\end{eqnarray}
\begin{eqnarray}
	C_3(u,\kappa) = - \frac{ (\kappa - u)}{3 (1 + u)^7} [\kappa^4 u(2 - 2u) + \kappa^3 u (2 - 2u)(3 + u)   \nonumber  \\
\label{C3_sun}
+ 2 \kappa^2 u(4  + u + u^3)  +  \kappa u(6  + 10 u + 10 u^2 + 8 u^3 + 2 u^4)  \\
	-(1 + u)^2 (1 + 2 u + 4 u^2 + 2 u^3 + u^4) ] \nonumber  \ ,
\end{eqnarray}
\begin{eqnarray}
	C_4 (u,\kappa) = -  \frac{ (\kappa- u)}{4 (1 + u)^{10}} [ \kappa^6 u (5 - 14 u + 5 u^2)  \nonumber  \\
	+ \kappa^5 u (4 + u) (5 - 14 u + 5 u^2)   \nonumber  \\
	+ \kappa^4 u (36 - 68 u - 11 u^2 + 14 u^3 - 7 u^4)  \nonumber  \\
\label{C4_sun}
	+ \kappa^3 u (38 - 32 u - 18 u^2 + 13 u^3 - 22 u^4 - 7 u^5)    \\
	+ \kappa^2 u (1 + u) (26 - 8 u + 30 u^2 + 8 x^3 - 5 u^4 + 3 u^5)   \nonumber  \\
	+ \kappa u (1 + u)^2 (12 + 10 u + 34 u^2 + 16 u^3 + 12 u^4 + 3 u^5)  \nonumber  \\
	-(1 + u)^3 (1 + 3 u + 9 u^2 + 9 u^3 + 9 u^4 + 3 u^5 + u^6)] \ ,  \nonumber
\end{eqnarray}
\begin{eqnarray}
	C_5 (u,\kappa) = -  \frac{  (\kappa -u )}{5 (1 + u)^{13} } \Big[  \kappa^8 u (14 - 74 u + 74 u^2 - 14u^3)  \nonumber  \\
	+ \kappa^7 u (5+u)(14 u - 74 u^2 + 74 u^3 - 14u^4 )   \nonumber \\
	+ 2 \kappa^6 u (80 - 365 u + 222 u^2 + 68 u^3 - 58 u^4 + 13 u^5)  \nonumber \\
	+ 2 \kappa^5 u (110 - 410 u + 125 u^2 + 112 u^3 - 172 u^4 + 22 u^5 + 13 u^6)   \nonumber \\
\label{C5_sun}
+2 \kappa^4 u (101 - 257 u - 17 u^2 + 66 u^3 - 249 u^4 - 25u ^5 + 29 u^6 - 8 u^7)    \\
	+ 2 \kappa^3 u (65 - 62 u - 16 u^2 + 44 u^3 - 177 u^4 - 88 u^5 - 4 u^6 -  34 u^7 - 8 u^8 ) \nonumber \\
	+ 2\kappa^2  u (1 + u)^2 (30 - 25 u + 98 u^2 - 27 u^3 + 30 u^4 + 10 u^5 -
	8 u^6 + 2 u^7)  \nonumber \\
	+ 2  \kappa u (1 + u)^3 (10 + 10 u + 55 u^2 + 33 u^3 + 55 u^4 + 20 u^5 +10 u^6
	+ 2 u^7)  \nonumber \\
- (1 + u)^4 (1 + 4 u + 16 u^2 + 24 u^3 + 36 u^4 + 24 u^5 + 16 u^6 + 4 u^7 + u^8) \Big] \ .  \nonumber
\end{eqnarray}


\begin{thebibliography}{99}
	
	\bibitem{baxter_book} 
	R.~J.~Baxter, 
	Exactly Solved Models in Statistical Mechanics,
	Dover Publications, Illustrated edition, 2008.
	%
	\bibitem{wilson_lgt} 
	K.~G.~Wilson, 
	Confinement of quarks, 
	Phys.Rev. D {\bf 10} (1974) 2445, 
	DOI:10.1103/PhysRevD.10.2445.
	%
	\bibitem{gross_witten} 
	D.~J.~Gross, E.~Witten, 
	Possible Third Order Phase Transition in the Large N Lattice Gauge Theory,
	Phys.Rev. D {\bf 21} (1980) 446, 
	DOI:10.1103/PhysRevD.21.446.
	%
	\bibitem{wadia} 
	S.~R.~Wadia, 
	N=Infinity Phase Transition in a Class of Exactly Soluble Model Lattice Gauge Theories, 
	Phys.Lett. B {\bf 93} (1980) 403, 
	DOI:10.1016/0370-2693(80)90353-6.
	%
	\bibitem{un_vs_sun} 
	O.~Borisenko, V.~Chelnokov, S.~Voloshyn, 
	The large N limit of SU(N) integrals in lattice models, 
	Nucl.Phys B {\bf 960} (2020) 115177,
	DOI:10.1016/j.nuclphysb.2020.115177, 
	[arXiv:2008.00773 [hep-lat]].
	%
	\bibitem{un_group_integral} 
	G.~Akemann, N.~Aygün, T.~R.~Würfel, 
	Generalised unitary group integrals of Ingham-Siegel and Fisher-Hartwig type, 
	J. Math. Phys. {\bf 65} (2024) 023501, 
	DOI:10.1063/5.0160923, 
	[arXiv:2305.19852 [math-ph]].
	%
	\bibitem{spin_flux1} 
	C.~Gattringer, 
	Flux representation of an effective Polyakov loop model for QCD thermodynamics,
	Nucl.Phys. B {\bf 850} (2011) 242, 
	DOI:10.1016/j.nuclphysb.2011.04.018, 
	[arXiv:1104.2503 [hep-lat]]. 
	%
	\bibitem{duals_lgt} 
	O.~Borisenko, V.~Chelnokov, S.~Voloshyn, 
	Dual formulations of Polyakov loop lattice models, 
	Phys.Rev. D {\bf 102} (2020) 014502,
	DOI:10.1103/PhysRevD.102.014502, 
	[arXiv:2005.11073 [hep-lat]].
	%
	\bibitem{dual_abelian} 
	O.~Borisenko, V.~Chelnokov, S.~Voloshyn, P.~Yefanov,
	Duals of lattice Abelian models with static determinant at finite density,
	Phys.Lett. B {\bf 827}  (2022) 137000, 
	DOI:10.1016/j.physletb.2022.137000, 
	[arXiv:2112.06002 [hep-lat]].
	%
	\bibitem{philipsen_14} 
	J.~Langelage, M.~Neuman and O.~Philipsen, 
	Heavy dense QCD and nuclear matter from an effective lattice theory, 
	JHEP {\bf 09} (2014) 131, 
	DOI:10.1007/JHEP09(2014)131, 
	[arXiv:1403.4162 [hep-lat]].
	%
	\bibitem{philipsen_quarkyon_19} 
	O.~Philipsen, J.~Scheunert, 
	QCD in the heavy dense regime for general Nc: On the existence of quarkyonic matter, 
	JHEP {\bf 11} (2019) 022,
	DOI:10.1007/JHEP11(2019)022, 
	[arXiv:1908.03136 [hep-lat]].
	%
	\bibitem{hv_lat21} 
	O.~Borisenko, V.~Chelnokov, S.~Voloshyn, 
	The 't Hooft-Veneziano limit of the Polyakov loop models, 
	PoS \textbf{LATTICE2021} 453, 
	DOI:10.22323/1.396.0453,
	[arXiv:2111.07103 [hep-lat]].
	%
	\bibitem{hv_pl21} 
	O.~Borisenko, V.~Chelnokov, S.~Voloshyn, 
	The Polyakov loop models in the large $N$ limit: Phase diagram at finite density, 
	Phys.Rev. D {\bf 105} (2022) 014501,
	DOI:10.1103/PhysRevD.105.014501, 
	[arXiv:2111.00474 [hep-lat]].
	%
	\bibitem{sign_langevin} 
	G.~Aarts, K.~Splittorff, 
	Degenerate distributions in complex Langevin dynamics: one-dimensional QCD at finite chemical potential, 
	JHEP {\bf 08} (2010) 017,
	DOI:10.1007/JHEP08(2010)017, 
	[arXiv:1006.0332 [hep-lat]].
	%
	\bibitem{sign_subset_13} 
	J.~Bloch, F.~Bruckmann, T.~Wettig, 
	Subset method for one-dimensional QCD,
	JHEP {\bf 10} (2013) 140, 
	DOI:10.1007/JHEP10(2013)140, 
	[arXiv:1307.1416 [hep-lat]].
	%
	\bibitem{sign_num} 
	A.~Ammon, T.~Hartung, K.~Jansen, H.~Leövey, J.~Volmer, 
	Overcoming the sign
	problem in 1-dimensional QCD by new integration rules with polynomial exactness,
	Phys.Rev. D {\bf 94} (2016) 114508, 
	DOI:10.1103/PhysRevD.94.114508,
	[arXiv:1607.05027 [hep-lat]].
	%
	\bibitem{sign_thimble} 
	F.~Di Renzo, G.~Eruzzi, 
	One-dimensional QCD in thimble regularization,
	Phys.Rev. D {\bf 97} (2018) 014503, 
	DOI:10.1103/PhysRevD.97.014503, 
	[arXiv:1709.10468 [hep-lat]].
	%
	\bibitem{ilgenfritz_85} 
	E.-M.~Ilgenfritz, J.~Kripfganz, 
	Dynamical Fermions at Nonzero Chemical Potential and Temperature: Mean Field Approach,
	Z. Phys. C {\bf 29} (1985) 79,
	DOI:10.1007/BF01571383.
	%
	\bibitem{Hooft_74} 
	G.~t' Hooft, 
	A planar diagram theory for strong interactions,
	Nucl.Phys. B {\bf 72} (1974) 461, 
	DOI:10.1016/0550-3213(74)90154-0.
	%
	\bibitem{Veneziano_76} 
	G.~Veneziano,  
	Some aspects of a unified approach to gauge, dual and Gribov theories, Nucl.Phys. B {\bf 117} (1976) 519, 
	DOI:10.1016/0550-3213(76)90412-0.
	%
	\bibitem{bilic_88} 
	N.~Bili\^{c}, K.~Demeterfi, 
	One-dimensional QCD with finite chemical potential, 
	Phys.Lett. B {\bf 212} (1988) 83, 
	DOI:10.1016/0370-2693(88)91240-3.
	%
	\bibitem{1d_qcd_07} 
	L.~Ravagli, J.J.M.~Verbaarschot, 
	QCD in One Dimension at Nonzero Chemical Potential, 
	Phys.Rev. D {\bf 76} (2007) 054506,
	DOI:10.1103/PhysRevD.76.054506, 
	[arXiv:0704.1111 [hep-th]].
	%
	\bibitem{qcd_cont_sphere} 
	S.~Hands, T.~J.~ Hollowood, J.~C.~Myers, 
	QCD with Chemical Potential in a Small Hyperspherical Box, 
	JHEP 2010, 86 (2010),
	DOI:10.1007/JHEP07(2010)086, 
	[arXiv:1003.5813v3 [hep-th]]. 
	%
	\bibitem{russo_20} 
	J.~G.~Russo, 
	Phases of unitary matrix models and lattice QCD2, 
	Phys.Rev. D {\bf 102} (2020) 105019, 
	DOI:10.1103/PhysRevD.102.105019,
	[arXiv:2010.02950 [hep-th]]. 
	%
	\bibitem{russo_tiers_20} 
	J.~G.~Russo, M.~Tierz,  
	Multiple phases in a generalized Gross-Witten-Wadia matrix model,
	JHEP, 2020, 81 (2020), 
	DOI:10.1007/JHEP09(2020)081,
	[arXiv:2007.08515 [hep-th]].
	%
	\bibitem{santilli_20} 
	L.~Santilli, M.~Tierz, 
	Exact equivalences and phase discrepancies between random matrix ensembles, 
	J. Stat. Mech. (2020) 083107,  
	DOI:10.1088/1742-5468/aba594,
	[arXiv:2003.10475 [math-ph]]. 
	%
	\bibitem{determinant_calc} 
	C.~Krattenthaler, 
	Advanced Determinant Calculus,
	In: Foata D., Han GN. (eds) The Andrews Festschrift. Springer, Berlin, Heidelberg,
	DOI:10.1007/978-3-642-56513-7\_17, 
	[arXiv:math/9902004v3 [math.CO]].
	%
	\bibitem{plane_part} 
	N.~Bogoliubov, C.~Malyshev, 
	The asymptotics of plane partitions with fixed volumes of diagonal parts, 
	Zap. Nauchn. Sem. POMI {\bf 487} (2019) 68,
	DOI:10.1007/s10958-021-05495-z.
	%
	\bibitem{sun_integral} 
	O.~Borisenko, V.~Chelnokov, S.~Voloshyn, 
	SU(N) polynomial integrals and some applications, 
	Rep. Mathematical Physics, V85 (2020) 129, 
	DOI:10.1016/S0034-4877(20)30015-X, 
	[arXiv:1812.06069 [hep-lat]].
	%
	\bibitem{roberge_weiss}
	A.~Roberge, N.~Weiss, 
	Gauge Theories With Imaginary Chemical Potential and the Phases of QCD, Nucl.Phys. B {\bf 275} (1986) 734,
	DOI:10.1016/0550-3213(86)90582-1.
	%
	\bibitem{imaginary_isospin_lattice} 
	A.~Chabane, G.~Endrődi, 
	Roberge-Weiss transitions at imaginary isospin chemical potential,
	PoS \textbf{LATTICE2021} 097,
	DOI:10.22323/1.396.0097, 
	[arXiv:2110.13536 [hep-lat]].
	%
	\bibitem{imaginary_isospin} 
	B.~Brandt, A.~Chabane, V.~Chelnokov, F.~Cuteri, G.~Endrődi, 
	Light Roberge-Weiss tricritical endpoint at imaginary isospin and baryon chemical potential,
	Phys. Rev. D {\bf 109} (2024) 034515,
	DOI:10.1103/PhysRevD.109.034515,
	[arXiv:2207.10117 [hep-lat]].
	%
	\bibitem{orthog_polynom} 
	Y.Y.~Goldschmidt, 
	1/N expansion in two-dimensional lattice gauge theory, 
	J.~Math.~Phys. {\bf 21} (1980) 1842, 
	DOI:10.1063/1.524600.
	
	
\end{thebibliography}
\end{document}